\pdfoutput=1
\documentclass[11pt,twoside,a4paper,cmspaper,final,collab]{cms-tdr}
\def\svnVersion{5cd7060}\def\svnDate{2025/12/24}\def\cmsCernNoTag{CERN-EP-2025-233}\def\cmsCernDate{\today}\def\cmsMessage{Published in the Journal of Instrumentation as \href{http://dx.doi.org/10.1088/1748-0221/20/12/P12032}{\doi{10.1088/1748-0221/20/12/P12032}.}}
\begin{document}\cmsNoteHeader{TAU-24-001}

\newcommand{\PNfix}    {\hspace{-.04em}}
\newcommand{\mumu}     {{\ensuremath{\PGm\PNfix\PGm}}\xspace}
\newcommand{\etau}     {{\ensuremath{\Pe\PNfix\PGt_\mathrm{h}}}\xspace}
\newcommand{\mutau}    {{\ensuremath{\PGm\PNfix\PGt_\mathrm{h}}}\xspace}
\newcommand{\mvis}     {{\ensuremath{m_\text{vis}}}\xspace}
\newcommand{\mWst}     {{\ensuremath{m_{\PWst}}}\xspace}
\newcommand{\FF}       {{\ensuremath{F_{F}}}\xspace}
\newcommand{\Wjets}    {\ensuremath{\PW+\text{jets}}\xspace}
\newcommand{\Zjets}    {\ensuremath{\PZ/\PGg^{*}+\text{jets}}\xspace}
\newcommand{\PWst}     {\ensuremath{\PW^{\ast}}\xspace}
\newcommand{\hpm}      {\ensuremath{\mathrm{h}^{\pm}}\xspace}
\newcommand{\oneProngZeroPizero}  {\ensuremath{\mathrm{h}^{\pm}}\xspace}
\newcommand{\oneProngOnePizero}   {\ensuremath{\mathrm{h}^{\pm}\PGpz}\xspace}
\newcommand{\oneProngTwoPizero}   {\ensuremath{\mathrm{h}^{\pm}\PGpz\PGpz}\xspace}
\newcommand{\threeProngZeroPizero}{\ensuremath{\mathrm{h}^{\pm}\mathrm{h}^{\mp}\mathrm{h}^{\pm}}\xspace}
\newcommand{\threeProngOnePizero} {\ensuremath{\mathrm{h}^{\pm}\mathrm{h}^{\mp}\mathrm{h}^{\pm}\PGpz}\xspace}
\newcommand{\twoProngZeroPizero}  {\ensuremath{\mathrm{h}^{\pm}\mathrm{h}^{\pm/\mp}}\xspace}
\newcommand{\twoProngOnePizero}   {\ensuremath{\mathrm{h}^{\pm}\mathrm{h}^{\pm/\mp}\PGpz}\xspace}
\newcommand{\DeepTau}{\textsc{DeepTau}\xspace}
\newcommand{\Djet}   {\ensuremath{D_\text{jet}}\xspace}
\newcommand{\De}     {\ensuremath{D_\Pe}\xspace}
\newcommand{\Dm}     {\ensuremath{D_\PGm}\xspace}
\newcommand{\yadv}   {\ensuremath{y_\text{adv}}}
\newcommand{\Ladv}   {\ensuremath{L_\text{adv}}\xspace}
\newcommand{\Lclass} {\ensuremath{L_\text{class}}\xspace}
\newcommand{\pass}   {\textsc{PASS}\xspace}
\newcommand{\fail}   {\textsc{FAIL}\xspace}
\newlength\cmsTabSkip\setlength{\cmsTabSkip}{1ex}
\newlength\cmsFigWidth\setlength\cmsFigWidth{0.45\textwidth}

\cmsNoteHeader{TAU-24-001}

\title{Identification of tau leptons using a convolutional neural network with domain adaptation}

\author*[cern]{The CMS Collaboration}

\date{\today}

\abstract{
A tau lepton identification algorithm, \DeepTau, based on convolutional neural network techniques, has been developed in the CMS experiment to discriminate reconstructed hadronic decays of tau leptons (\tauh) from quark or gluon jets and electrons and muons that are misreconstructed as \tauh candidates. The latest version of this algorithm, v2.5, includes domain adaptation by backpropagation, a technique that reduces discrepancies between collision data and simulation in the region with the highest purity of genuine \tauh candidates. Additionally, a refined training workflow improves classification performance with respect to the previous version of the algorithm, with a reduction of 30--50\% in the probability for quark and gluon jets to be misidentified as \tauh candidates for given reconstruction and identification efficiencies. This paper presents the novel improvements introduced in the \DeepTau algorithm and evaluates its performance in LHC proton-proton collision data at $\sqrt{s}=13$ and 13.6\TeV collected in 2018 and 2022 with integrated luminosities of 60 and 35\fbinv, respectively. Techniques to calibrate the performance of the \tauh identification algorithm in simulation with respect to its measured performance in real data are presented, together with a subset of results among those measured for use in CMS physics analyses.
}

\hypersetup{%
  pdfauthor={CMS Collaboration},%
  pdftitle={Identification of tau leptons using a convolutional neural network with domain adaptation},%
  pdfsubject={CMS},%
  pdfkeywords={CMS,tau,hadronic,identification,convolutional,backpropagation,domain adaptation,energy scale}
}

\maketitle

\section{Introduction} \label{sec:introduction}

Tau leptons are the heaviest leptons in the standard model and therefore play an important role in Higgs physics~\cite{Sirunyan:2017khh,Sirunyan:2017djm,ATLAS:2018ynr,ATLAS:2018uni,ATLAS:2020evk}, where the scalar couplings to fermions are proportional to their masses. In addition, tau leptons are also crucial for physics analyses searching for particles beyond the standard model (\PWpr, \PZpr, leptoquarks, Higgs-like bosons, tau lepton supersymmetric partners, etc.) and studying anomalous couplings~\cite{Sirunyan:2018zut,ATLAS:2018gfm,Sirunyan:2018pzn,Sirunyan:2019xjg,Sirunyan:2019shc,CMS:2019hvr,Sirunyan:2019hkq,ATLAS:2020zms,CMS:2019eln,Sirunyan:2018vhk,ATLAS:2019qpq,cp_paper}. Since tau leptons decay to hadrons and neutrinos in about 65\% of the cases, these analyses require the efficient reconstruction and identification of hadronic tau lepton decays (\tauh).

In the CMS experiment~\cite{CMS:2008xjf,CMS:2023gfb}, hadronic decays of tau leptons are reconstructed with the hadrons-plus-strips (HPS) algorithm~\cite{HPS1}, which identifies the decay products and assigns a decay mode based on the number of charged hadrons and neutral pion candidates.
The primary challenge is the distinction between genuine \tauh decays and the major background from quark or gluon jets, which are copiously produced by many processes in proton-proton ($\Pp\Pp$) collisions.
In addition, light charged leptons can also be misidentified as \tauh decays. Electrons can mimic the signature of charged hadrons and bremsstrahlung radiation can be seen as the product of neutral pion decays. Muons have a chance to be reconstructed as \tauh candidates as well, primarily in the decay mode with only one charged hadron, although with a much lower probability.
To address these challenges, machine-learning techniques are exploited to build the \DeepTau identification algorithm~\cite{DeepTau_v2p1} that can efficiently identify genuine \tauh candidates and reduce the background of jets, electrons, and muons misidentified as \tauh candidates by the HPS algorithm.

In this paper, we describe the new version of the \DeepTau algorithm, v2.5, which achieves improved classification performance with respect to the previous version, v2.1~\cite{DeepTau_v2p1}, and includes domain adaptation by backpropagation~\cite{domainadapt} to reduce data-to-simulation performance discrepancies, originating from potential inaccuracies in simulation. While the classification network architecture has remained largely unchanged, the training dataset has been updated and the hyperparameters have been fine-tuned, leading to a reduction of 30--50\% in the probability for quark and gluon jets to be misidentified as \tauh candidates.
In order to implement domain adaptation, a new adversarial subnetwork has been added and trained on a mixture of $\Pp\Pp$ collision data and simulation with the purpose of discriminating between the two datasets. 
The goal of the algorithm tuning is to maximize the classification performance while minimizing the discrepancies in performance between data and simulation.

The new identification algorithm was trained on real $\Pp\Pp$ collision data collected in 2018 and on simulated events produced using the 2018 data-taking conditions. The algorithm was then validated and calibrated using simulation and data collected by the CMS detector in 2018 at a 13\TeV centre-of-mass energy and in 2022 at 13.6\TeV, corresponding to integrated luminosities of 60 and 35\fbinv, respectively. These years had an average of 37 and 46 $\Pp\Pp$ interactions per bunch crossing. The algorithm has been introduced for use in CMS physics analyses using data recorded from 2022 onwards.

The paper is structured as follows. After an overview of the CMS detector, event samples, and event reconstruction in Sections~\ref{sec:detector}--\ref{sec:objects}, the new version of the \DeepTau identification algorithm is described, and its performance is presented in Section~\ref{sec:deeptau}. Section~\ref{sec:calibration} gives an overview of the measurements of \tauh reconstruction and identification performance. The paper concludes with a summary in Section 7.

\section{The CMS detector} \label{sec:detector}

The CMS apparatus~\cite{CMS:2008xjf,CMS:2023gfb} is a multipurpose, nearly hermetic detector, designed to trigger on~\cite{CMS:2020cmk,CMS:2016ngn,CMS:2024aqx} and identify~\cite{CMS:2020uim,CMS:2018rym,CMS:2014pgm} electrons, muons, photons, and (charged and neutral) hadrons. Its central feature is a superconducting solenoid of 6\unit{m} internal diameter, providing a magnetic field of 3.8\unit{T}. Within the solenoid volume are a silicon pixel and strip tracker, a lead tungstate crystal electromagnetic calorimeter (ECAL), and a brass and scintillator hadron calorimeter (HCAL), each composed of a barrel and two endcap sections. Forward calorimeters extend the pseudorapidity coverage provided by the barrel and endcap detectors. Muons are reconstructed using gas-ionization detectors embedded in the steel flux-return yoke outside the solenoid.

Events of interest are selected using a two-tiered trigger system. The first level (L1), composed of custom hardware processors, uses information from the calorimeters and muon detectors to select events at a rate of around 100\unit{kHz} within a fixed latency of about 4\mus~\cite{CMS:2020cmk}. The second level, known as the high-level trigger (HLT), consists of a farm of processors running a version of the full event reconstruction software optimized for fast processing, and reduces the event rate to a few kHz before data storage~\cite{CMS:2016ngn,CMS:2024aqx}.

A more detailed description of the CMS detector, together with a definition of the coordinate system used and the relevant kinematic variables, can be found in Refs.~\cite{CMS:2008xjf,CMS:2023gfb}.

\section{Simulated event samples} \label{sec:samples}

 There are two sets of Monte Carlo simulated event samples used in this analysis. One is used for performance studies, whereas the other as input for the neural network training. 
Unless explicitly mentioned, the simulated event samples include decays to all lepton flavours, \ie to electrons, muons, and tau leptons. The samples listed in the following are produced separately for conditions corresponding to the 2018 and 2022 data-taking periods.

The standard model processes considered for the measurements are \PZ (or $\PGg^{*}$) and \PW boson production in association with jets, denoted as ``\Zjets'' and ``\Wjets'', diboson ($\PW\PW$, $\PW\PZ$, $\PZ\PZ$) production, pair production of top quarks ($\ttbar$), and single top quark production. Additionally, the $\PWst\to\PGt\PGnGt$ process, where a highly virtual \PW boson (\PWst, $m_{\PWst}> 200\GeV$) decays in a \PGt lepton and neutrino, is used for high-\pt measurements.
The \Zjets and \Wjets processes are simulated using the \MGvATNLO~\cite{aMCatNLO} generator v2.6.1 at leading order (LO) precision with the MLM jet matching and merging scheme~\cite{MLM}. Additional \Zjets samples generated with \POWHEG~\cite{POWHEG0,POWHEG1,POWHEG2} at next-to-leading order (NLO) precision are used for a subset of our measurements.    
The \MGvATNLO generator is employed for diboson production simulated at NLO precision in perturbative quantum chromodynamics (QCD) with the FxFx jet matching and merging scheme~\cite{FxFx}, whereas \POWHEG v2 is used for \ttbar~\cite{POWHEG_TT,POWHEG_TT2} and single top quark~\cite{POWHEG_ST_st,POWHEG_ST_tW} production at NLO precision.
The LO \PYTHIA 8.230~\cite{Pythia8_2015} event generator is used for $\PWst\to\PGt\PGnGt$ simulation.
The \Zjets, \ttbar, and single top quark processes are normalized using cross sections computed at next-to-next-to-leading order (NNLO) in perturbative QCD~\cite{sigma_DY,sigma_TT,sigma_ST}.

Additional samples are generated only for the neural network training. Events composed of jets produced through the strong interaction, referred to as QCD multijet events, are generated at LO using \MGvATNLO as well as \PYTHIA. The \PYTHIA event generator is also used to produce event samples of heavy gauge bosons ($\PZpr \to \Pe\Pe$), with the mass of the boson ranging from 1 to 5\TeV. The production of a 125\GeV Higgs boson (\PH) via gluon-gluon fusion, vector-boson fusion and associated production with a vector boson, and with the Higgs boson decaying to tau leptons ($\PH\to\PGt\PGt$) is simulated with the \POWHEG generator. Finally, a sample of events with a single tau lepton is simulated with \PYTHIA, where the pseudorapidity $\eta$ and the transverse momentum \pt of the tau lepton are uniformly distributed in the $-2.5<\eta<2.5$ and $15<\pt<3000\GeV$ ranges. This ``\PGt gun'' sample includes the tau lepton decay, but no $\Pp\Pp$ interaction.

The event generators are interfaced with \PYTHIA to model the parton showering and fragmentation, as well as the decay of the tau lepton. The \PYTHIA parameters affecting the description of the underlying event are set to the {CP5} tune~\cite{CUETP8M1}. 
For the $\PZpr \to \Pe\Pe$ and $\PWst\to\PGt\PGnGt$ samples, \TAUOLA~\cite{Davidson:2010rw} is used instead to simulate tau lepton decays.
Generated events are processed through a simulation of the CMS detector based on \GEANTfour~\cite{GEANT4} and are reconstructed with the same algorithms as those used for recorded data.
The simulated samples include additional $\Pp\Pp$ interactions from the same or nearby bunch crossings, referred to as ``pileup''~\cite{CMS:2020ebo}.
The effect of pileup is taken into account by generating concurrent inelastic collision events with \PYTHIA. The simulated events are weighted such that the distribution of the number of pileup interactions matches that in recorded data.

\section{Event reconstruction} \label{sec:objects}

The reconstruction of observed and simulated events relies on the particle-flow (PF) algorithm~\cite{PF2017}, which optimally combines information from the various elements of the CMS detector to reconstruct and identify the particles emerging from the $\Pp\Pp$ collisions: charged and neutral hadrons, photons, muons, and electrons. The energy of photons is obtained from the ECAL measurement. The energy of electrons is determined from a combination of the electron momentum at the primary interaction vertex (PV) as determined by the tracker, the energy of the corresponding ECAL cluster, and the energy sum of all bremsstrahlung photons spatially compatible with originating from the electron track. The energy of muons is obtained from the curvature of the corresponding track. The energy of charged hadrons is determined from a combination of their momentum measured in the tracker and the matching ECAL and HCAL energy deposits, corrected for the response function of the calorimeters to hadronic showers. Finally, the energy of neutral hadrons is obtained from the corresponding corrected ECAL and HCAL energies.

For each event, hadronic jets are clustered from all the PF candidates using the infrared- and collinear-safe anti-\kt algorithm~\cite{antikT,fastjet1,fastjet2} with a distance parameter of 0.4. Jet momentum is determined as the vectorial sum of all particle momenta in the jet, and is found from simulation to be, on average, within 5 to 10\% of the true momentum over the entire \pt spectrum and detector acceptance. Pileup can contribute additional tracks and calorimetric energy depositions, increasing the apparent jet momentum. To mitigate this effect, tracks identified to be originating from pileup vertices are discarded, and an offset correction is applied to correct for remaining contributions~\cite{CMS:2020ebo}.

Jet energy corrections are derived from simulation studies so that the average measured energy of jets becomes identical to that of particle level jets. In situ measurements of the momentum balance in dijet, $\text{photon} + \text{jet}$, \Zjets, and multijet events are used to determine any residual differences between the jet energy scale in data and in simulation, and appropriate corrections are made~\cite{CMS:2016lmd}. Additional selection criteria are applied to each jet to remove jets potentially dominated by instrumental effects or reconstruction failures.

The missing transverse momentum vector \ptvecmiss is computed as the negative vector sum of the transverse momenta of all the PF candidates in an event, and its magnitude is denoted as \ptmiss~\cite{CMS:2019ctu}. The \ptvecmiss is modified to account for corrections to the energy scale of the reconstructed jets in the event.

Muons are measured in the pseudorapidity range $\abs{\eta} < 2.4$, with detection planes made using three technologies: drift tubes, cathode strip chambers, and resistive-plate chambers. The single-muon trigger efficiency exceeds 90\% over the full $\eta$ range, and the efficiency to reconstruct and identify muons is greater than 96\%. Matching muons to tracks measured in the silicon tracker results in a relative transverse momentum resolution, for muons with \pt up to 100\GeV, of 1\% in the barrel and 3\% in the endcaps. The \pt resolution in the barrel is better than 7\% for muons with \pt up to 1\TeV~\cite{CMS:2018rym}.

The electron momentum is estimated by combining the energy measurement in the ECAL with the momentum measurement in the tracker. The momentum resolution for electrons with $\pt \approx 45\GeV$ from $\PZ \to \Pe \Pe$ decays ranges from 1.6 to 5\%. It is generally better in the barrel region than in the endcaps, and also depends on the bremsstrahlung energy emitted by the electron as it traverses the material in front of the ECAL~\cite{CMS:2020uim,CMS-DP-2020-021}.

\begin{figure}[bth!]
  \centering
  \includegraphics[width=1\textwidth]{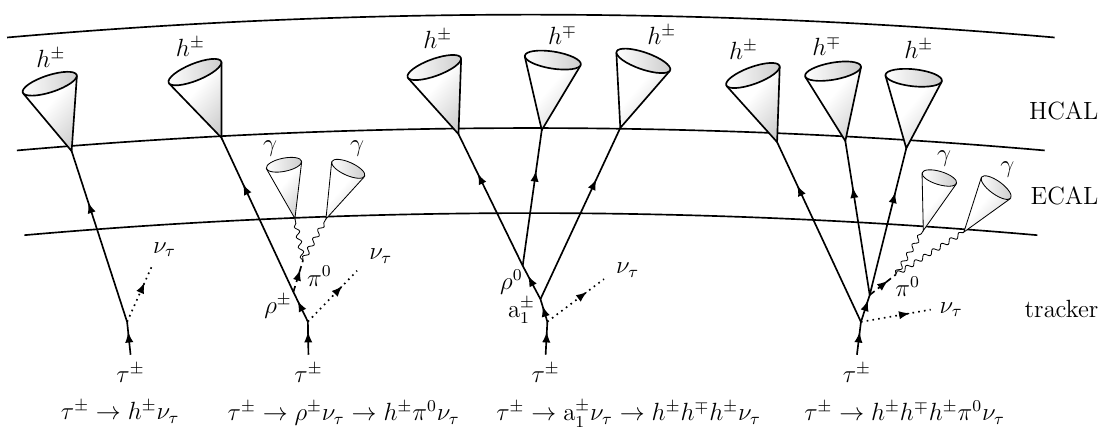}
  \caption{Schematic illustration of the signatures of the \oneProngZeroPizero, \oneProngOnePizero, \threeProngZeroPizero, and \threeProngOnePizero decay modes of the tau lepton in the CMS detector. Charged hadrons are reconstructed by the PF algorithm by matching tracks with energy deposits in the ECAL and HCAL, whereas the HPS algorithm aims to reconstruct each $\PGpz\to\PGg\PGg$ decay as a single ``strip'' of energy clusters in ECAL.}
  \label{fig:HPS_dm}
\end{figure}

The \tauh candidates are reconstructed with the HPS algorithm~\cite{HPS2,DeepTau_v2p1}, which is seeded with anti-\kt jets. The algorithm uses the PF candidates near the seeding jet direction of flight to reconstruct the neutral pions that are present in most \tauh decays. The high probability for photons originating from $\PGpz\to\PGg\PGg$ decays to convert to $\EE$ pairs is accounted for by collecting photons and electrons into clusters (``strips''). The \tauh candidates are then formed by combining the strips with the charged-particles (``prongs'') present in the seeding jet and its immediate surroundings. Based on the observed number of strips and charged particles, each \tauh candidate is assigned to one of the following decay modes (the main ones are illustrated in Fig.~\ref{fig:HPS_dm}):
\begin{itemize}
    \item a single charged particle without any strips, \oneProngZeroPizero;
    \item combination of one charged particle and one strip, \oneProngOnePizero;
    \item combination of one charged particle with two strips, \oneProngTwoPizero;
    \item combination of two charged particles without any strips, \twoProngZeroPizero;
    \item combination of two charged particles and one strip, \twoProngOnePizero;
    \item combination of three charged particles, \threeProngZeroPizero; 
    \item combination of three charged particles and a strip, \threeProngOnePizero. 
\end{itemize}

The \oneProngOnePizero and \oneProngTwoPizero decay modes are effectively merged, since neutral pion candidates reconstructed close to each other are clustered together and count as one strip.
The \twoProngZeroPizero and \twoProngOnePizero decay modes recover the three-prong-decays where a charged particle has been lost, but are usually not considered in the main reconstruction routine because of their large charge misassignment probability.

\section{The \texorpdfstring{\tauh}{tauh} identification using a deep neural network with domain adaptation}\label{sec:deeptau}

The \tauh candidates that have been reconstructed by the HPS algorithm are required to pass an identification step, using \DeepTau, a deep convolutional neural network, first described as v2.1 in Ref.~\cite{DeepTau_v2p1}.
The \DeepTau algorithm simultaneously discriminates \tauh candidates against quark and gluon jets, electrons and muons.
The algorithm uses a combination of high-level input variables and information from particles in the vicinity of the candidate. 
The neural network outputs an estimation of the probabilities that a candidate is a genuine \tauh, or a quark and gluon jet, electron, or muon.

The latest iteration of \DeepTau, v2.5, incorporates domain adaptation by backpropagation into the training workflow, to reduce performance discrepancies when the algorithm is applied to collision data.
The algorithm was trained on a balanced mix of simulated events, and on $\Pp\Pp$ collision data collected by the CMS detector in 2018 at $\sqrt{s}=13\TeV$.

In addition to achieving better data-to-simulation agreement, \DeepTau~v2.5 also demonstrates improved classification performance compared to its predecessor, v2.1, with a larger and more balanced training dataset, optimized hyperparameters, and a revised selection of inputs.

\subsection{Input variables} \label{sec:input}

\begin{figure}[!ht]
    \centering
    \includegraphics[width=0.7\textwidth]{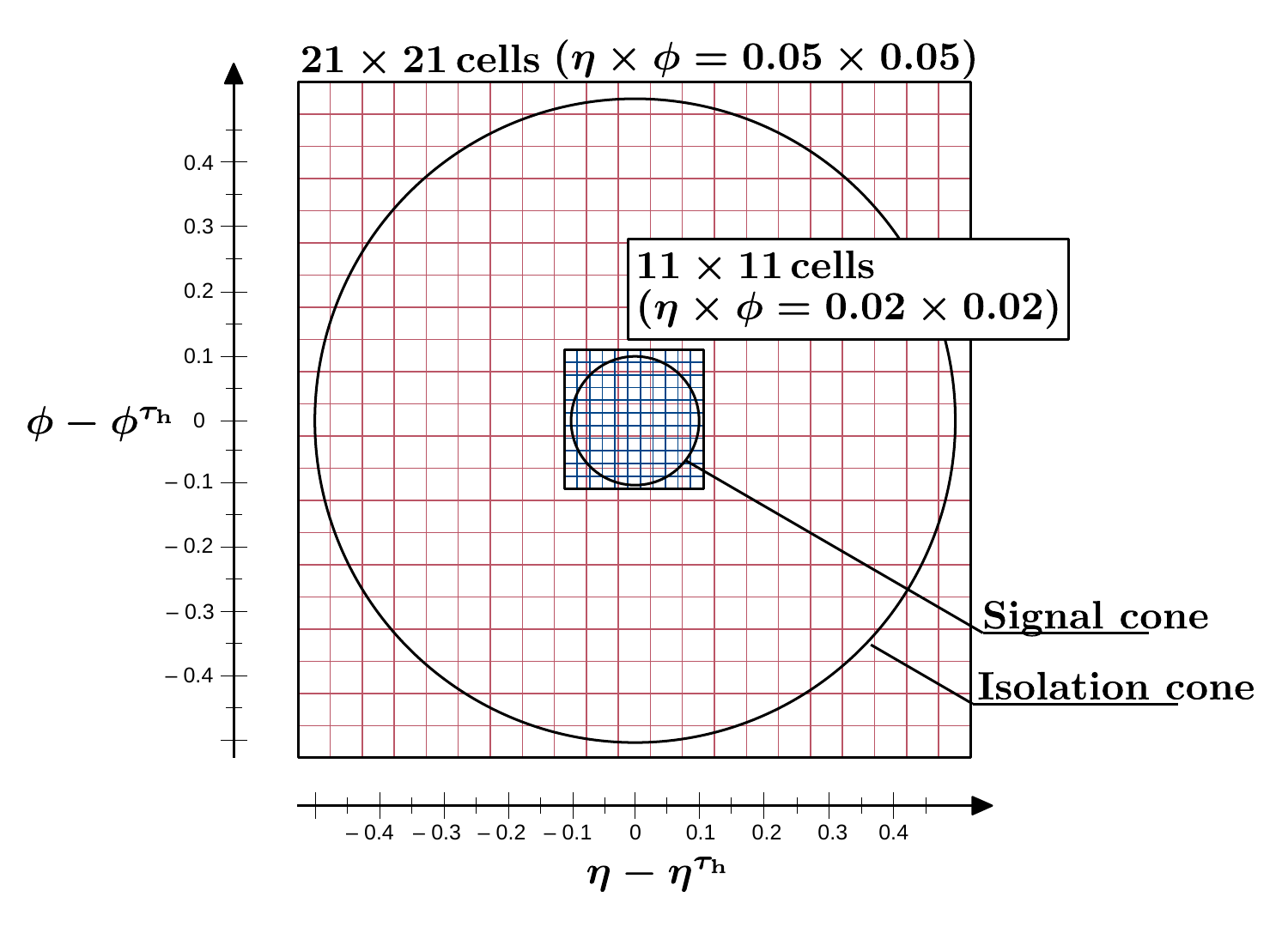}
    \caption{Inner and outer grid layout in $\eta$-$\phi$ space~\cite{DeepTau_v2p1}. The inner grid encapsulates the signal cone of maximal radius 0.1, which contains the $\hpm$ and $\PGpz$ candidates, and consists of $11{\times}11$ cells with a size of $0.02{\times}0.02$ each. The outer grid contains the isolation cone of radius 0.5, and consists of $21{\times}21$ cells with a size of $0.05{\times}0.05$ each.}
    \label{fig:grid}
\end{figure}

Particle-level inputs are stored in two overlapping grids in pseudorapidity-azimuth ($\eta$-$\phi$) space, centred on the \tauh candidate axis, as shown in Fig.~\ref{fig:grid}.
The inner grid, encapsulating the signal cone, which contains the $\hpm$ and $\PGpz$ candidates, comprises $11{\times}11$ cells of size $0.02{\times}0.02$. The outer grid of $21{\times}21$ cells of size $0.05{\times}0.05$, contains the isolation cone.

Properties of seven different types of reconstructed particles can be stored in each cell: those reconstructed by the PF algorithm (electrons, muons, photons, charged hadrons, and neutral hadrons), as well as electrons and muons reconstructed using dedicated standalone algorithms~\cite{electron,CMS:2018rym}. 

The particle-level inputs are identical to those used in Ref.~\cite{DeepTau_v2p1}. They include basic kinematic properties of each object: \pt, distance from the \tauh candidate axis in $\eta$-$\phi$ space ($\Delta\eta$ and $\Delta\phi$), and reconstructed charge. 
Track quality information, and compatibility with the PV, or possibly a secondary vertex (SV), is included, as well as characteristics of energy deposits in the detector (ECAL, HCAL, muon station hits). The estimated probability that the particle comes from another pileup interaction, is computed using the pileup-per-particle identification (PUPPI) algorithm~\cite{PUPPI}.

There are 43 high-level input variables, which correspond to those used in Ref.~\cite{DeepTau_v2p1}, except for four that were removed in this work: the azimuthal angle $\phi$ of the \tauh candidate, to reduce dependence on detector conditions, and the absolute coordinates of the point of closest approach of the leading charged track because of mismodelling. The high-level variables used are mostly those which were found useful in previous multivariate analysis classifiers~\cite{HPS2}.

The high-level variables include \tauh candidate kinematic quantities ($\eta$, \pt, and energy) and charge, the number of charged and neutral hadrons associated with the \tauh candidate by the HPS decay mode reconstruction, and characteristics of energy deposits from various particle types in the isolation cone. Information about the tracks is included: compatibility with the PV and properties of the SV, if reconstructed in a multiprong decay mode. Additionally, observables related to the $\eta$ and $\phi$ distributions of the reconstructed energy in the $\PGpz$ strips, estimated pileup density, and calorimeter variables that provide good discrimination against electrons are used as input variables.

Integer inputs and variables with finite domain are transformed linearly to limit their values to the $[-1,1]$ range. Other inputs are standardized using their mean values ($\mu_x$) and uncertainty ($\sigma_x$),
\begin{equation}
    x_\text{std} = \frac{x_\text{orig}-\mu_x}{\sigma_x},
\end{equation}
where $x_\text{orig}$ is the original input, and $x_\text{std}$ is the standardized input, which is then restricted to $[-5,5]$ to remove outliers.

The reconstructed \tauh candidates are assigned a type: genuine \tauh, misidentified electron (\Pe), and misidentified muon (\PGm), or misidentified quark or gluon jet (jet).
Candidates matched to generated electrons and muons with $\pt>8\GeV$, including those originating from leptonic tau decays, are assigned the \Pe or \PGm classes, respectively. Leptons below these low \pt thresholds are not considered, as they are expected to be misidentified jet components (given that the HPS tau candidate must have $\pt>20\GeV$).
Those matched to a generated \tauh with visible $\pt>15\GeV$ are assigned genuine \tauh. In all of these cases the matching is performed with a cone $\DR = \sqrt{\smash[b]{\Delta\eta^2 + \Delta\phi^2}} < 0.2$.  
If a candidate is not matched to a generated lepton or genuine \tauh, it is assigned to the jet class. 
Only generated leptons and genuine \tauh originating from the PV are considered for \tauh candidate type assignment. 

Approximately 100 million candidates, a balanced mix of the different types of \tauh candidates from various simulated events, are used to train the \DeepTau algorithm in its default configuration. These events are simulated and reconstructed according to 2018 data-taking conditions. All types of reconstructed \tauh candidates are sourced from \Zjets, \ttbar (semileptonic and fully hadronic final states), and \Wjets events. Additional genuine \tauh candidates are obtained from $\PH\to\PGt\PGt$ and $\PGt$ gun samples, and additional misidentified jets are obtained from simulated QCD multijet samples. Additional misidentified electrons are obtained from $\PZpr \to \Pe\Pe$ decays. 

A loose selection is applied, with the candidate reconstructed by the HPS algorithm required to have a transverse momentum $20<\pt<1000\GeV$. Additionally, limits are imposed on the pseudorapidity $\abs{\eta}<2.5$ and longitudinal impact parameter $\abs{d_z}<0.2\cm$ (distance between the leading charged track and PV).
Weights are applied in the \pt and $\eta$ bins to ensure a uniform distribution between the classes across the various training samples in all kinematic regions.

\subsection{Classification architecture and loss function}\label{sec:arch}

The architecture used for \tauh type classification in this work is similar to the one described in Ref.~\cite{DeepTau_v2p1}.
The input variables (high-level features, and all the inner/outer grid cells) are initially processed separately using fully connected layers for feature extraction. 
Convolutional layers are then used to reduce the dimensionality of the grids to $1{\times}1$.
The processed features are then concatenated and passed through a final set of fully connected layers for \tauh type classification. A softmax activation function is then applied to yield probability estimates that the \tauh candidate is a genuine \tauh, jet, electron, or muon.
The predicted output takes the form: $\textbf{y}^\text{pred}= \left(y_\Pe, y_\PGm, y_\PGt, y_{\text{jet}}\right)$.

Batch normalization~\cite{ioffe2015batch} and dropout regularization are applied after each fully connected and convolutional layer. Nonlinearities are introduced using the PReLu activation function~\cite{DBLP:journals/corr/HeZR015}.

The classification loss function used during the training of the neural network is a sum of a cross-entropy term~\cite{GoodBengCour16} for the genuine \tauh target class, a focal loss component~\cite{focal_loss} for classification against all backgrounds combined, and three individual components for classification as electrons, muons, or jets. The latter three components are activated only when the genuine \tauh probability is high, at which point the distinction between different background sources becomes important. The binary cross-entropy term assigns greater importance to very high genuine \tauh identification efficiencies, which typically have higher misidentification rates. The addition of the focal loss terms improves the classification performance for genuine \tauh identification efficiencies in the 50--80\% range, for which most physics analyses involving \tauh candidates show the highest sensitivity. Furthermore, in regions where the genuine \tauh identification efficiency is low, binary classification to separate signal from the combined backgrounds is more important than distinguishing between background types. 
The full definition of the loss function is given in Appendix~\ref{loss_appendix}.

The loss function is minimized with the adaptive momentum estimation (\textit{Adam}) algorithm~\cite{Adam} and the Nesterov-momentum accelerated variant (\textit{NAdam})~\cite{NAdam}.
The setup uses the \textsc{TensorFlow v2.5.0}~\cite{tensorflow2015-whitepaper} Python library  with \textsc{Keras v2.5.0}~\cite{chollet2015keras} as an interface. Training was performed using \textsc{NVIDIA Tesla V100} and \textsc{T4}, as well as \textsc{GeForce GTV 1080 Ti} GPUs.

The discriminators against electrons, muons, and quark or gluon jets are defined as
\begin{equation}
    D_\alpha (\textbf{y}) = \frac{y_\PGt}{y_\PGt + y_\alpha},
\end{equation}
where $y_\alpha$ is the predicted probability that the $\tauh$ candidate belongs to the target class $\alpha \in \{\Pe, \PGm, \text{jet}\}$. The discrimination of a genuine \tauh against a particular background improves as the corresponding discriminator score approaches 1.

\subsection{Domain adaptation by backpropagation}\label{sec:domain_adapt}

\DeepTau~v2.1 was trained exclusively on simulated events. While these samples generally provide a good representation of the $\Pp\Pp$ collision data, some of the features used as inputs are not perfectly modelled. As a consequence, the previous setup shows increasing discrepancies between observed data and expectations from simulations for high \Djet scores. This is particularly problematic since the affected region is the most important for analyses, as it has the highest genuine \tauh purity.

Previously, this mismodelling was only corrected using a set of dedicated calibration measurements, where simulations were fitted to the observed data, to determine the identification efficiency scale factors.
However, such corrections lead to larger uncertainties, and do not correct the shape of the discriminator output. The raw \DeepTau~v2.1 score could therefore not be reliably used as an input for analyses.

Simply removing any variables which are not perfectly described in simulation is not optimal, as they remain important for achieving good classification performance. Furthermore, studies have shown that the mismodelling is a highly multidimensional effect, making it very difficult to identify the affected set of inputs and correct them with traditional methods. A better approach is domain adaptation, which can discourage the algorithm from using combinations of features that are not well modelled in simulation by identifying differences between data and simulation in the hidden layers of the neural network during training.

The performance discrepancies between data and simulated samples are reduced at the training level by implementing domain adaptation by backpropagation in \DeepTau~v2.5. This involves simultaneously training two subnetworks with competing goals. In this case, one is used for \tauh type classification, and the other for domain discrimination between data and simulated events. The goal is to maximize the classification performance while minimizing data-simulation discrimination. 
A mathematical description of the gradient reversal technique used to achieve this is available in Ref.~\cite{domainadapt}. This technique has previously been used for displaced jet tagging~\cite{dispjettag}.

The advantages of this method are that the scale factors to correct residual differences can be brought closer to unity. Additionally, the network optimization algorithm has the opportunity to find trainable parameter values that ensure good \tauh type discrimination while being less sensitive to mismodelling.

As the neural network training with domain adaptation requires collision data, the main challenge of this method is to define a set of collision events with sufficient genuine \tauh purity and control over background.

\subsubsection{Event selection and mixing}
Training the domain adaptation subnetwork requires a control region dataset containing a mixture of data and simulated events, along with a high purity of genuine \tauh candidates.
The control region is a sample of $\PZ\to\PGt\PGt$ events in which one tau lepton decays to a muon, and the other to hadrons (\mutau). This decay channel is chosen as there is good control over genuine \tauh purity and background.

The selection requirements on the muon and \tauh candidates are summarized in Table~\ref{tab:selection_summary}. 
These are applied to data collected during 2018 by the CMS detector using a single-muon trigger with a nominal \pt threshold of 24\GeV, as well as \Zjets, \ttbar, and \Wjets simulated events.

The transverse mass \mT of the muon \pt plus the missing transverse momentum is defined as
\begin{equation}
  \mT(\pt^\PGm,\ptmiss)=\sqrt{2\pt^\PGm \ptmiss(1-\text{cos}(\Delta \phi))},
\end{equation}
where $\Delta \phi$ is the azimuthal separation between the muon \pt and \ptvecmiss.

\begin{table}[htb]
    \centering
    \renewcommand*{\arraystretch}{1.25}
    \topcaption{Selection requirements for the domain adaptation dataset. The impact parameters for the muon (or \tauh candidate), $d_z$ and $d_{xy}$, are defined as the distances between the muon track (or leading charged-hadron track) and the PV. The medium muon identification is defined in Ref.~\cite{CMS:2018rym}. The previous \DeepTau discriminator scores described in Ref.~\cite{DeepTau_v2p1} against quark and gluon jets, electrons, and muons, are denoted $D_\text{jet}^\text{v2.1}$, $D_\Pe^\text{v2.1}$, and $D_\PGm^\text{v2.1}$. The transverse mass of the muon \pt and the missing transverse momentum system is denoted as $\mT(\pt^\PGm,\ptmiss)$. Working points Tight and VVLoose are defined in Table~\ref{tab:wp-definition}.}
    \begin{tabular}{ll}
    \textbf{Object} & \textbf{Selection requirement} \\
    \hline
    Muon & $\pt^\PGm > 25\GeV$ \\ 
         & $\abs{\eta^\PGm}<2.1$ \\ 
         & $\abs{d_z}<0.2\cm$, $\abs{d_{xy}}<0.045\cm$ \\ 
         & Relative isolation $I_\text{rel}^\PGm<0.15$ \\ 
         & Pass medium muon identification \\
         & $\mT(\pt^\PGm, \ptmiss) < 30\GeV$ \\
    [\cmsTabSkip]
    \tauh & $\pt^\PGt > 20\GeV$ \\ 
          & $\abs{\eta^\PGt}<2.3$ \\ 
          & $\abs{d_z}<0.2\cm$ \\ 
          & HPS decay mode with 1 or 3 prongs \\
          & $D_\text{jet}^\text{v2.1}>0.9$ \\ 
          & $D_\Pe^\text{v2.1}>0.168$ (VVLoose) \\
          & $D_\PGm^\text{v2.1}>0.875$ (Tight) \\
    [\cmsTabSkip]
    Pair & $\DR(\PGm,\tauh)>0.5$ \\ 
         & Opposite electric charge \\
    \end{tabular}
    \label{tab:selection_summary}
\end{table}

If there is more than one muon or \tauh candidate fulfilling the criteria in Table~\ref{tab:selection_summary}, the most isolated candidate is selected, unless they are equal, in which case the highest \pt candidate is chosen. A veto is imposed on events containing loosely identified additional electrons and muons.

In order to improve the modelling of events involving a nongenuine \tauh, a looser selection is applied to increase the number of available events from simulated QCD multijet samples and misidentified muons originating from  \Zjets processes. Only the \tauh candidate and pair selections in Table~\ref{tab:selection_summary} are applied, and the \pt spectra are then reweighted to the expectation for the nominal selection yields.

The resulting purity in the domain adaptation dataset, defined as the fraction of events where the \tauh candidate originates from a tau lepton, is estimated from simulation to be 76\%.

\subsubsection{Domain adaptation subnetwork and backpropagation}

\begin{figure}[!tb]
    \centering
    \includegraphics[width=\textwidth]{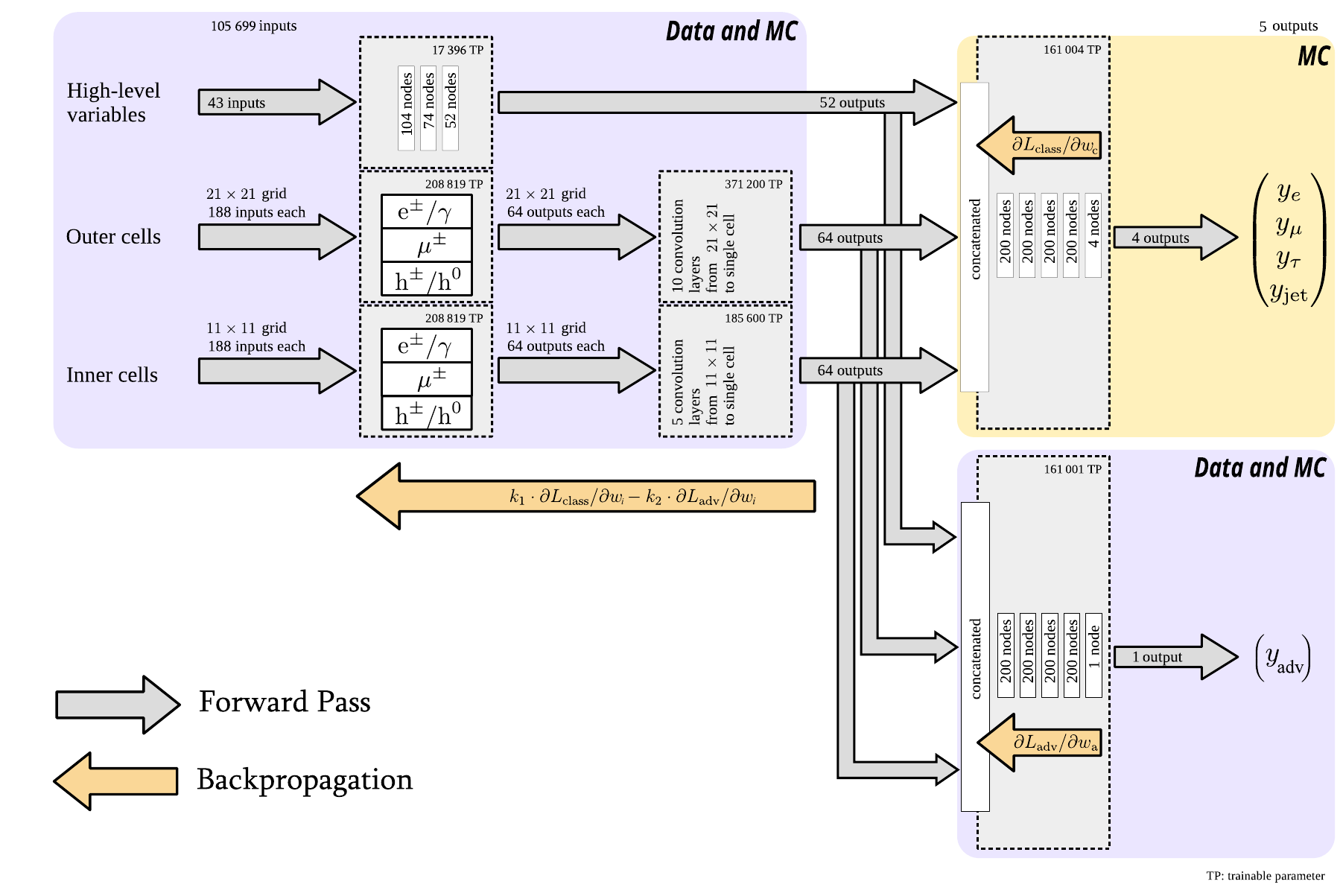}
    \caption{The \DeepTau architecture with the domain adaptation configuration~\cite{DeepTau_thesis1}. A set of final domain adaptation layers was introduced for data-simulation discrimination, consisting of several dense layers followed by a softmax layer that yields an output $\yadv$ between zero and one. The backpropagation is modified to include the ``adversarial loss'', as described in the text.}
    \label{fig:adv_archi}
\end{figure}

A domain adaptation block was introduced to the network, which attempts to discriminate between observed and simulated \tauh candidates.
Similarly to the final classification layers described in Section~\ref{sec:arch}, this takes the processed high-level and particle-level variables as inputs, before passing them through fully connected layers. A sigmoid activation layer is then used to cast the output to a value between 0 and 1, denoted as $\yadv$. The architecture of \DeepTau in the domain adaptation configuration is shown in Fig.~\ref{fig:adv_archi}.

The domain adaptation loss function (``adversarial loss'') used to compare this prediction to the labels is defined as
\begin{equation}
    \Ladv = H_\text{bin}( \yadv^\text{true}, \yadv^\text{pred}),
\end{equation}
where $H_\text{bin}$ is a binary cross-entropy loss function~\cite{GoodBengCour16}.
A binary accuracy metric was introduced to evaluate the fraction of candidates for which the network successfully predicts the domain (data or simulation).
The layers that process the high-level variables, as well as the inner and outer grids, are referred to as common layers, since inputs for both the final domain adaptation and classification layers pass through these.

The model is first trained with only the classification architecture described in Section~\ref{sec:arch}, in order to obtain a good \tauh candidate classification performance baseline before applying domain adaptation methods. 
The training is then continued with the domain adaptation control dataset and subnetwork introduced.
For this step, the network backpropagation procedure is modified such that the gradients are passed to the optimization algorithm for the common layers.

The classification loss is denoted as \Lclass, and is computed on the output score of the \tauh classification, with only the \tauh candidates from the standard training dataset. The adversarial loss is computed on the domain adaptation output with only the \tauh candidates from the control region dataset.
The gradients of \Lclass and \Ladv are referred to as classification and adversarial gradients. 

In order to prevent data-simulation discrimination in the feature extraction layers, the signs of the adversarial gradients are reversed and combined linearly with the classification gradients. The combined gradient is expressed in the form
\begin{equation}\label{grad}
    G = k_1\frac{\partial \Lclass}{\partial w_{i}} - k_2\frac{\partial \Ladv}{\partial w_{i}},
\end{equation}
where $w_i$ are the weights of the feature extraction layers and $k_1$ and $k_2$ are the domain adaptation hyperparameters that determine the relative importance of the \tauh candidate classification and prevent data-simulation discrimination. The sign reversal of the adversarial component of the gradients in the common layers means that the optimizer partially attempts to adjust the layer weights in the direction opposite to the one that improves data-simulation discrimination.

The gradients that are passed to the optimizer therefore encourage minimization with reduced sensitivity to mismodelling. Two Adam optimizers are used, each with a different learning rate. The principal optimizer targets the common layers and final classification layers, whereas the adversarial optimizer targets the final domain adaptation layers. 

There is no gradient sign reversal in the final domain adaptation layers, as this provides a good measure of how much the network can actively discriminate data and simulated events with the information available at the output level of the common layers.  
An optimization of the domain adaptation hyperparameters was performed by comparing the distributions of $\yadv$ for data and simulated events using a $\chi^2$ test, while monitoring the overall performance of the \tauh type classification. The optimal choice was $k_1=1$ and $k_2=10$.

\begin{figure}[!ht]
    \centering
    \includegraphics[width=\cmsFigWidth]{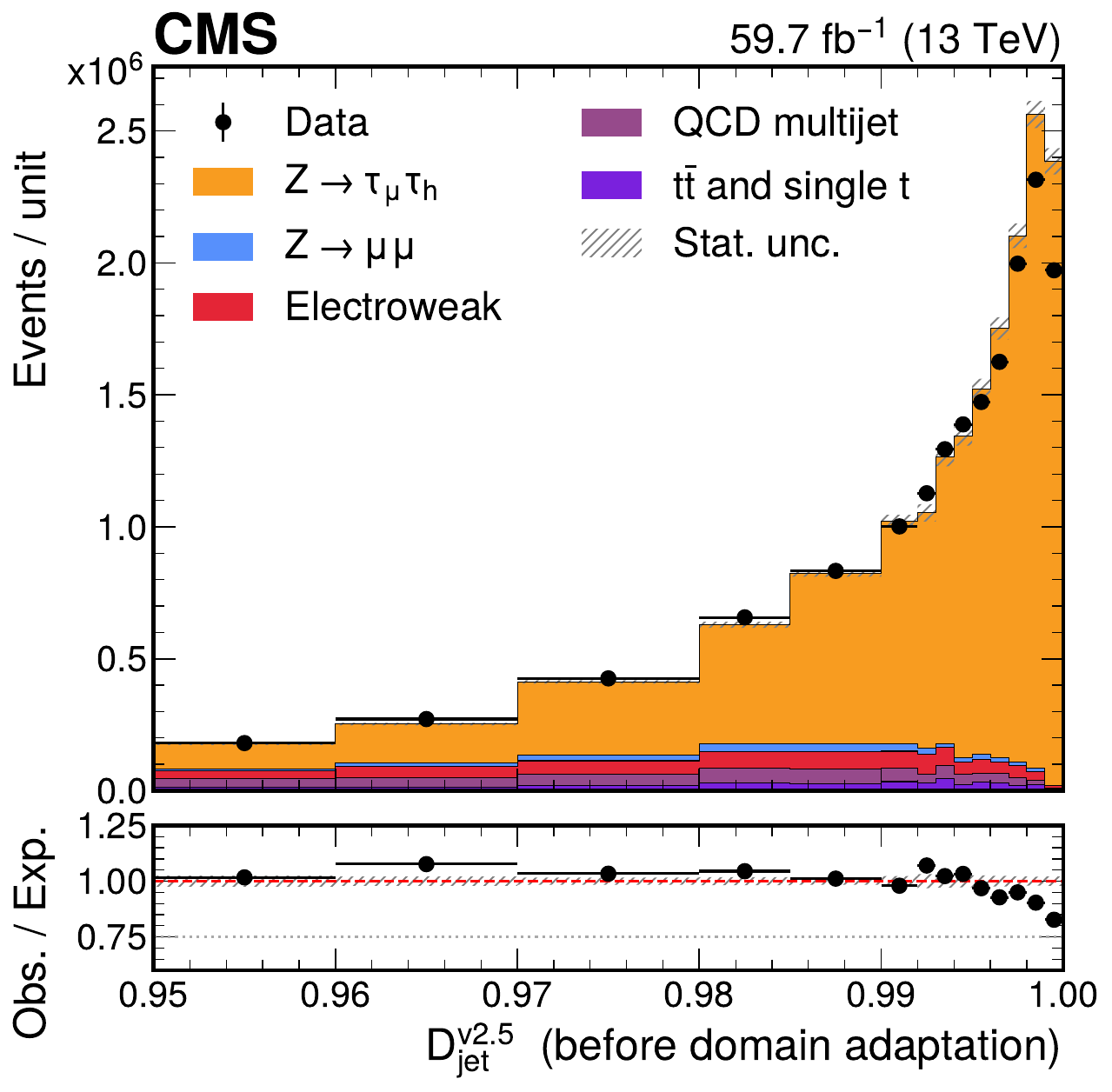}
    \includegraphics[width=\cmsFigWidth]{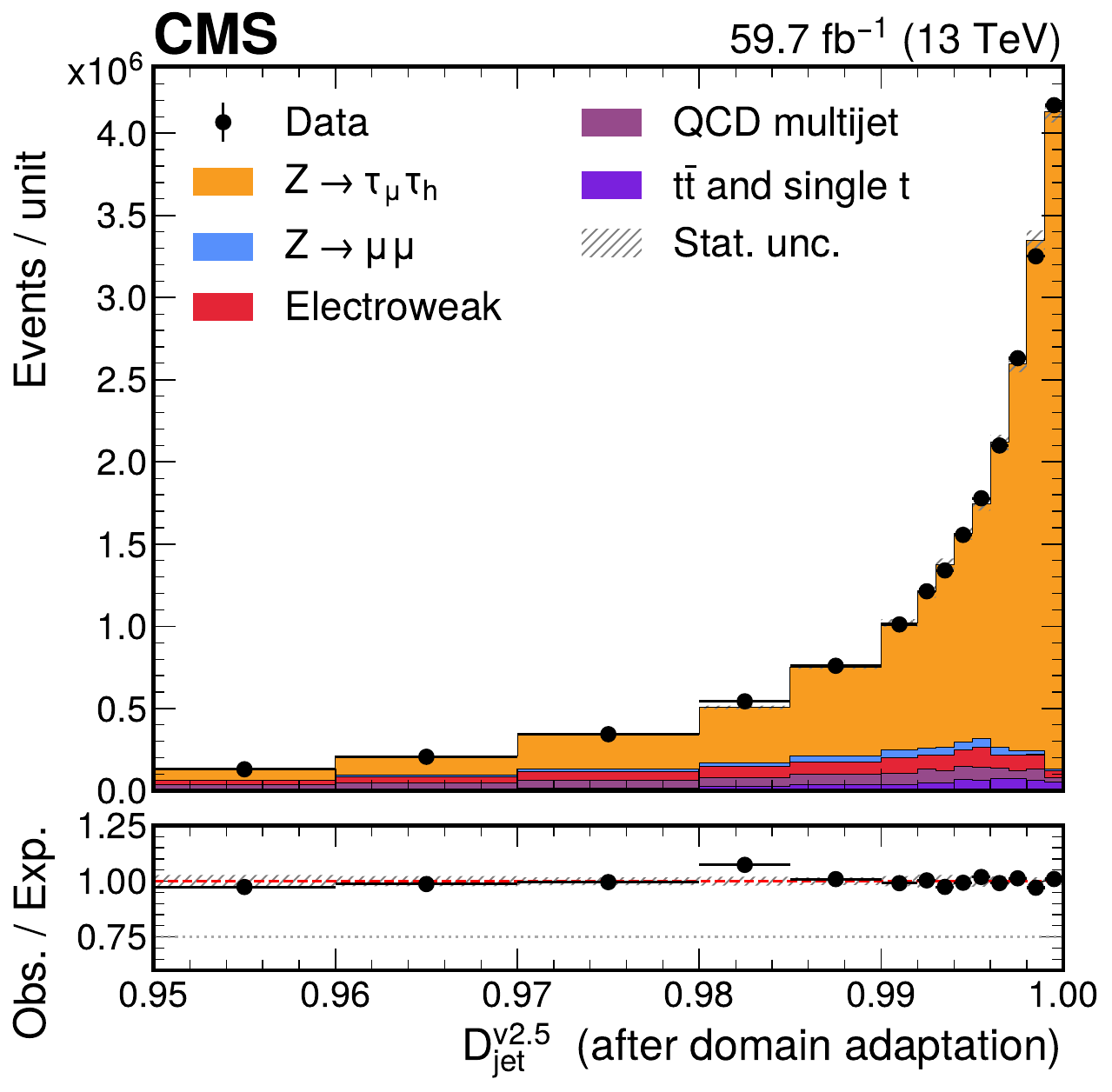}
    \caption{Distribution of the \DeepTau discriminator against quark and gluon jets before (left) and after (right) domain adaptation, for the 2018 dataset used for domain adaptation training. There is significant improvement in data-simulation agreement in the control region, with the discrepancies in the final bin reduced to 0.9\%. The vertical bars on the data points indicate the statistical uncertainty; on most points the bars are smaller than the marker size.}
    \label{fig:def_djet}
\end{figure}

The distributions of the \DeepTau discriminator against quark and gluon jets for the model before and after the domain adaptation are shown in Fig.~\ref{fig:def_djet}.
The impact of the domain adaptation training on the \DeepTau discriminator distribution against quark and gluon jets for the final model is visible on the right plot. The model was evaluated on events passing the control region selection in simulation and in data. 
There is a significant improvement in data-simulation discrepancies in the highest discriminator score bins after domain adaptation training. The relative differences in the final bin are reduced from 17.4 to 0.9\%. The genuine \tauh purity in this region is estimated from the fractions of different simulated samples in the final model distribution to be above 96\%. Agreement in the control region overall is very good, with the data and simulated yields compatible within statistical uncertainties in the large majority of bins.
The domain adaptation training therefore successfully reduced the effects of simulation mismodelling in the \Djet distribution for the data-taking conditions described in the domain adaptation dataset.

\begin{figure}[!ht]
    \centering
    \includegraphics[width=\cmsFigWidth]{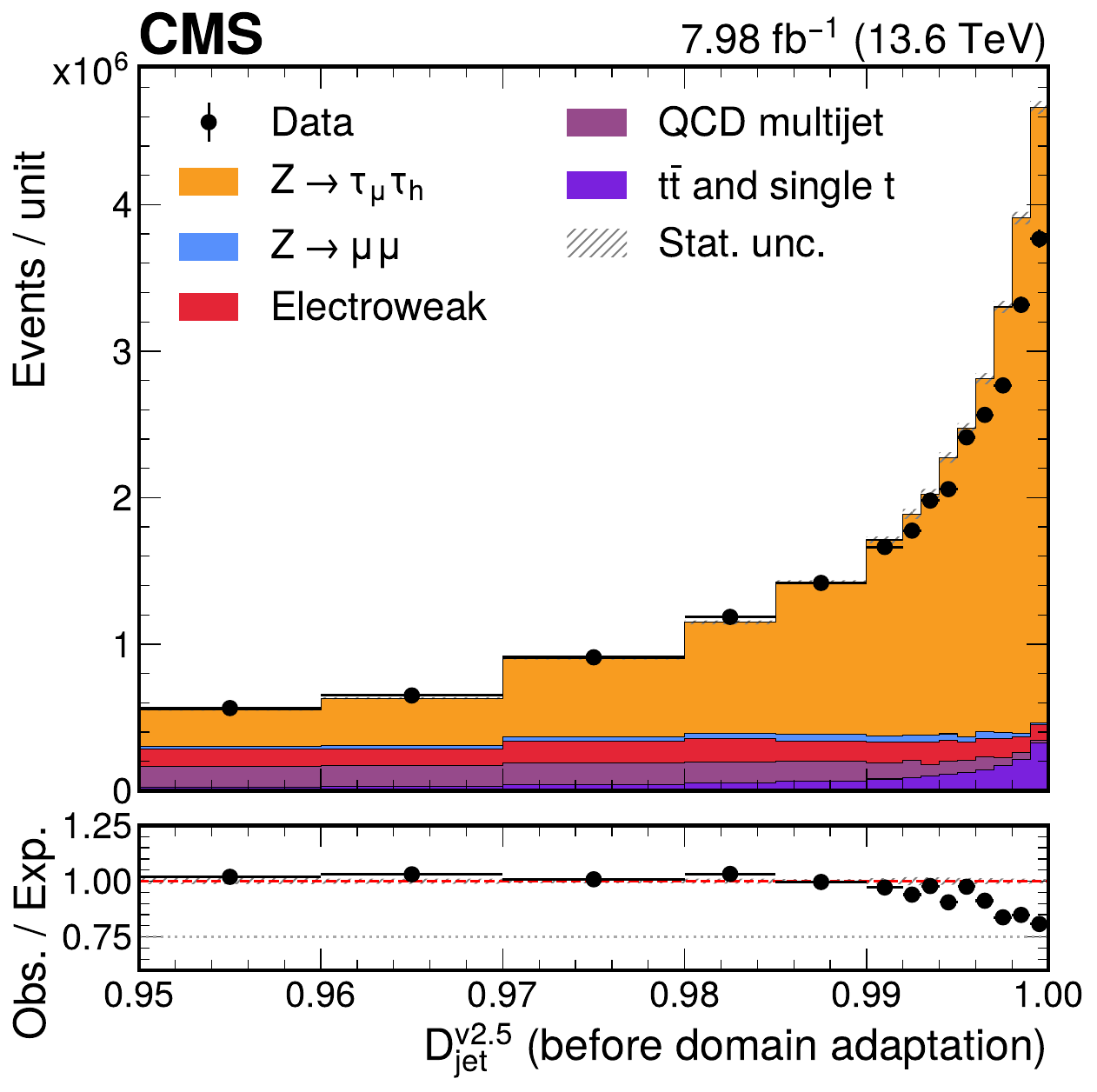}
    \includegraphics[width=\cmsFigWidth]{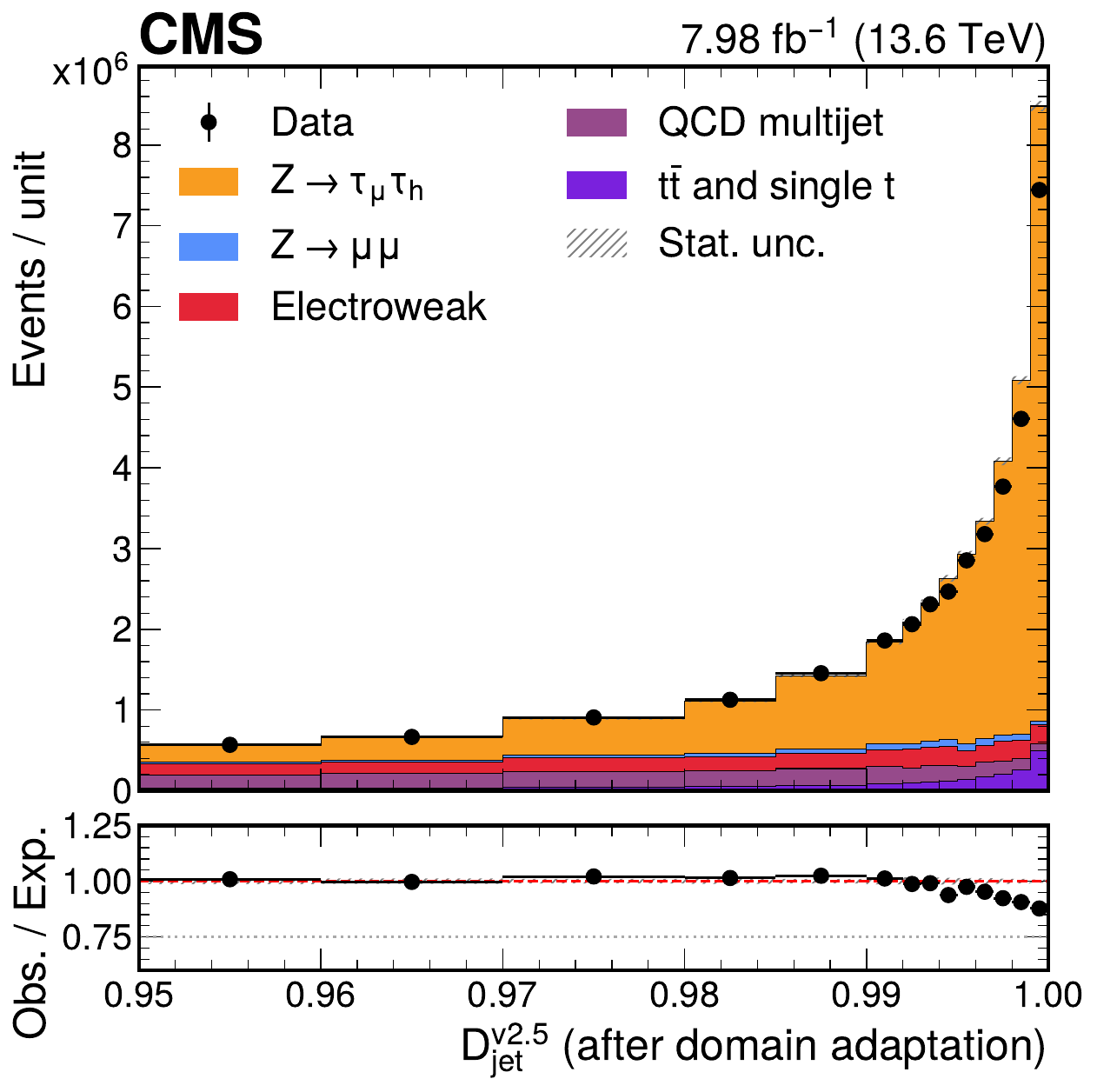}
    \caption{Distribution of the \DeepTau discriminator against quark and gluon jets before (left) and after (right) domain adaptation, for the early 2022 dataset. While data-to-simulation differences remain, there is an appreciable improvement  in the final bins with the inclusion of domain adaptation, despite \DeepTau being trained on 2018 data and simulation. The vertical bars on the data points indicate the statistical uncertainty; on most points the bars are smaller than the marker size.}
    \label{fig:def_djet_22}
\end{figure}

The 2022 datasets with $\sqrt{s}=13.6\TeV$ are expected to show weaker data-to-simulation agreement, as the data-taking conditions are not described in the domain adaptation dataset. 
Differences in the centre-of-mass energy, pileup, detector performance, and missing transverse momentum reconstruction with respect to 2018 are contributing factors.
The distribution of the \DeepTau discriminator against quark and gluon jets in the early 2022 dataset is shown for $\Djet>0.95$ in Fig.~\ref{fig:def_djet_22} before and after the domain adaptation training. While data-to-simulation differences remain, the inclusion of domain adaptation results in an appreciable improvement in the final bins, despite the training dataset not corresponding to equivalent detector conditions or collision energies. It is expected that if the domain adaptation algorithm were to be retrained with the inclusion of these new datasets, agreement would improve to a similar level as seen in the 2018 dataset.

\subsection{Expected performance}\label{sec:perf_deeptau}

Working points are used to guide the usage of the \DeepTau discriminators in physics analyses, with suitable corrections applied.
The target genuine \tauh identification efficiencies, reported in Table~\ref{tab:wp-definition}, are defined as the efficiency for genuine \tauh in the $\PH\to\PGt\PGt$ event sample that are reconstructed as \tauh candidates with $30 < \pt < 70\GeV$ to pass the given discriminator.

\begin{table}[htb]
    \centering
	\topcaption{Target genuine \tauh identification efficiencies for the different working points defined for the three discriminators. The target efficiencies are evaluated with the $\PH\to\PGt\PGt$ event sample for \tauh candidates with $\pt \in [30, 70]\GeV$.}
	
	\begin{tabular}{lcccccccc}
		& VVTight & VTight & Tight & Medium & Loose & VLoose & VVLoose & VVVLoose \\
		\hline
		\De & 60\% & 70\% & 80\% & 90\% & 95\% & 98\% & 99\% & 99.5\% \\
		\Dm &  &  & 99.5\% & 99.8\% & 99.9\% & 99.95\% &  & \\
		\Djet & 40\% & 50\% & 60\% & 70\% & 80\% & 90\% & 95\% & 98\% \\
	\end{tabular}
    \label{tab:wp-definition}
\end{table}

\begin{figure}[!tb]
  \centering
  \includegraphics[width=\cmsFigWidth]{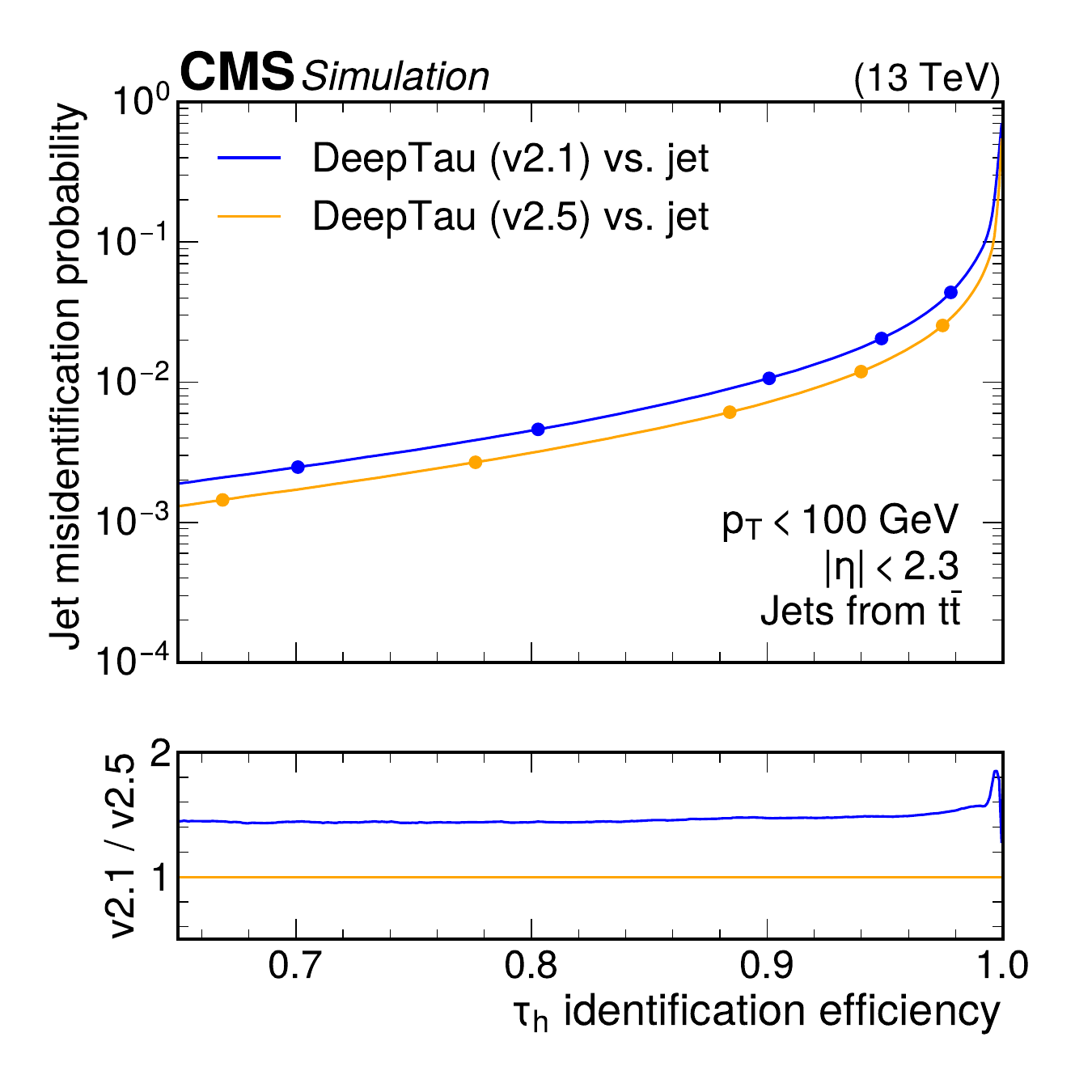}
  \includegraphics[width=\cmsFigWidth]{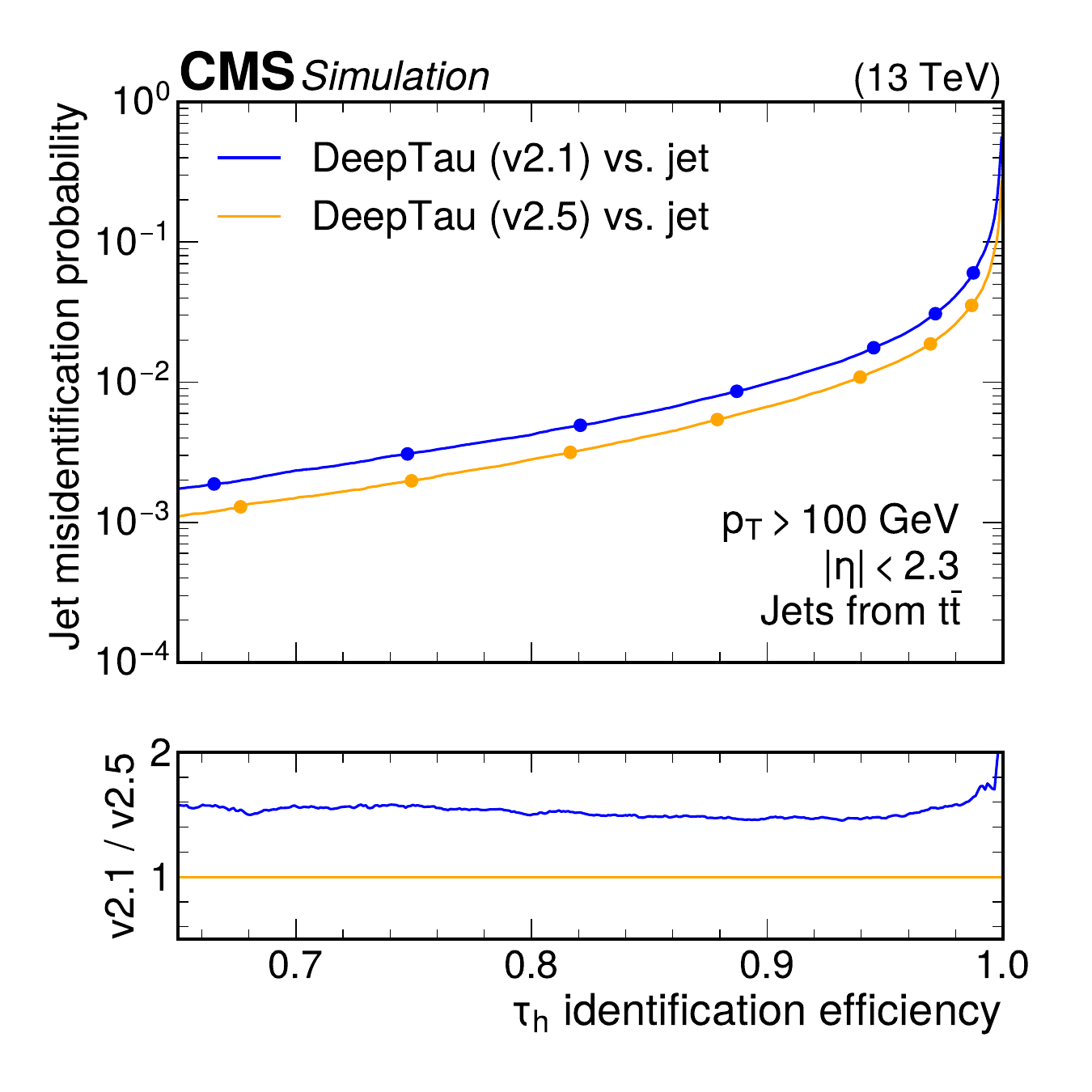}
  \caption{Jet misidentification probability versus genuine \tauh identification efficiency for low-\pt (left) and high-\pt (right) \tauh candidates, evaluated on 2018 simulated datasets. The genuine \tauh identification efficiency is estimated from $\PH\to \PGt \PGt$ simulations using reconstructed \tauh candidates that match generator-level \tauh objects. The jet misidentification probability is estimated from \ttbar simulations using reconstructed \tauh candidates that do not match prompt electrons, muons, or products of \tauh decays at the generator level. The defined working points of the discriminator are indicated as filled circles.
 }
  \label{fig:perf_vsjet}
\end{figure}

\begin{figure}[!htb]
  \centering
  \includegraphics[width=\cmsFigWidth]{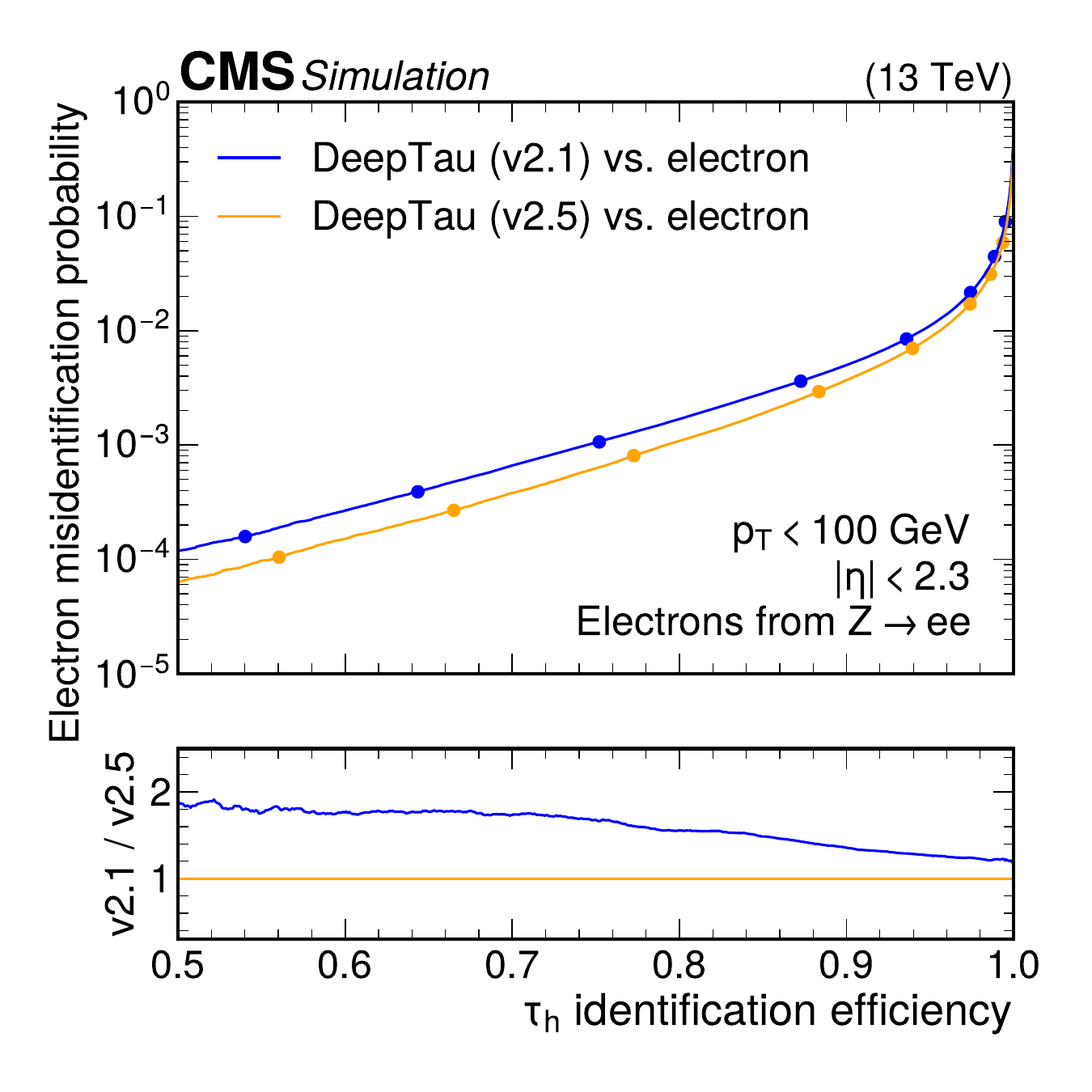}
  \includegraphics[width=\cmsFigWidth]{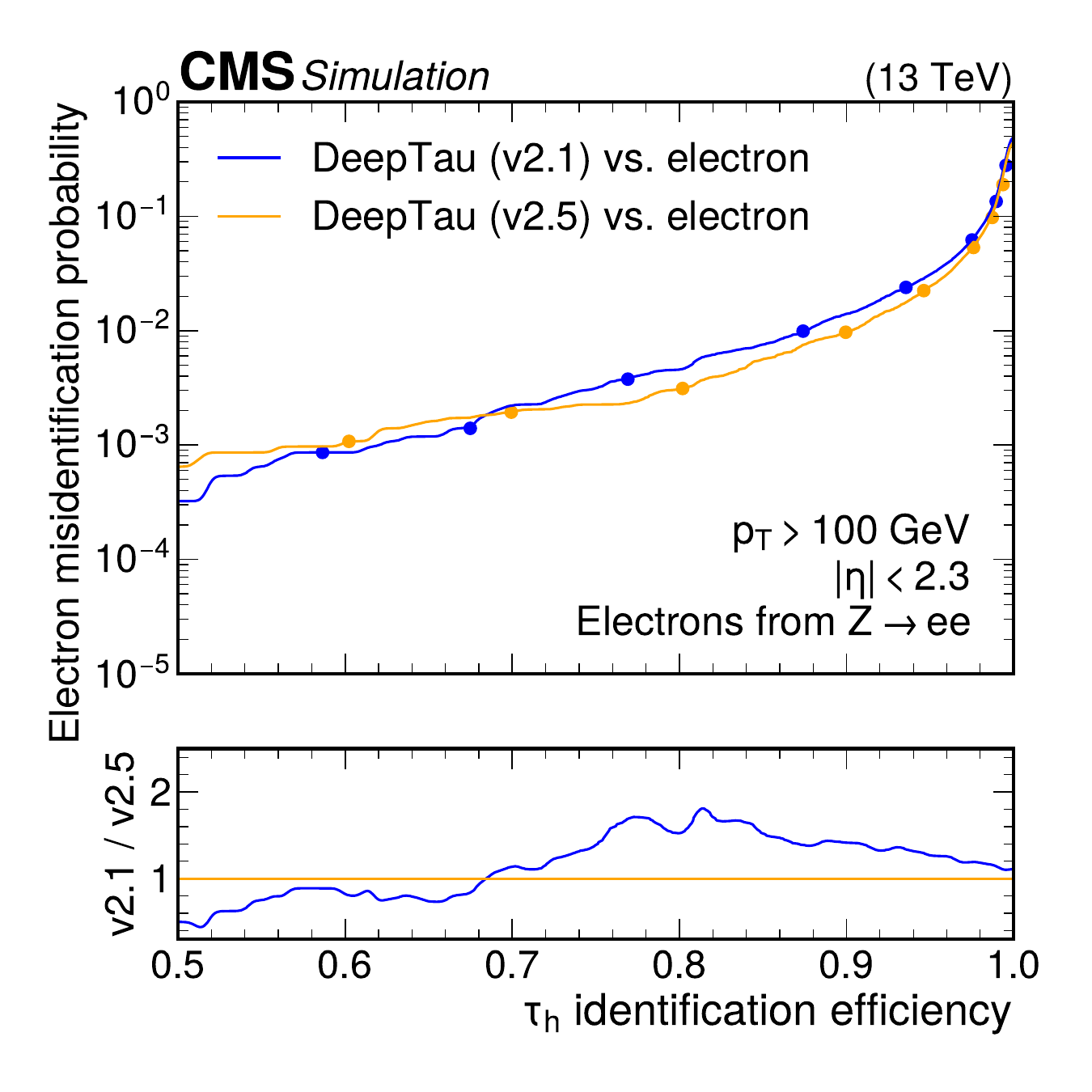}
  \caption{Electron misidentification probability versus genuine \tauh identification efficiency for low-\pt (left) and high-\pt (right) \tauh candidates, evaluated on 2018 simulated datasets. The genuine \tauh identification efficiency is estimated from $\PH \to \PGt\PGt$ simulations using reconstructed \tauh candidates that match generator-level \tauh objects. The electron misidentification probability is estimated from \Zjets simulation using reconstructed \tauh candidates that match electrons at the generator level. The defined working points of the discriminator are indicated as filled circles.
 }
  \label{fig:perf_vse}
\end{figure}

\begin{figure}[!htb]
  \centering
  \includegraphics[width=\cmsFigWidth]{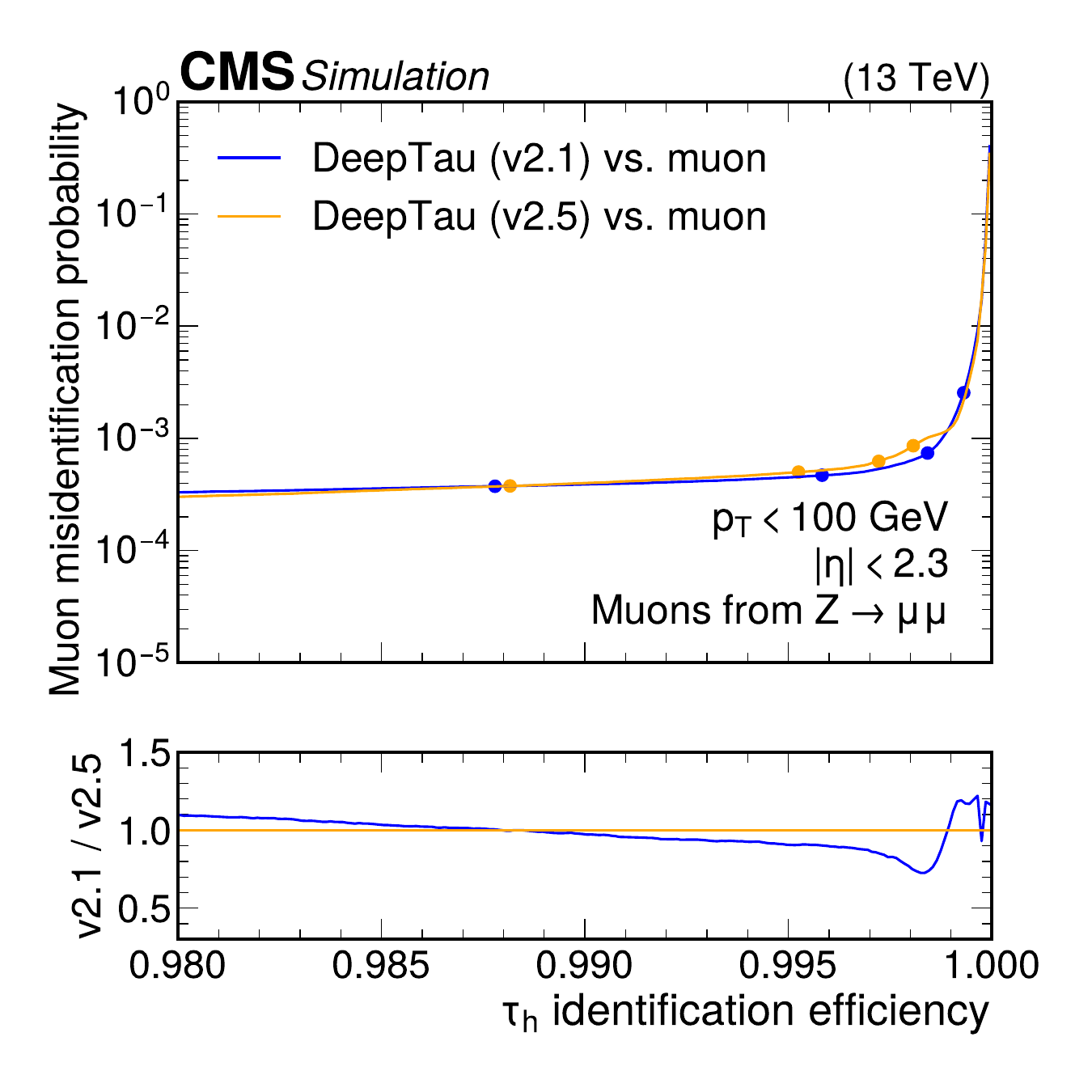}
  \includegraphics[width=\cmsFigWidth]{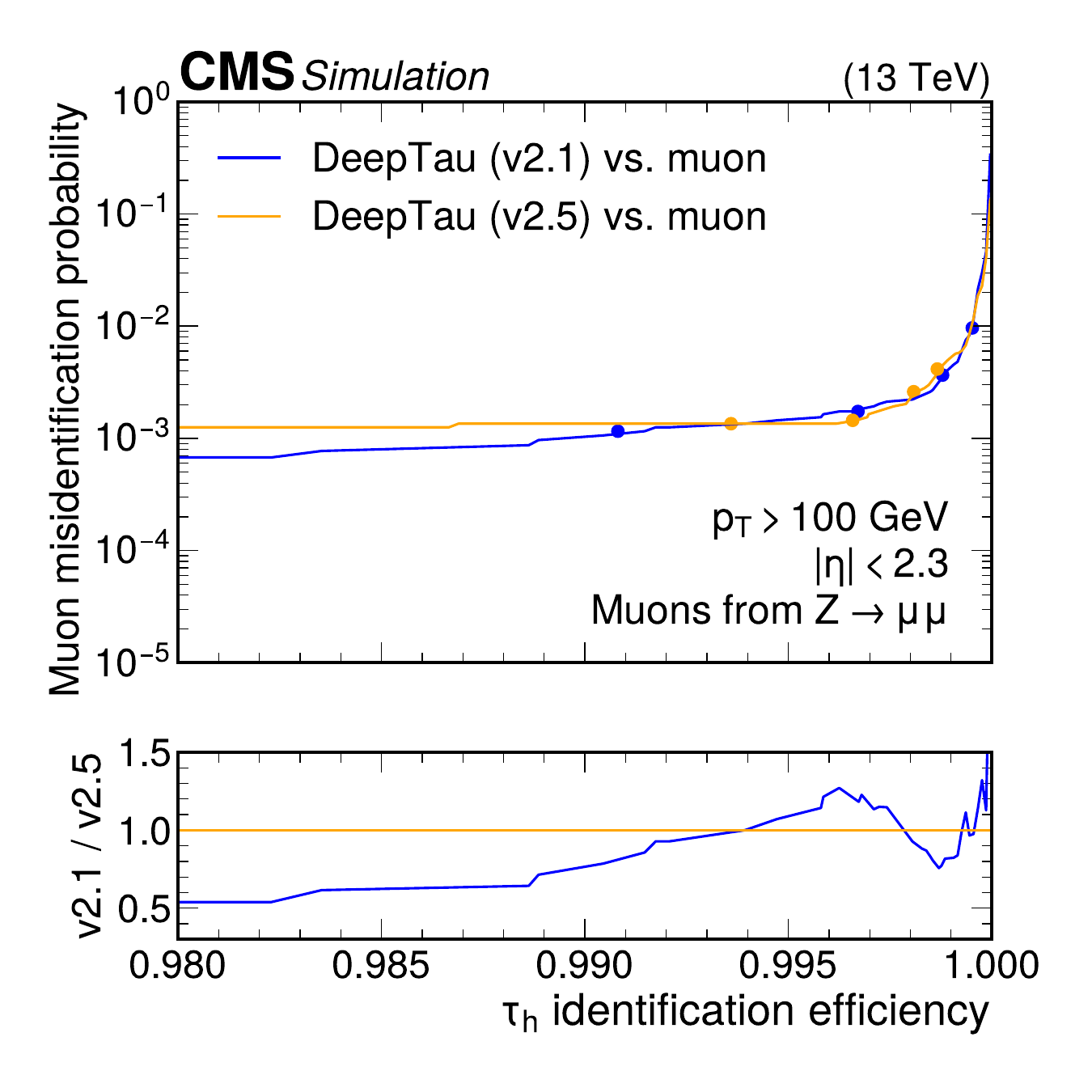}
  \caption{Muon misidentification probability versus \tauh identification efficiency for low-\pt (left) and high-\pt (right) \tauh candidates, evaluated on simulated 2018 datasets. The \tauh identification efficiency is estimated from $\PH \to \PGt\PGt$ simulations using reconstructed \tauh candidates that match generator-level \tauh objects. The muon misidentification probability is estimated from \Zjets simulation using reconstructed \tauh candidates that match muons at the generator level. The defined working points of the discriminator are indicated as filled circles.
 }
  \label{fig:perf_vsmu}
\end{figure}

The resulting tagger demonstrates improved performance with respect to its predecessor, as is visible in Figs.~\ref{fig:perf_vsjet}--\ref{fig:perf_vsmu}, which show the misidentification probability as a function of the genuine \tauh identification efficiency. These are inclusive plots that combine all commonly reconstructed decay modes. They are shown for the central pseudorapidity region ($\abs{\eta}<2.3$), and are separated into low-\pt (20--100\GeV) and high-\pt (100--1000\GeV) regions.
The filled circles identify the discriminator working points, which do not match exactly the targeted efficiencies in Table~\ref{tab:wp-definition} because of the different samples considered. 
It can be observed that the jet misidentification probability is reduced by $\sim$50\% across all defined working points. The improvement in electron rejection with respect to the previous version of the algorithm is particularly pronounced for the tightest working points, where the misidentification probability is reduced by almost a factor of two. The muon rejection performance is compatible between the two versions of the identification algorithm. The slightly worse performance of v2.5 in the high-\pt, low efficiency region for the $\De$ and  $\Dm$ discriminators is likely caused by domain adaptation, which can prevent the use of certain feature combinations that are useful for lepton discrimination but exhibit significant differences between data and simulation. 

\begin{figure}[!htb]
  \centering
  \includegraphics[width=\cmsFigWidth]{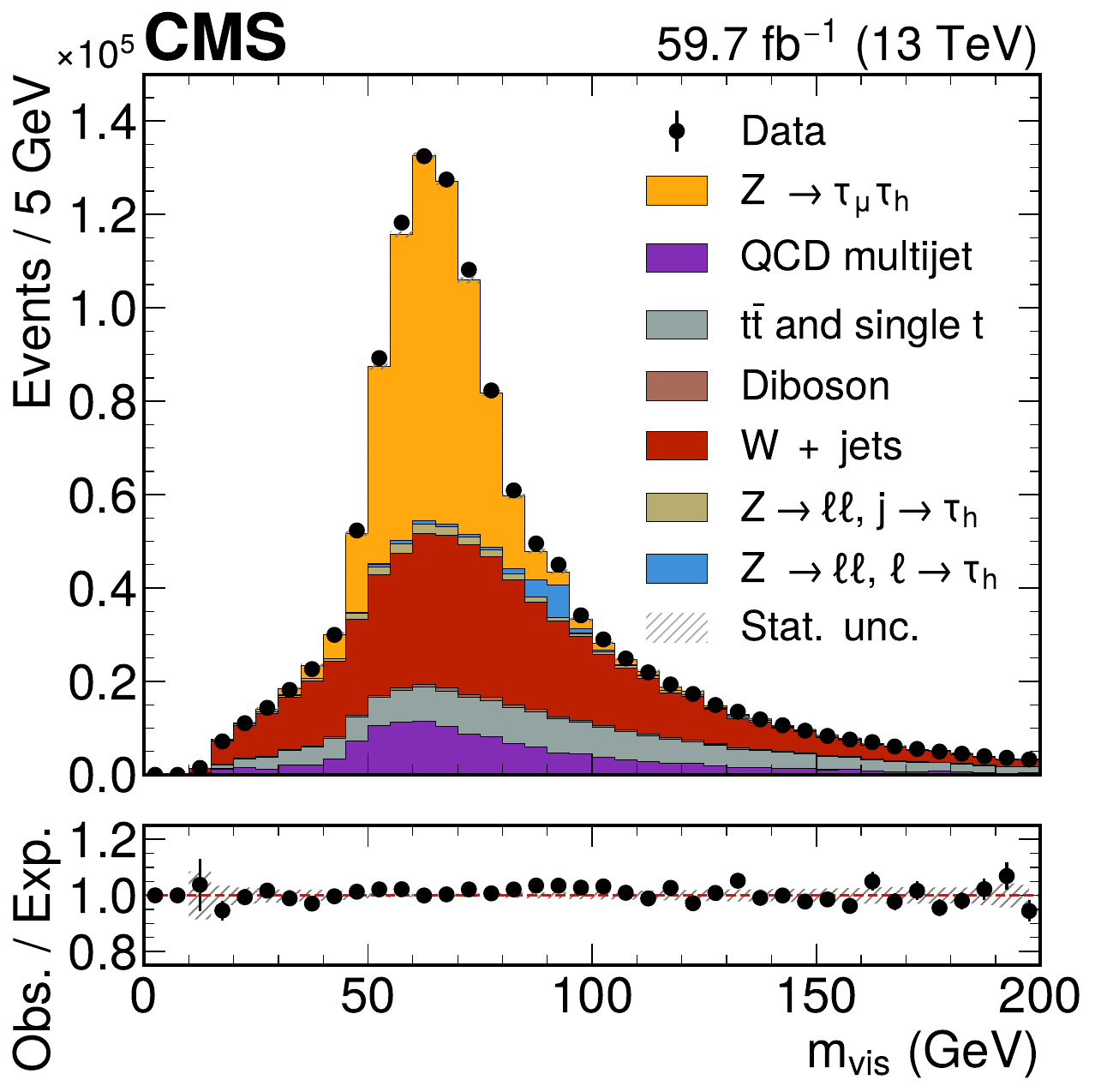}
\includegraphics[width=\cmsFigWidth]{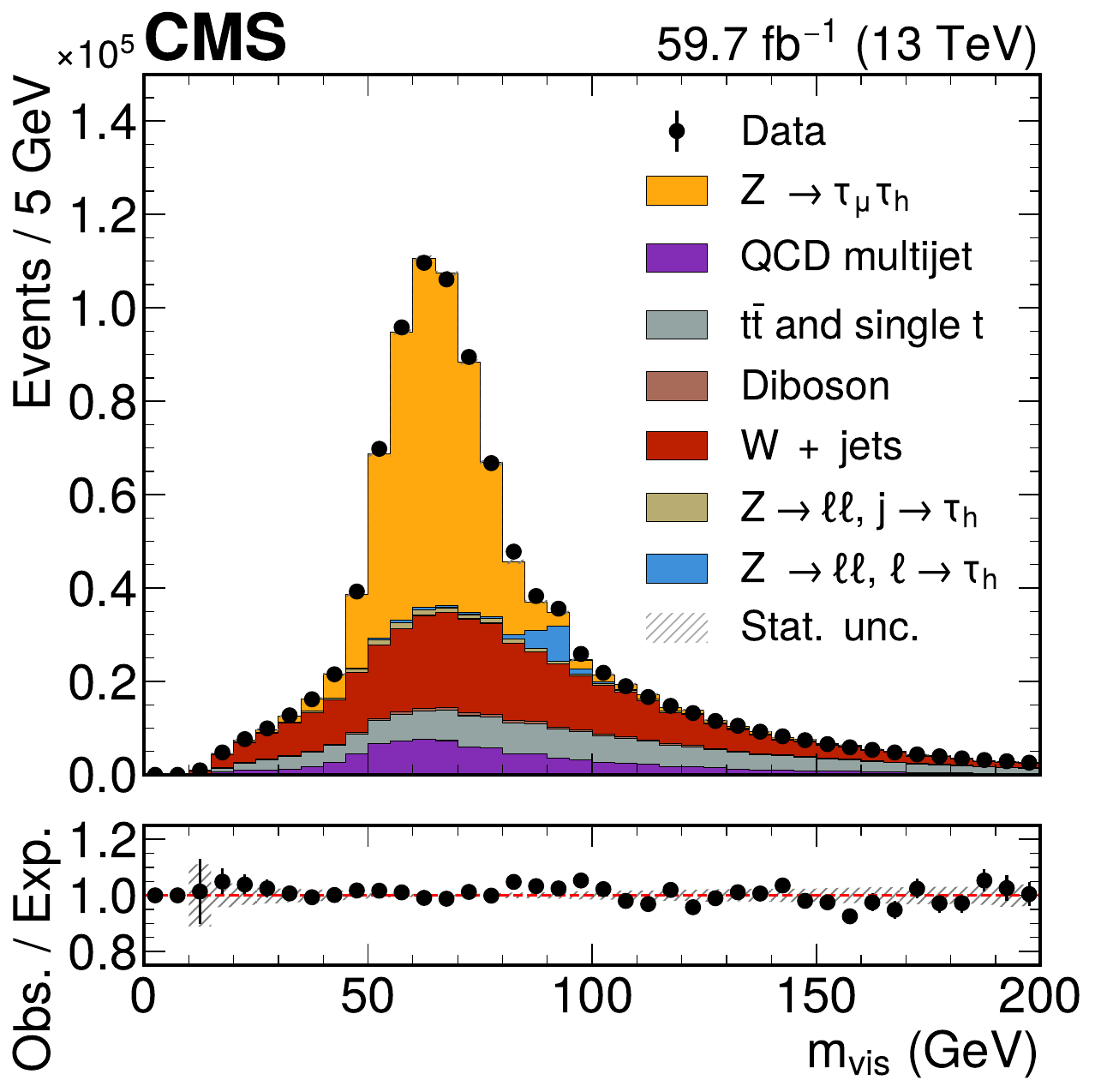}
    \caption{Distribution of the invariant mass of the reconstructed \mutau system when using \DeepTau~v2.1 (left) and v2.5 (right) for discrimination in the 2018 dataset. The \DeepTau working points used are: Medium for \Djet, VVLoose for \De and, Tight for \Dm (see Table~\ref{tab:wp-definition}). The correction factors are applied in both cases. The vertical bars correspond to the statistical uncertainties in the observed event yields.}
  \label{fig:mvis}
\end{figure}

Figure~\ref{fig:mvis} shows the distribution of the visible invariant mass \mvis of the reconstructed \mutau system when applying \DeepTau~v2.5, compared to the application of the previous version. The working points chosen are Tight for \Dm, Medium for \Djet, and VVLoose for \De. A reduction of the background from misidentified jets is estimated to be $\sim$30\%, especially visible in the decrease of the \Wjets process.

\section{Performance with \texorpdfstring{$\sqrt{s}=13\text{ and }13.6\TeV$}{sqrt(s)=13 and 13.6 TeV} data}
\label{sec:calibration}

The calibration of the \tauh identification algorithm consists of measuring identification (or misidentification) rate scale factors and energy scale corrections. 
The scale factors are multiplicative correction factors applied to simulated events to account for differences in the performance of the detector and of the reconstruction and identification algorithms between data and simulation. They are typically defined as the ratio of an efficiency (or misidentification rate) measured in real data to that measured in simulation, and are used to improve the agreement between predicted and observed data.

The \tauh identification efficiency scale factors and energy scale corrections are measured separately for the two data-taking years using a tag-and-probe method~\cite{CMS:2010svw} on $\PZ\to\PGt\PGt$ events in the \mutau final state.
For high-\pt \tauh candidates a separate measurement is performed using $\PWst \to \PGt\PGn$ events. 
In addition, we measured the scale factors for the misidentification rate of electrons and muons. All of these measurements are essential ingredients for physics analyses involving \tauh objects.

The scale factors are generally measured for all of the most used combinations of the \DeepTau \Djet, \Dm and \De discriminator working points. We discuss a subset of measurements here only for representative working points. Consistent results have been obtained for the other working points. We have not performed such efficiency measurements for the decay modes with missing charged hadrons (described in Section~\ref{sec:objects}); future analyses including these decay modes will need to derive appropriate corrections. 

The following results have been determined using the CMS statistical analysis tool \textsc{Combine}~\cite{CAT-23-001}.

\subsection{The \texorpdfstring{\tauh}{tauh} energy scale correction and identification scale factors}\label{tau_id_es}

The events with a \mutau final state are selected by requiring at least one well-identified and isolated muon, referred to as the ``tag'', alongside one \tauh candidate that meets loose preselection criteria, termed the ``probe''. The muon candidate must have $\pt > 25\GeV$, $\abs{\eta} < 2.4$, and a relative isolation of $I_{\PGm} < 0.15$. The \tauh candidate is required to have $\pt > 20\GeV$, $\abs{\eta} < 2.3$, and must pass a specified threshold on the \Djet discriminator. Additionally, the \tauh candidate must satisfy the VVLoose working point of the \De discriminator and the Tight working point of the \Dm discriminator to mitigate background from muons or electrons misidentified as \tauh candidates.
Only decay modes \oneProngZeroPizero, \oneProngOnePizero, \threeProngZeroPizero, and \threeProngOnePizero are considered.
In cases where multiple lepton or \tauh candidates are identified, the one with the highest \pt is selected. The selected muon and \tauh candidates must be separated by $\DR > 0.5$ and opposite-sign (OS) charges. The difference in $\eta$ between the reconstructed muon and the \tauh candidate is required to be $\abs{\Delta\eta} < 1.5$.

In addition to the \mutau event sample, a $\mumu$ event sample is defined to normalize the $\PZ \to \PGt\PGt$ event yields. This event sample adheres to the same trigger and muon selection criteria as the \mutau event sample, ensuring that related uncertainties partially cancel in the normalization scale factor.

The \tauh energy scale correction is defined as the factor that is applied to the reconstructed \tauh energy in simulation to bring it into agreement with the \tauh energy observed in real data. A \tauh identification efficiency scale factor is defined as the ratio of the number of \tauh candidates passing the selection criteria in real data to that in simulation. The scale factors are derived from a maximum likelihood fit to the \mvis distribution, alongside the expected and observed event yields in the \mumu control region. All known sources of systematic uncertainties are incorporated into the fit as nuisance parameters. Some of these uncertainties affect the yields of the involved processes, including integrated luminosity, muon identification, isolation, and trigger efficiencies, as well as uncertainties in the normalization of \ttbar, QCD multijet, and \Zjets backgrounds, and of quark and gluon jets misidentified as \tauh candidates. The remaining systematic uncertainties affect the shape of the \mvis distribution, including the energy scale for jets and leptons misidentified as \tauh candidates, the reweighting of the \pt distribution of the \PZ boson, and the uncertainties associated with the limited size of the simulated event samples.
Two distinct methods have been employed to derive the scale factors for 2022, yielding compatible outcomes: a separate fit for the identification efficiency scale factor and a combined fit for both the scale factor and the \tauh energy scale. These methods are described in the following sections.

\subsubsection{Individual fit} \label{sec:individual_fit}
The individual fit uses the baseline selection outlined above in Section~\ref{tau_id_es} with an additional $\mT<65\GeV$ requirement to improve the signal-to-background ratio. The first step is to measure the number of $\PZ\to\PGt_\PGm\tauh$ events in real data. This is done by estimating the number of $\PZ\to\PGt_\PGm\tauh$ events from simulation and estimating the contribution of background due to misidentified jets from observed events, along with other small backgrounds containing genuine hadronic taus from simulation, and then fitting the visible mass, \mvis, distribution with the $\PZ\to\PGt_\PGm\tauh$ normalization floating. The number of $\PZ\to\PGt_\PGm\tauh$ events in real data depends on the \PZ boson cross section, muon identification efficiency, trigger efficiency, integrated luminosity, and the \tauh identification efficiency. If the same procedure is followed for $\PZ \to \PGm\PGm$ events and a ratio is taken between $\PZ\to\PGt_\PGm\tauh$ and $\PZ\to\PGm\PGm$ events, then the \tauh identification efficiency can be determined. The \tauh identification efficiency scale factors can be derived using a binned maximum likelihood fit as described in Ref.~\cite{DeepTau_v2p1}.

The dominant backgrounds in the measurement of the \tauh identification efficiency scale factors are QCD multijet and \Wjets events. The QCD multijet background is estimated using a data-driven approach with same-sign (SS) events. Lepton pairs with the same charge sign are chosen from the selected events, and the QCD multijet yield is defined as the difference between the SS observed data events and the sum of all known non-QCD backgrounds estimated from simulation in the SS region. The \Wjets background is estimated from a high-\mT control region, while smaller backgrounds, such as the diboson, \Zjets, and top quark processes, are estimated from simulation.
 
Considering a new uncertainty in the QCD multijet background helps to estimate the precision of the OS/SS ratio used to extrapolate the QCD yield. This was validated using a simulated sample of QCD multijet events to derive central values and uncertainties for the ratio in the regions where both leptons are isolated. Because of the small sample size, the Loose \Djet working point was used for all scale factor derivations. A control region for the \Wjets background, including events at high \mT, was introduced to improve background modelling and to prevent the fit from compensating for mismodelling by artificially increasing the \Wjets yields, which could otherwise lead to an underestimation of the $\PZ\to\PGt_\PGm\tauh$ yield and thus the scale factor. Additionally, for the 2018 measurement, a control region enriched in QCD, defined by events with an anti-isolated lepton, was introduced to better constrain the QCD yield. This was not included for the 2022 measurement due to mismodelling in the control region.

In the scale factor measurement for \DeepTau~v2.5, we explore how the \tauh identification efficiency scale factors are impacted by the choice of the nominal \tauh energy scale value and associated uncertainties. Studies such as the investigation detailed in Section~\ref{sec:combined_fit} have demonstrated that variations in \tauh energy scale values can significantly affect the efficiency scale factors. Moreover, the \tauh energy scale has been observed to depend on factors such as \tauh identification requirements, \pt, and decay modes, suggesting that the current uncertainty estimates may not fully capture all relevant variations. To address this, we developed an alternative method for handling \tauh energy scale uncertainties by externalizing them from the \tauh identification efficiency scale factor determination.
In this approach, the \tauh energy scale is fixed to the nominal value and a fit is performed to determine the \tauh identification efficiency scale factors without allowing the fit to overconstrain the \tauh energy scale uncertainties.
The \tauh energy scale is then set as its postfit value and is shifted by $\pm$1 standard deviations in a second fit used to measure the corresponding variations in terms of efficiency scale factors and, hence, estimate the \tauh energy scale related uncertainty.

This method provides a more reliable treatment of \tauh energy scale uncertainties, ensuring that they are not overly reduced by the fit, and allowing for more consistent application across different analyses. This approach is particularly relevant for the $\tauh\tauh$ channel, where mismodelling effects are more pronounced and proper \tauh energy scale uncertainty treatment is critical for accurate results.

The scale factors are extracted in different \tauh \pt bins in order to take into account any \pt dependence. The lower bin edges 20, 25, 30, 35, 40, 50, 60, 80, and 100\GeV are used, where the last bin contains the scale factors for \tauh objects with $\pt>100\GeV$. The scale factors are derived for each \pt bin of each decay mode of each data-taking period. A Laurent polynomial function is used to describe the trend in the scale factors, as it offers a better fit than a linear function. The Laurent polynomial is more flexible and reduces the $\chi^2$ per degree of freedom compared to a linear fit. Unlike standard polynomials, Laurent polynomials can include terms with negative powers, making them more suitable in our case.

\begin{figure}[bt!]
    \centering
    \includegraphics[width=\cmsFigWidth]{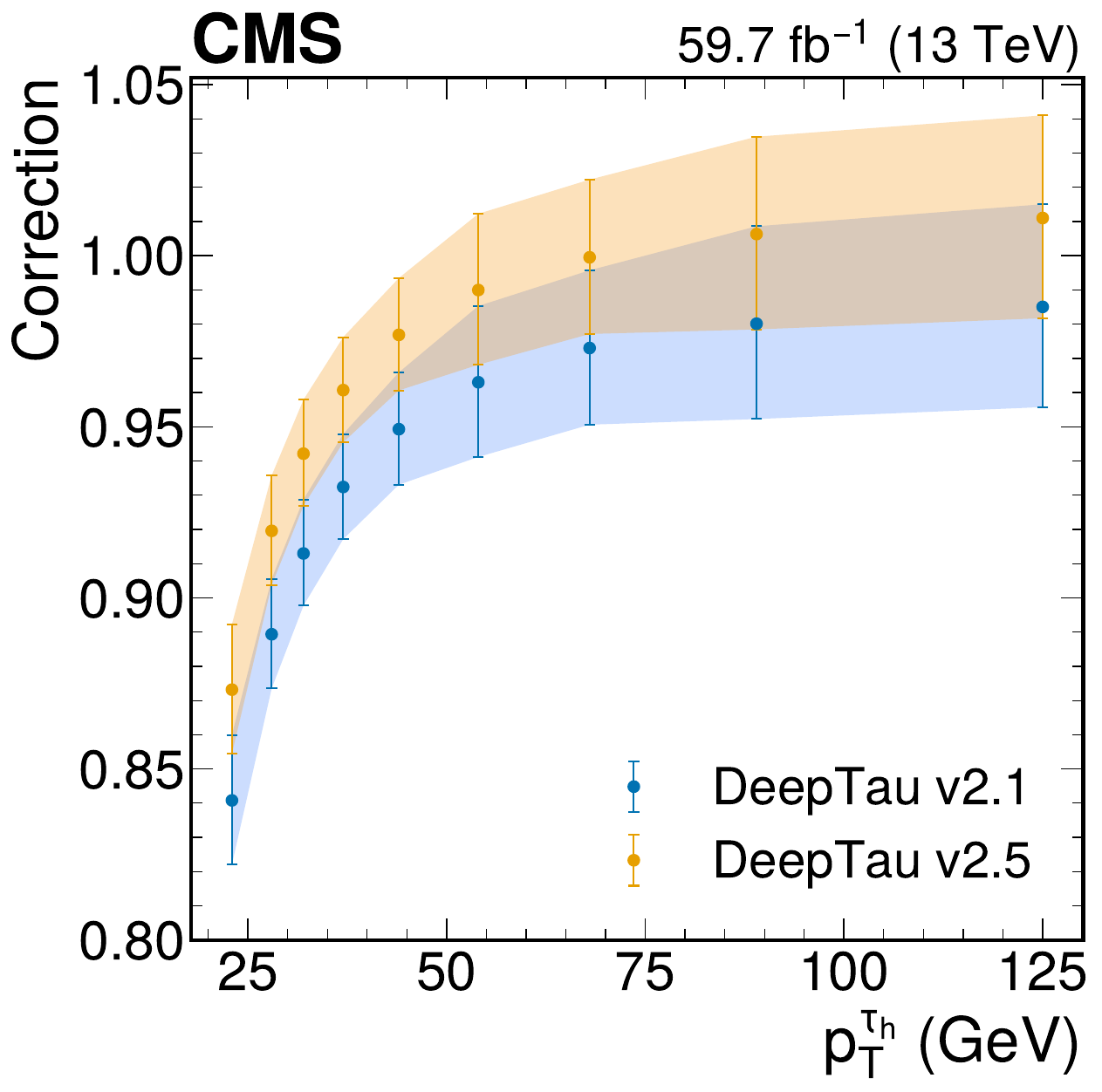}
    \includegraphics[width=\cmsFigWidth]{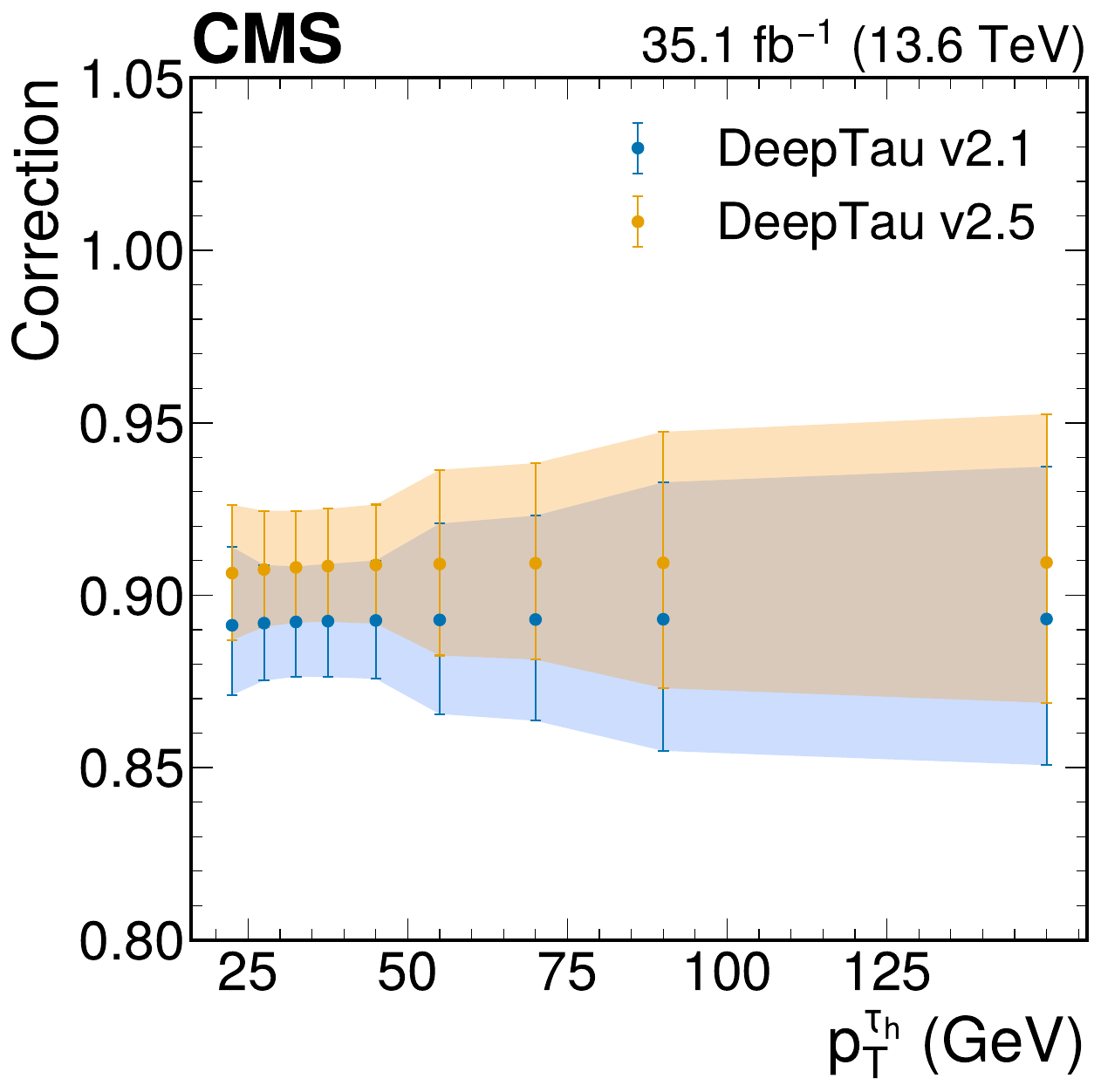}
    \caption{The data-to-simulation scale factors of the \tauh identification efficiency as functions of \pt in the 2018 (left) and 2022 (right) data-taking periods, including all \tauh decay modes, and requiring the \Djet Medium working point (see Table~\ref{tab:wp-definition}) and $\mT(\pt^\PGm,\ptmiss) < 65\GeV$. The vertical bars correspond to the combined statistical and systematic uncertainties in the individual scale factors. For a fair scale factor comparison in 2022, the tau energy scale have been fixed to the one measured for \DeepTau~v2.5 which showcases higher genuine \tauh purity.}
    \label{fig:individual_scalefactors}
\end{figure}

Figure~\ref{fig:individual_scalefactors} shows the \tauh identification efficiency scale factors for the 2018 and 2022 data-taking periods. 
There is a better understanding of the data collected during 2018, thanks to extensive studies that allowed for optimized reprocessing with improved algorithm calibrations. The 2022 dataset is therefore likely to yield weaker agreement with simulation.
The scale factors are derived for each \pt bin of each decay mode, but the summary plot in Fig.~\ref{fig:individual_scalefactors} combines the different decay modes according to their branching fractions and reconstruction efficiencies. The Medium \Djet, VVLoose \De, and Tight \Dm working points are used.

The scale factors are generally a bit smaller than 1, typically within 20\% of unity. Compared to the previous version of \DeepTau, the scale factors measured for v2.5 are systematically closer to one for both 2018 and 2022 dataset, demonstrating that the domain adaptation approach leads to better agreement between data and simulation. The scale factors below 1 can be attributed to various factors, such as inaccuracies in the modelling of hadronization and imperfections in the simulation of the detector alignment and track hit reconstruction efficiencies.

\begin{figure}[bt!]
    \centering
    \includegraphics[width=\cmsFigWidth]{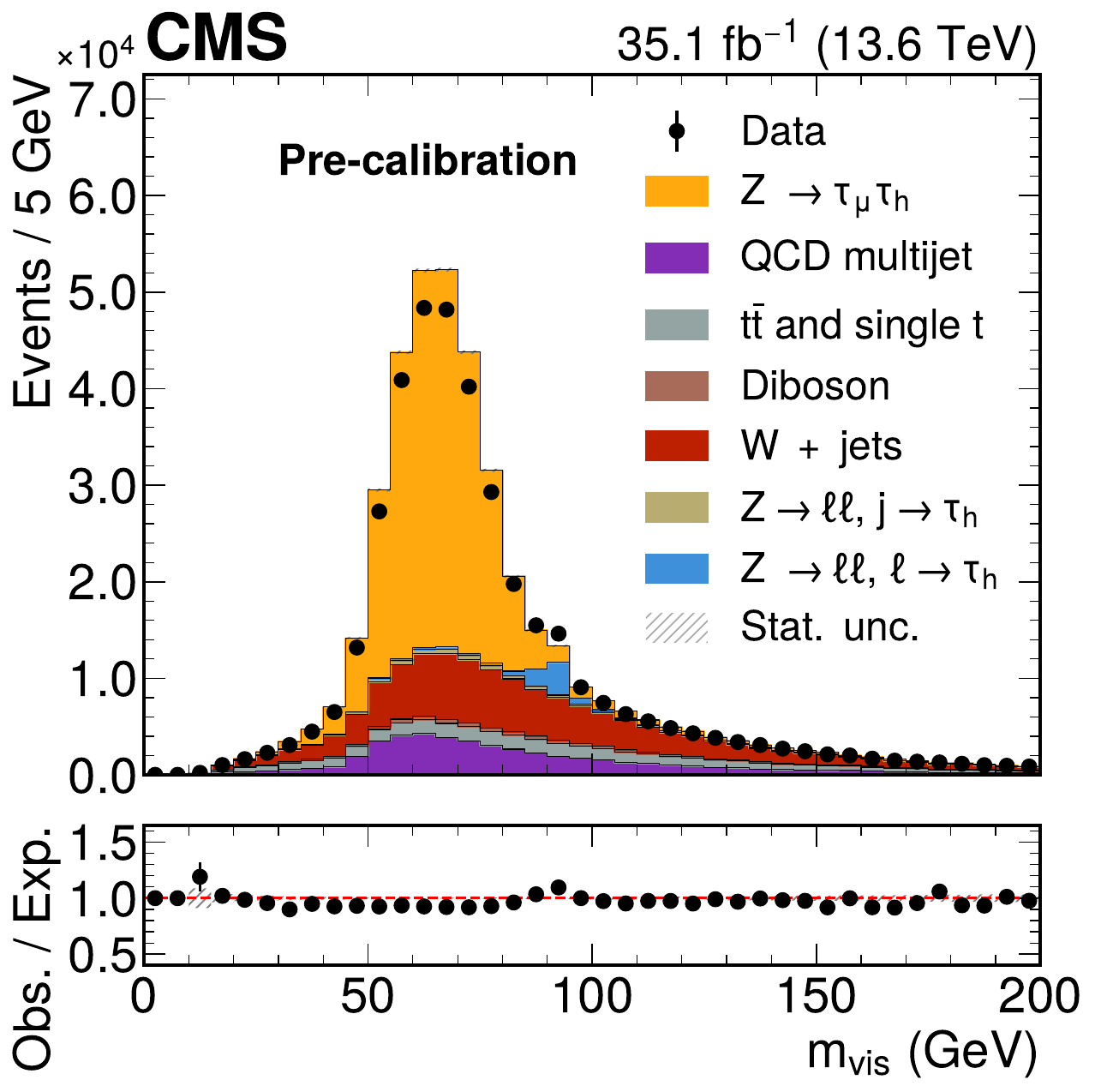}
    \includegraphics[width=\cmsFigWidth]{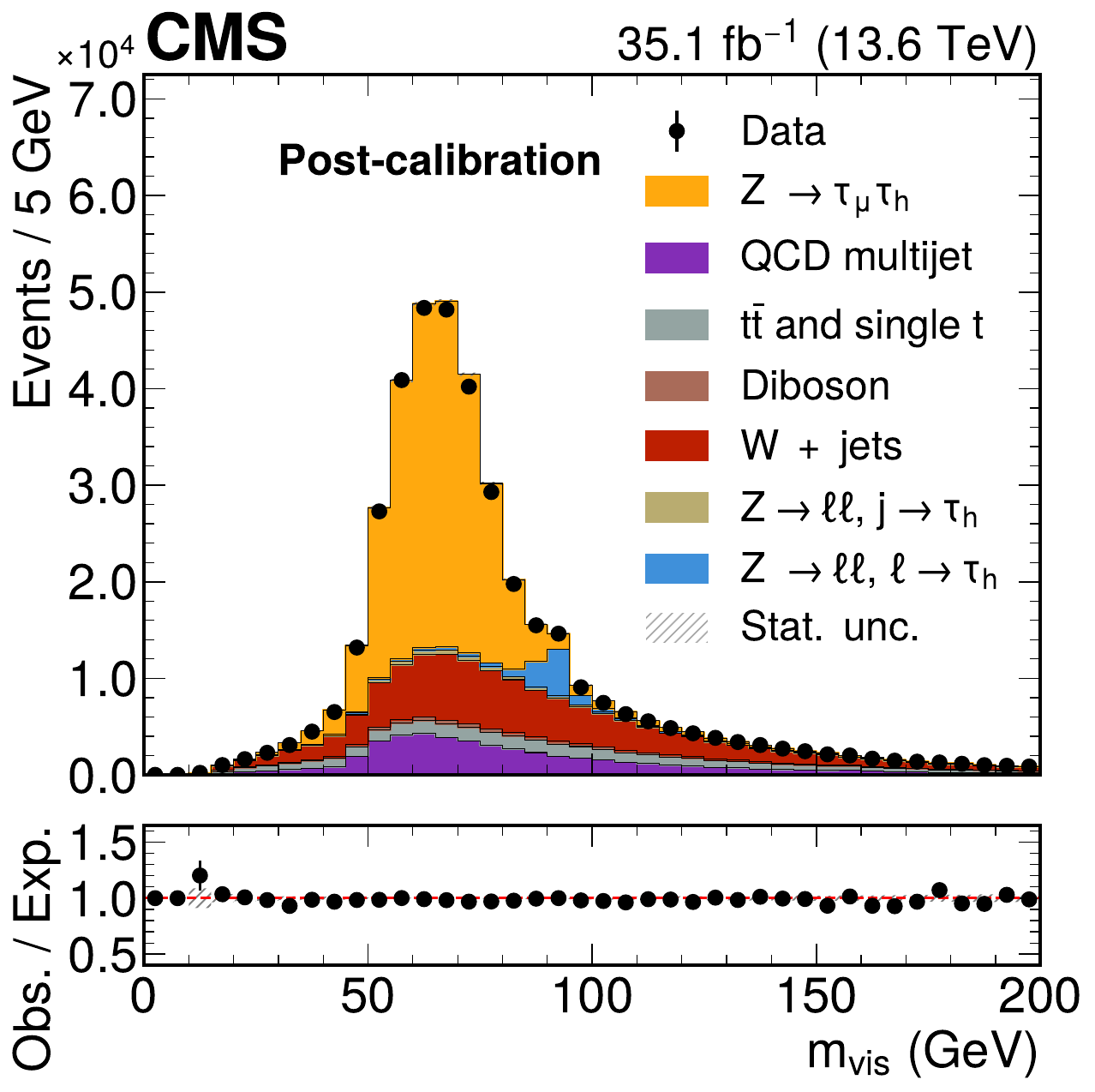}
     \caption{
     The \mvis distribution in the $\PZ\to\PGt_\PGm\tauh$ channel for the 2022 dataset before (left) and after (right) the full calibration. The \DeepTau working points used are: Medium for \Djet, VVLoose for \De and, Tight for \Dm (see Table~\ref{tab:wp-definition}).
     The application of correction factors improves the agreement between data and simulation.}
    \label{fig:individual_control_plots_mutau}
\end{figure}

\begin{figure}[hbt!]
    \centering
    \includegraphics[width=\cmsFigWidth]{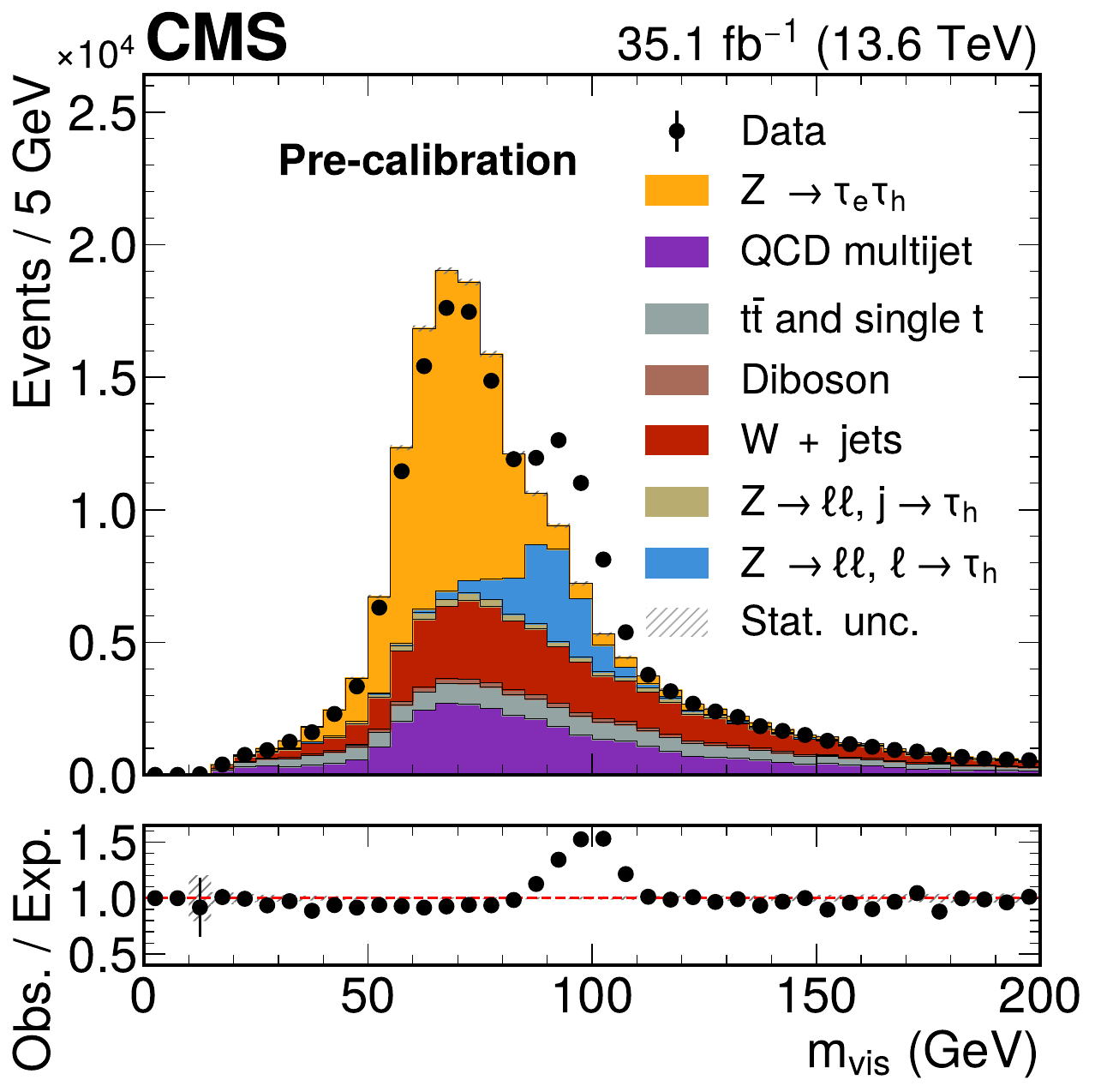}
    \includegraphics[width=\cmsFigWidth]{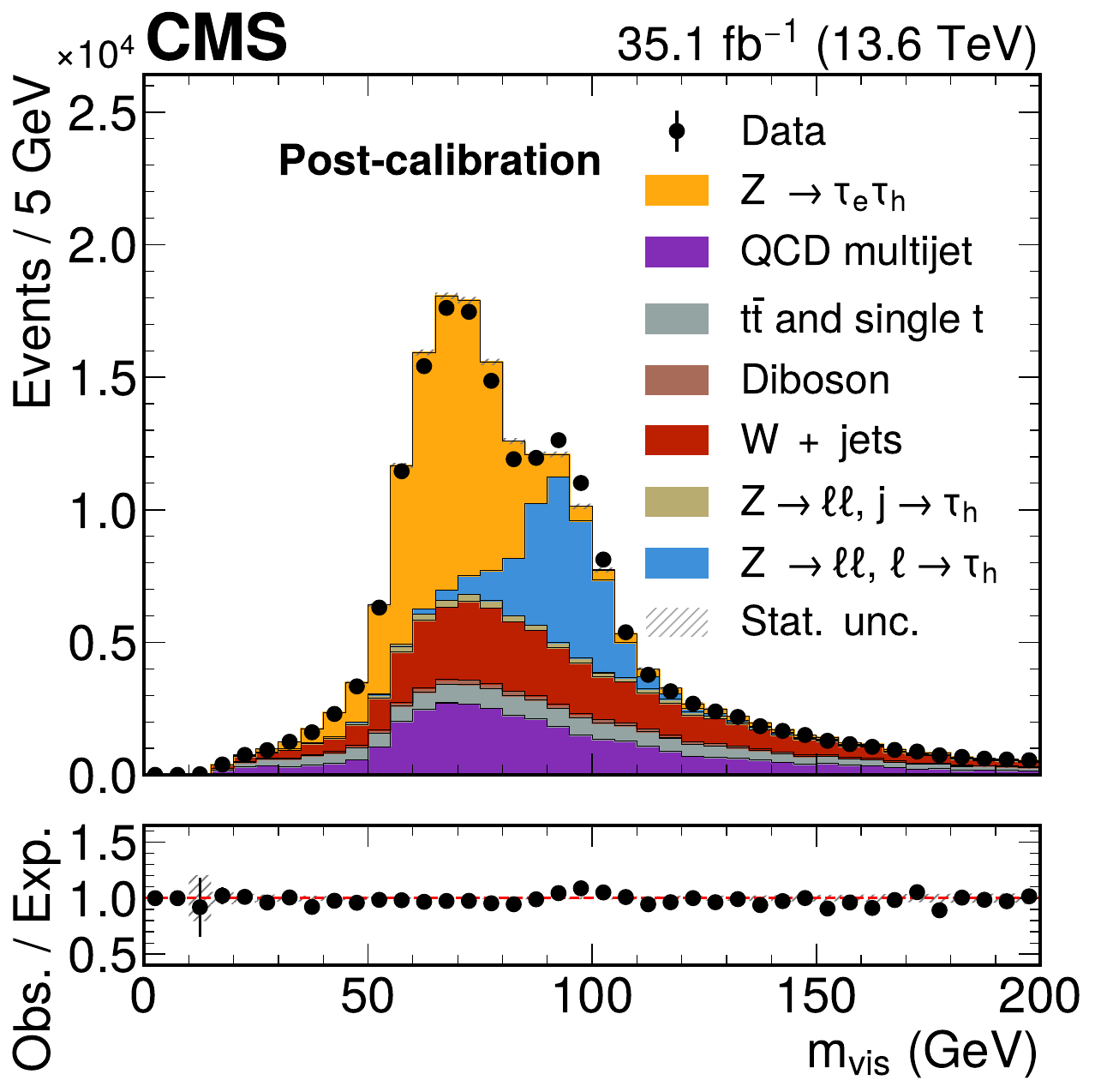}
    \caption{
    The \mvis distribution in the $\PZ\to\PGt_\Pe\tauh$ channel for the 2022 dataset before (left) and after (right) the full calibration. The \DeepTau working points used are: Medium for \Djet, Tight for \De and, Tight for \Dm (see Table~\ref{tab:wp-definition}). Specific 2022 detector conditions that affected electron reconstruction are not perfectly modelled in the simulation. As a result, the amount of electrons misidentified as \tauh is enhanced in data with respect to simulated events.
    The application of correction factors improves the agreement between data and simulation.}
    \label{fig:individual_control_plots_etau}
\end{figure}

To verify the performance of the derived scale factors a set of control plots are produced. Figures~\ref{fig:individual_control_plots_mutau} and~\ref{fig:individual_control_plots_etau} show the 2022 \mvis distribution in the $\PZ\to\PGt_\PGm\tauh$ and $\PZ\to\PGt_\Pe\tauh$ channels, respectively, before (left) and after (right) the application of the \tauh identification efficiency scale factors and the misidentification rate scale factors that will be later described in Section~\ref{sec:leptonFR}. The data-to-simulation agreement is highly improved by the application of these scale factors.

\subsubsection{Combined fit} \label{sec:combined_fit}

The newly proposed technique combines the \tauh identification efficiency and energy scales estimation into a single fit, simplifying the measurement by applying a unified adjustment to the same distribution, using consistent systematics, and accounting for correlations between the two factors. This method was developed with 2018 data, where the limited size of the dataset prevented the measurement of identification scale factors and energy correction in both decay mode and multiple transverse momentum bins. When applied to the 2022 dataset, which had an even smaller event sample size, the fit was therefore performed only as a function of decay mode, with the \tauh energy correction profiled and the identification scale factor treated as a free parameter. Despite these limitations, the method remains reliable and is presented here, as it is expected to fully demonstrate its potential with the large dataset to be collected between 2024 and 2025. 
This section details the combined fit method which was developed and validated with 2018 data, and applied to determine the \tauh energy scale corrections for the 2022 dataset.

In the fit, the ratio of the observed to expected efficiency for a genuine \tauh candidate to pass the selection is treated as a free parameter, allowing for the extraction of an identification efficiency scale factor relative to the simulated efficiency. Similarly, the \tauh energy scale is included as a free parameter in the fit.

Binned \mvis distributions are generated in simulation for \tauh energy scale variations ranging from $-3$ to 3\% for the 2018 dataset and from $-10$ to 10\% for the 2022 datasets, in steps of 0.1\%. Interpolation between the discrete histograms is performed using the method described in Ref.~\cite{Baak_2015}, enabling a smooth variation of the energy scale. This approach improves upon standard interpolation methods that rely solely on fixed "up" and "down" variations, providing finer resolution and enhancing both the accuracy and robustness of the model.

For each energy scale shift, a maximum likelihood fit is performed, incorporating the combined expectations from $\PZ \to \PGt\PGt$ events in the \mutau final state with the relevant background processes and systematic uncertainties outlined in previous sections. The profile likelihood scans exhibit significant fluctuations due to limited event sample size. To robustly determine the minimum of the negative log-likelihood (NLL) and to extract the corresponding uncertainties, the NLL profiles are fitted using asymmetric parabolic functions. 

For the 2018 dataset, a selection criterion of $\mT(\pt^\PGm,\ptmiss) < 65\GeV$ was applied, and scale factors are provided for the \Djet Medium working point. The fit was performed across the \tauh decay modes and two transverse momentum intervals of the \tauh candidate: $[20, 40]$ and $[40, 200]\GeV$. Although a finer granularity, similar to the division used in the separate fit, was technically feasible and underwent preliminary testing for \tauh energy scales, the small size of the event sample resulted in significant instabilities in the fit. Consequently, these measurements are based on a two-region division for \pt to ensure a more stable fit.

\begin{figure}[bt!]
    \centering
    \includegraphics[width=\cmsFigWidth]{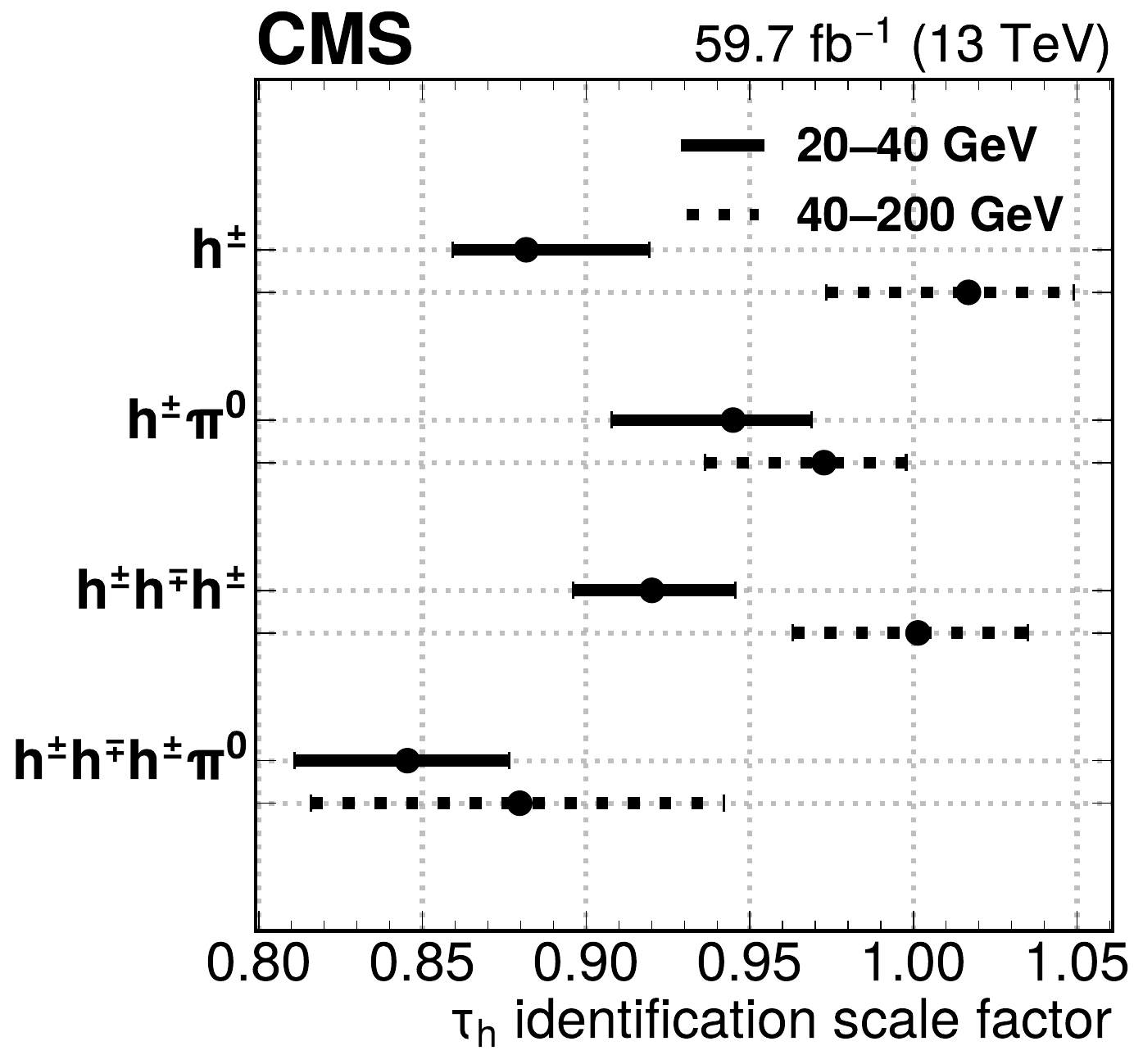}
    \includegraphics[width=\cmsFigWidth]{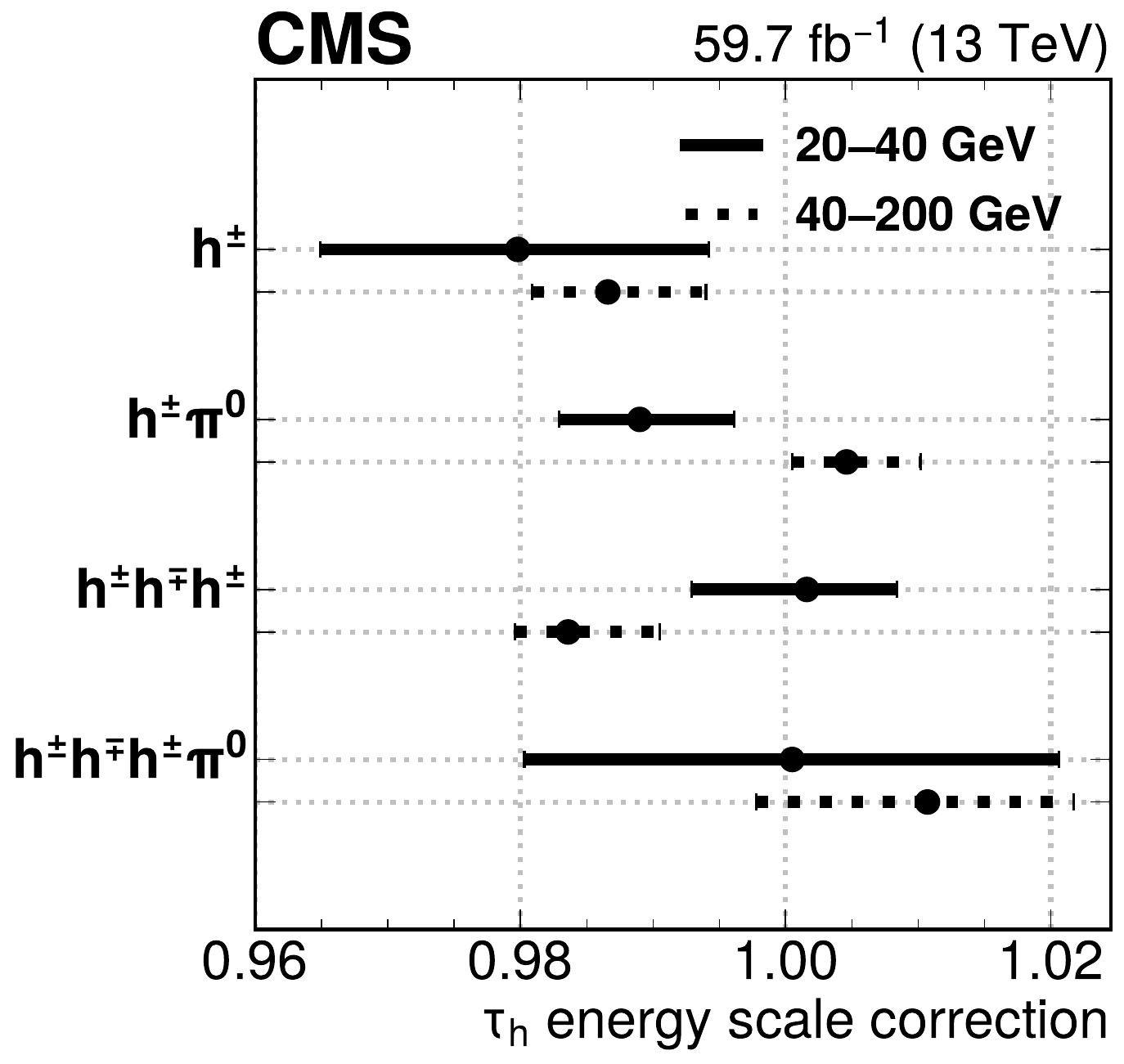}
    \caption{Summary of \tauh identification efficiency (left) and \tauh energy scale corrections (right) across the \tauh decay modes and \pt regions for 2018 with $\mT(\pt^\PGm,\ptmiss) < 65\GeV$ and the \Djet Medium working point (see Table~\ref{tab:wp-definition}). The horizontal bars represent the total uncertainty on the measurements, combining both statistical and systematic contributions.}
    \label{fig:IDSF_TES_DM_results}
\end{figure}

A summary of the results as a function of \pt is presented in Fig.~\ref{fig:IDSF_TES_DM_results}. A general trend of higher identification scale factors in the high-\pt region is observed, attributed to improved modelling in these regions. A low dependence between the scale factors in the lower \pt range is observed, while there is higher dependence in the higher-\pt range.
This dependence is attributed to bin migration in the visible mass distribution of each \pt region when varying the \tauh energy scale. This migration leads to normalization changes, which, in turn, induce a dependence between the \tauh identification efficiency and the \tauh energy scales. These findings highlight the necessity of a simultaneous fit for both scale factors to accurately capture their dependence. However, this approach is currently constrained by the number of \pt regions that can be effectively used, and further division will require a larger event sample, which could also help mitigate bin migration effects.

\begin{figure}[!tb]
    \centering
    \includegraphics[width=\cmsFigWidth]{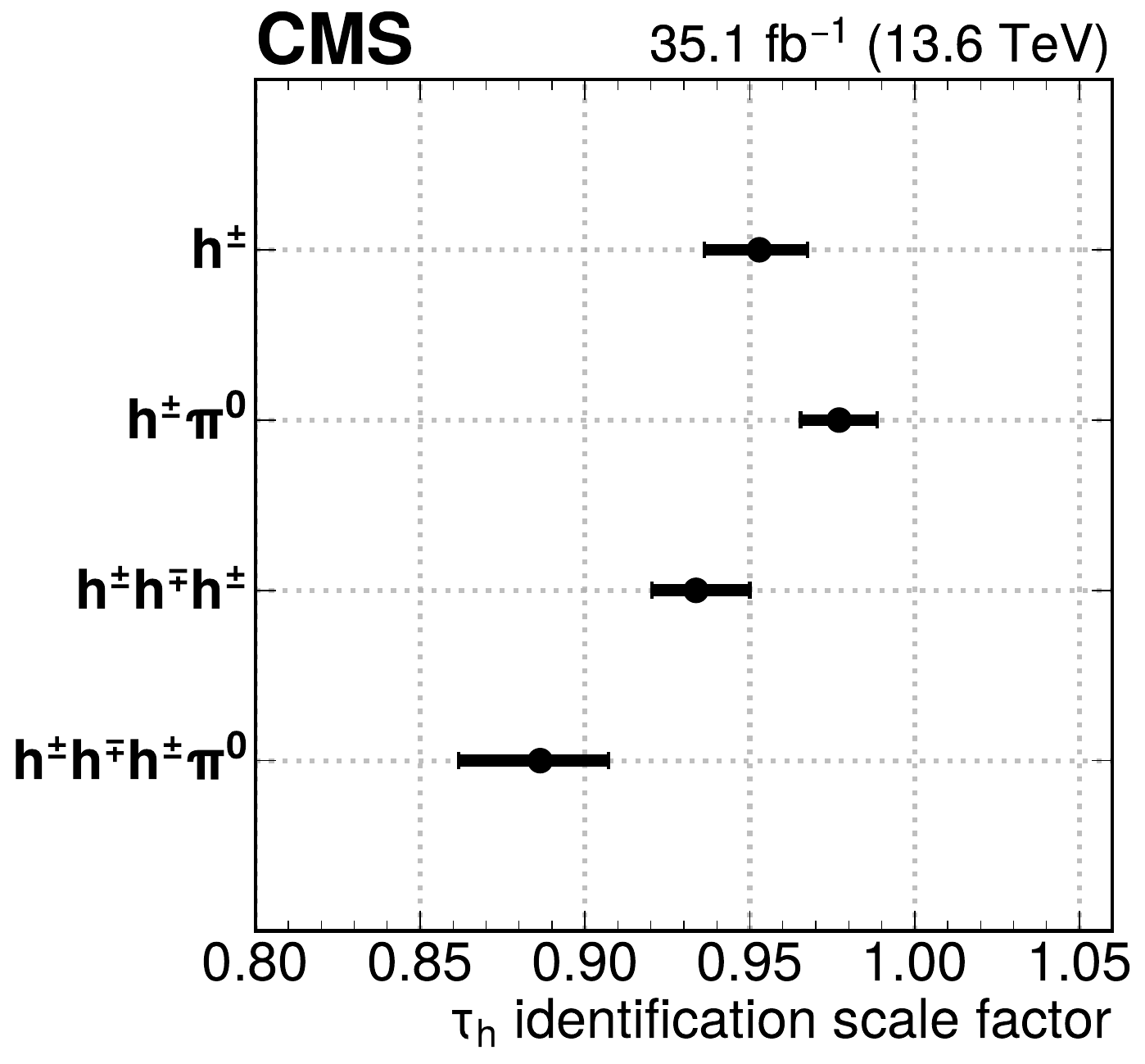}
    \includegraphics[width=\cmsFigWidth]{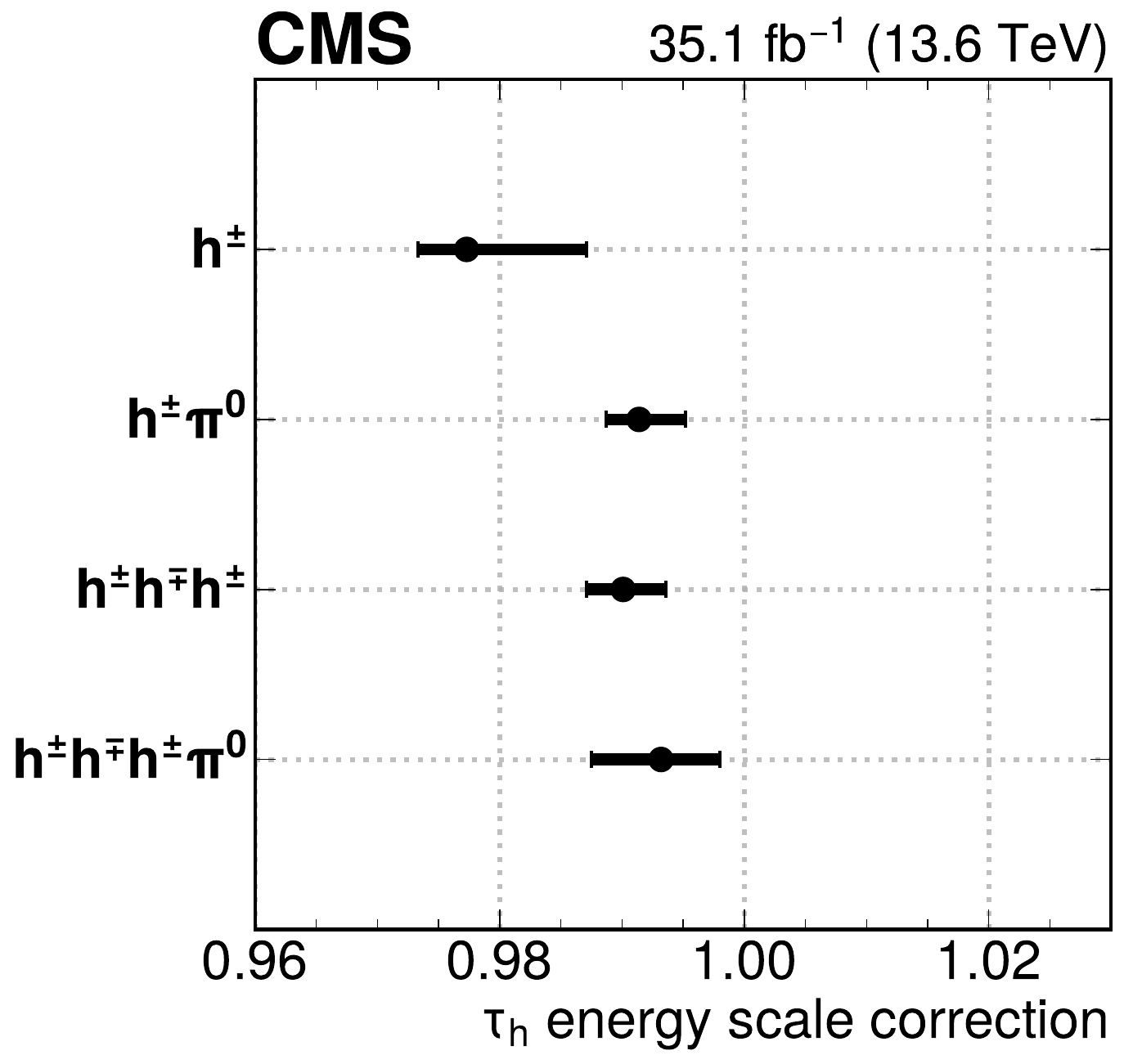}
    \caption{Summary of \tauh identification efficiency (left) and \tauh energy scale corrections (right) across the \tauh decay modes for 2022 with $\mT(\pt^\PGm,\ptmiss) < 40\GeV$ and \Djet Medium working point (see Table~\ref{tab:wp-definition}). The horizontal bars represent the total uncertainty on the measurements, combining both statistical and systematic contributions.}
   \label{fig:combined_fit_2022preEE}
\end{figure}

The results of the combined fit for the \tauh identification efficiency and energy scale for the 2022 dataset are presented in Fig.~\ref{fig:combined_fit_2022preEE} for the \Djet Medium working point. The fit is only performed across the \tauh decay modes because of the smaller dataset size. 

The \tauh energy scale results are all consistent with unity within 3\%. The \tauh identification efficiency scale factors are slightly closer to unity compared to those obtained in the 2018 measurement, likely due to the use of the \POWHEG generator in this specific case, as opposed to \MGvATNLO used in the others. 

Additional studies have been conducted, including variations in \mvis binning and range, optimization of the $\mT(\pt^\PGm,\ptmiss)$ requirement, and testing with control regions. To account for these differences, an additional uncertainty of 1\% is applied to the \tauh energy scales across all decay modes regions. Additionally, combined fits were performed across regions while constraining nuisance parameters, yielding results consistent within 5\%.

For the 2022 datasets, the combined fit shows a low dependency between the scale factors, with the results from the separate and combined fits being compatible. However, statistical limitations restrict the ability to divide the fit into \pt regions.

The choice between the combined fit and separate fit methods for deriving scale factors involves balancing the benefits and challenges of each approach. The combined fit method allows for simultaneous determination of the \tauh energy scale and identification efficiency scale factors, accounting for potential correlations and offering a comprehensive treatment of uncertainties. However, because it requires fitting not only the normalization of the \mvis distribution but also its shape, it demands a large dataset to achieve a stable fit. Additionally, statistical limitations prevent dividing the dataset into finer \pt regions, that poses a challenge for analyses sensitive to the \pt dependence of \tauh. In contrast, the separate fit method provides enhanced flexibility by allowing independent determination of each scale factor, making it well-suited for analyzing different kinematic regions. The separate fit does, however, assume low correlation between the factors, which must be verified for validity, and may offer a slightly less comprehensive uncertainty treatment if correlations are overlooked. To address these challenges, an alternative method for handling \tauh energy scale uncertainties was proposed by externalizing them from the \tauh identification scale factor determination, as described in the previous section. Ultimately, the selection of the method relies on statistical constraints of the analysis and on the specific requirements, particularly in regard to kinematic sensitivity. 

\subsection{The lepton misidentification rate scale factors}
\label{sec:leptonFR}

The performance of the \DeepTau~v2.5 algorithm in rejecting light leptons (\Pell) mimicking \tauh candidates is studied by measuring the muon and electron misidentification scale factors in $\PZ\to\Pell\Pell$ events.
We exploit a modified tag-and-probe method where a well-identified and isolated light lepton serves as a ``tag'' and a loosely selected \tauh candidate acts as a ``probe''. Events where the \tauh candidate satisfies a specific working point of the $D_{\Pell}$ discriminator form the \pass region, whereas the residual events define the \fail region.
The misidentification rate scale factors derived from the measurement account for event migration between the \pass and \fail regions. In addition, a normalization correction that affects both the \pass and \fail regions equally is applied.

The following two sections explain in detail the muon and electron misidentification rate scale factor measurements.

\subsubsection{The muon misidentification rate scale factors}
The measurement of scale factors of the muon misidentification rate requires OS and well-separated \mutau pairs, whose selection follows the one outlined in Section~\ref{tau_id_es}. The measurement is performed for multiple bins of the $\abs{\eta}$ of the \tauh and for all \Dm working points, while \De and \Djet working points are fixed at VVLoose and Medium, respectively.
The $\abs{\eta}$ bin edges of 0, 0.4, 0.8, 1.2, 1.7, and 2.5 are used to cover the barrel and endcap regions of the CMS detector, including their overlap region.

Muon misidentification rate scale factors are derived from a simultaneous fit of simulation to data. The fit model includes two parameters of interest, applied to the $\PZ\to\Pell\Pell$ process in the \pass region: a normalization scale factor and a migration scale factor. An additional dependent parameter is introduced in the \fail region to propagate any event migration. The final result of the measurement is the product of the two scale factors from the \pass region.

Systematic variations for the measurement are consistent between the 2018 and 2022 datasets and include a $\pm$3\% \tauh energy scale variation, a $\pm$10\% muon energy scale variation, and a $\pm$10\% muon energy resolution. The latter is a conservative estimate introduced to account for the observed disagreement in the shape of visible mass distribution between data and simulation in the early 2022 dataset.

\begin{figure}[h!]
    \centering
    \includegraphics[width=\cmsFigWidth]{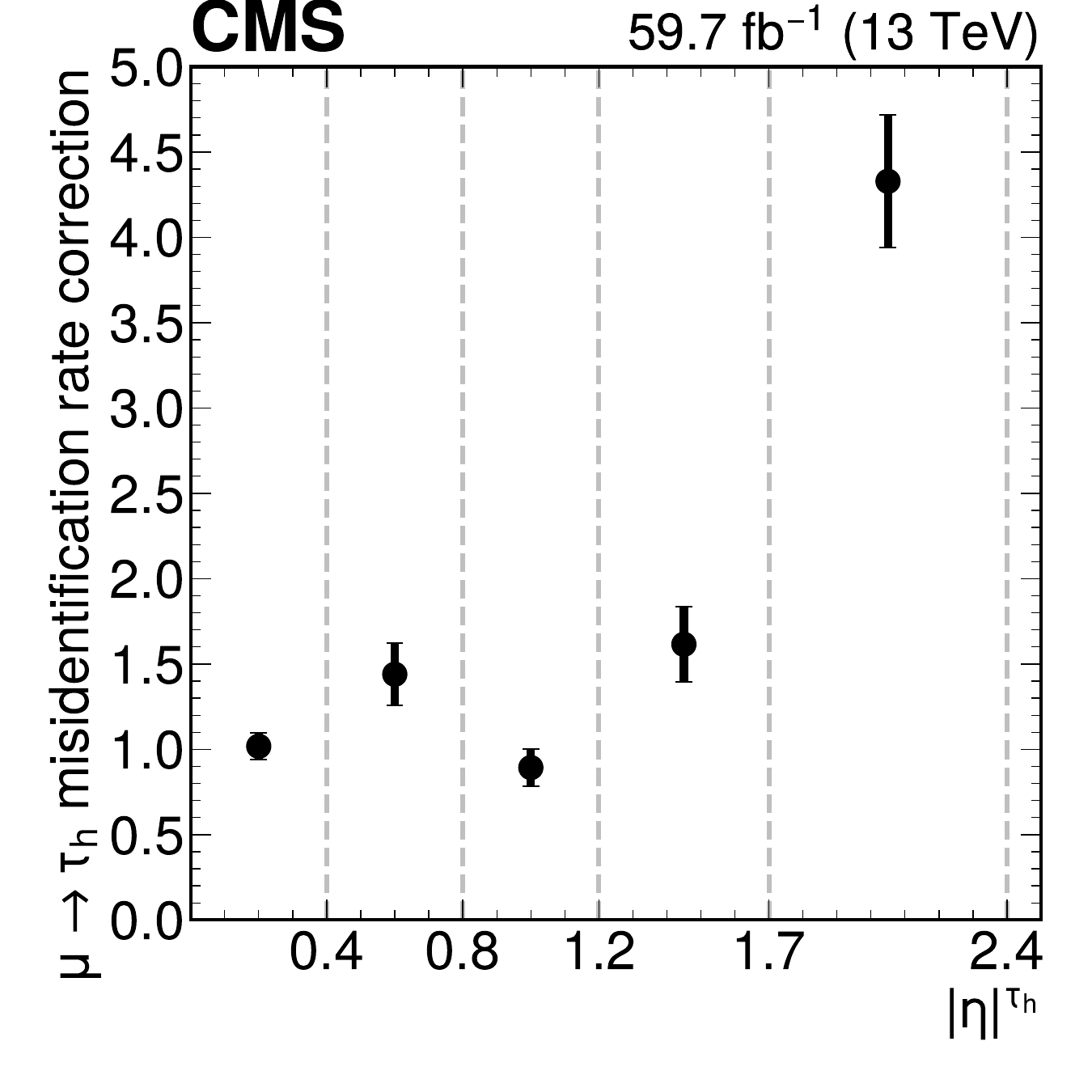}
    \includegraphics[width=\cmsFigWidth]{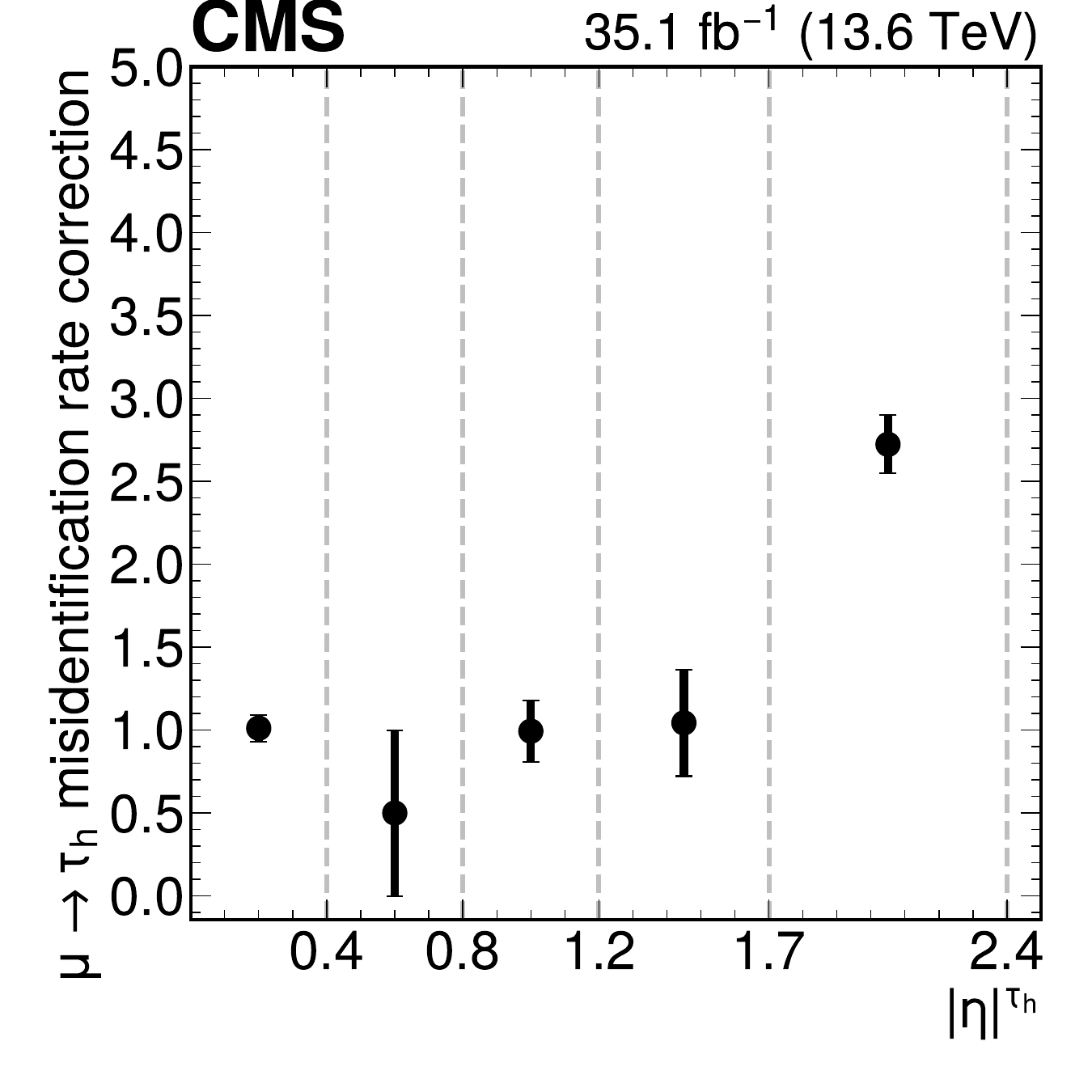}
    \caption{Muon misidentification rate scale factors binned by the \tauh $\abs{\eta}$ for the Medium \Dm working point (see Table~\ref{tab:wp-definition}). Measurement for the 2018 dataset is shown on the left and for the 2022 dataset on the right. The dashed lines indicate the boundaries of the \tauh $\abs{\eta}$ bins. The vertical bars represent the total uncertainty on the measurements, combining both statistical and systematic contributions.}
    \label{fig:mutau_res}
\end{figure}

Measurement results are summarized in Fig.~\ref{fig:mutau_res}, which shows the muon misidentification rate scale factors for the Medium working point of the \Dm discriminator. These results include measurements done for both the 2018 and the 2022 datasets. The overall behaviour of the scale factor values is similar between different \Dm working points, except for tighter working points, which tend to have slightly higher scale factors values. This stems from divergence between data and simulation in the \Dm output score, which is not uniform across the score range and is most pronounced in the region most pure in genuine \tauh. Choosing a tighter working point can therefore increase the data-simulation disagreement. The application of the misidentification rate scale factor reduces the difference between data and simulation, as visible in the enhancement of the simulated $\PZ\to\PGm\PGm$ peak around 90\GeV in the right panel of Fig.~\ref{fig:individual_control_plots_mutau}.

A significant deviation of the scale factors from unity is observed in the high-$\abs{\eta}$ region. Muons traversing this detector section may be more poorly reconstructed and consequently misidentified as a \tauh candidate. This effect is not properly captured by simulations.

\subsubsection{The electron misidentification rate scale factors}

The $\etau$ events for the measurement of scale factors of the electron misidentification rate are selected requiring a well-isolated electron triggering the event with \pt greater than 33\GeV and a \tauh candidate with a minimum \pt of 20\GeV passing the working points Medium for \Djet, Tight for \Dm and VVLoose or Tight for \De. The scale factors are derived separately in the ECAL barrel region ($\abs{\eta} < 1.46$) and in the ECAL endcap region ($\abs{\eta} > 1.56$) and in different \tauh decay modes.

The method is similar to the one outlined for muons, except for how the energy scale of leptons misidentified as \tauh candidates is treated. In this fit, the energy scale of electrons misidentified as genuine \tauh candidates is included as a parameter of interest alongside the misidentification rate scale factor, allowing both corrections to be extracted simultaneously. This was not attempted in the measurement of the misidentified muon energy scale because of the smaller contribution of the misidentified-muon background in the \mutau channel.

\begin{figure}[h!]
    \centering
    \includegraphics[height=0.4\textwidth]{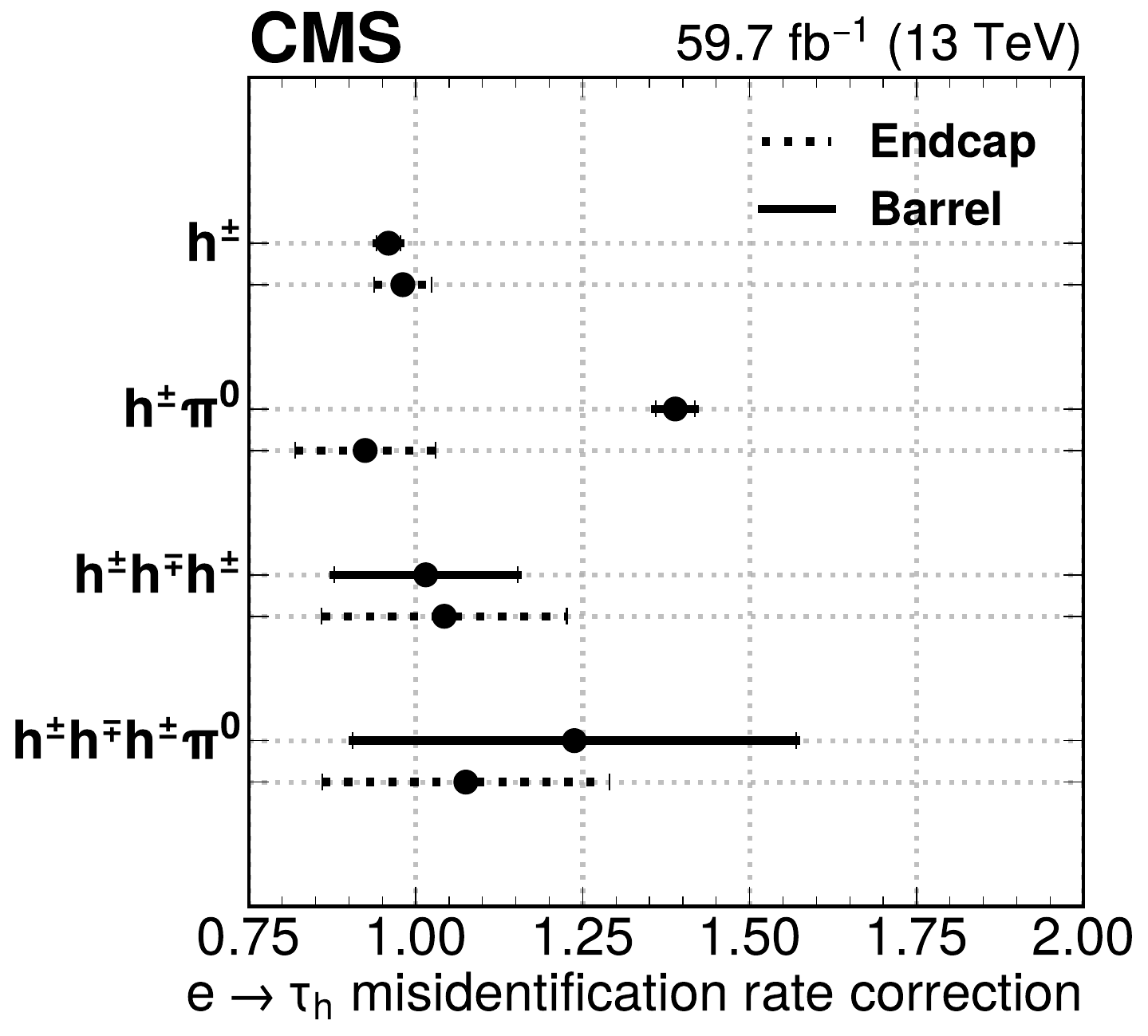}
    \includegraphics[height=0.4\textwidth]{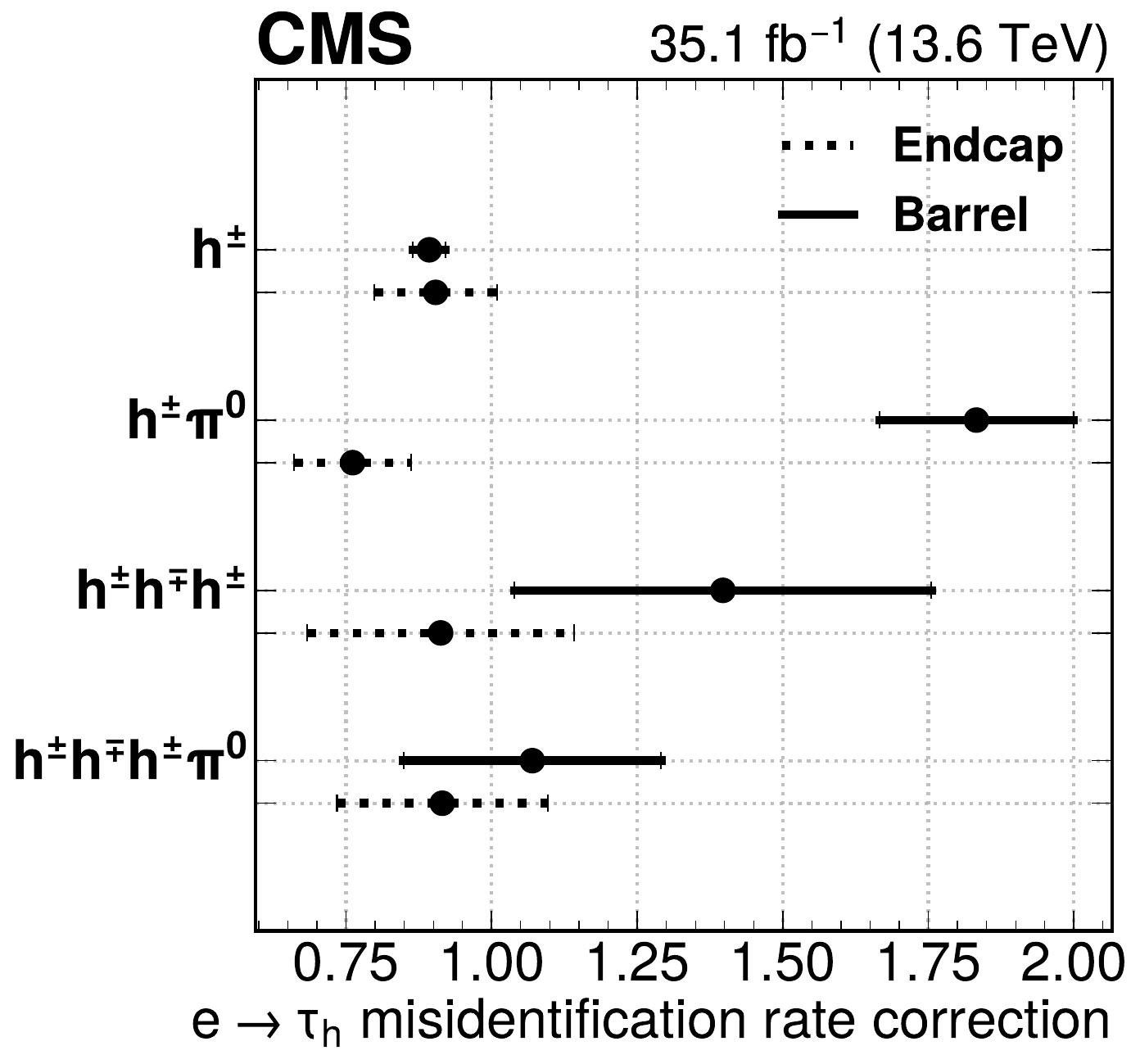}
    \caption{Summary plots of results for electron misidentification rate scale factors divided in decay modes and $\eta$ regions for the VVLoose \De working point (see Table~\ref{tab:wp-definition}). The corrections are shown for the 2018 (left) and 2022 (right) datasets. The horizontal bars represent the total uncertainty on the measurements, combining both statistical and systematic contributions.}
    \label{fig:etau_res}
\end{figure}

Binned \mvis distributions are generated in simulation for the energy scale variations of the electron misidentified as a genuine \tauh candidate ranging from $-10$ to 10\% for 2018 dataset, and from $-25$ to 25\% for 2022.
Summary plots of the misidentification rate scale factors are presented in Fig.~\ref{fig:etau_res} for the VVLoose working point of the \De discriminator, for 2018 (left) and 2022 (right). For the 2018 dataset, the corrections are all compatible with unity, except for the \oneProngOnePizero decay mode in the barrel. We can see a similar trend in the 2022 dataset, but with scale factors on average farther from unity. The extracted energy scales are always within $\simeq$5\% of unity for 2018 and $\simeq$10\% for 2022. For tighter working points, the same considerations hold with slightly higher values for the barrel scale factor for the \oneProngOnePizero decay mode (up to $\simeq$30\% increase for 2022).

The effect of applying the scale factors is visible in the enhancement of the simulated \PZ boson peak at 90\GeV in Fig.~\ref{fig:individual_control_plots_etau} (right). The derived scale factors correct the simulation for the observed increase of misidentified electrons in collision data, improving the general data-to-simulation agreement. 

\subsection{The \texorpdfstring{\tauh}{tauh} identification scale factors at high \texorpdfstring{\pt}{pt}}

Many analyses at the LHC deal with final states involving tau leptons with high transverse momentum. One example is searches for heavy neutral and charged Higgs bosons, predicted by models with an extended Higgs sector. The identification efficiency of genuine \tauh candidates with high transverse momentum ($\pt > 100\GeV$) can be measured using a sample of events in which a highly virtual \PW boson ($\mWst > 200\GeV$) is produced at small transverse momentum (and thus has little hadronic activity) and decays into a tau lepton and a tau neutrino, $\PWst\to\PGt\PGnGt$.

The measurement of the \tauh identification efficiency scale factors relies on two samples.
The data-simulation corrections for the \tauh identification efficiency are extracted from a measurement region that is enriched in $\PWst\to\PGt\PGnGt$ events. Additionally, a control region dominated by $\PWst\to\PGm\PGnGm$ events is used to constrain the normalization of the \PWst process in the $\mWst > 200\GeV$ phase space. The measurement is performed in two bins of \tauh transverse momentum: $100<\pt<200\GeV$ and $\pt>200\GeV$.

The $\PWst\to\PGt\PGnGt$ sample is selected with the trigger requiring reconstructed \ptmiss to be greater than 120\GeV, whereas the control sample of $\PWst\to\PGm\PGnGm$ decays is selected with a single-muon trigger with \pt threshold of 27\GeV. Selection of the $\PWst\to\PGt\PGnGt$ sample requires a \tauh candidate with $\pt>100\GeV$ and $\abs{\eta}<2.3$. Events selected in the $\PWst\to\PGm\PGnGm$  sample must contain an isolated prompt muon with $\pt>120\GeV$ and $\abs{\eta}<2.1$.
In both the measurement and control regions, the reconstructed \ptmiss must exceed 130\GeV and $\mT(\pt^{\Pell},\ptmiss)$ must be greater than 120\GeV, where \Pell denotes a \PGm or \PGt, respectively. The azimuthal angle difference between \ptvecmiss and the \ptvec of the muon or \tauh candidate must be greater than 2.8 to select events in which the lepton and neutrino are expected to be back-to-back. Finally, events with additional light leptons and jets are vetoed.

Modelling of the physics processes contributing to the control and measurement regions with genuine leptons (muon or \tauh) is based on simulation. Backgrounds with jets misidentified as genuine \tauh are estimated with the fake factor (\FF) method. The \FF is defined as the ratio of the probability for a jet to pass nominal \tauh identification criteria over the probability for it to pass some relaxed criteria but fail the nominal selection. We measure \FF as a function of the \tauh candidate \pt, as well as the \pt ratio between the \tauh candidate and the jet seeding the \tauh reconstruction. The measurement is performed with control samples of $\PW(\to\PGm\PGnGm)+\text{jet}$ and dijet events. Each \FF is used to extrapolate the misidentified jets background from a dedicated control region defined using the same selection as in the measurement region with the exception that the \tauh candidate is required to pass a relaxed identification criterion, but fail the nominal one. Estimate of the background from jets misidentified as \tauh is obtained by applying \FF's as weights to events in this control region.

The \tauh identification scale factors are extracted from the binned maximum likelihood fit applied to distributions of $\mT(\pt^{\Pell},\ptmiss)$ simultaneously in the control and two measurement regions, split by \tauh candidate \pt. The fit is performed with three unconstrained rate parameters. One of them simultaneously scales the yields of simulated $\PWst\to\PGm\PGnGm$ and $\PWst\to\PGt\PGnGt$ events in the control and measurement regions respectively. This rate parameter accounts for the fiducial cross section of \PWst production in the phase space probed by the measurement. The other two rate parameters scale the yield of simulated events with genuine \tauh separately in two measurement regions, $100<\pt<200\GeV$ and $\pt>200\GeV$. These two parameters are interpreted as the \tauh identification efficiency scale factors for two \pt bins.

Theoretical and instrumental uncertainties affecting the measurement are incorporated in the fit via penalty terms for additional nuisance parameters in the likelihood function.
The largest theoretical uncertainty is related to the modelling of the differential mass lineshape in the \PWst production. It is estimated following the prescription in Ref.~\cite{CMS:2015hmx}. The variation in the simulated shape of the \mWst distribution ranges from 2\% at $\mWst=200\GeV$ to 6\% for $\mWst>1\TeV$. Uncertainties in \FF modify yields of the misidentified jets background sample by between 10 and 20\% across different bins of the $\mT(\pt^{\tauh}, \ptmiss)$ distribution in the low-\pt measurement bin, and by between 20 and 30\% in the high-\pt measurement bin.

A conservative uncertainty of 5\% is assigned to the \tauh momentum scale and decorrelated across different \tauh decay modes. Uncertainties in the jet and unclustered energy scales affect the efficiency of the jet veto and reconstruction of \ptmiss. These uncertainties lead to variations in the simulated $\mT(\pt^{\tauh}, \ptmiss)$ distributions by between 5 and 10\%. Other uncertainties have a smaller impact on the measurement.

\begin{figure}
    \centering
    \includegraphics[width=\cmsFigWidth]{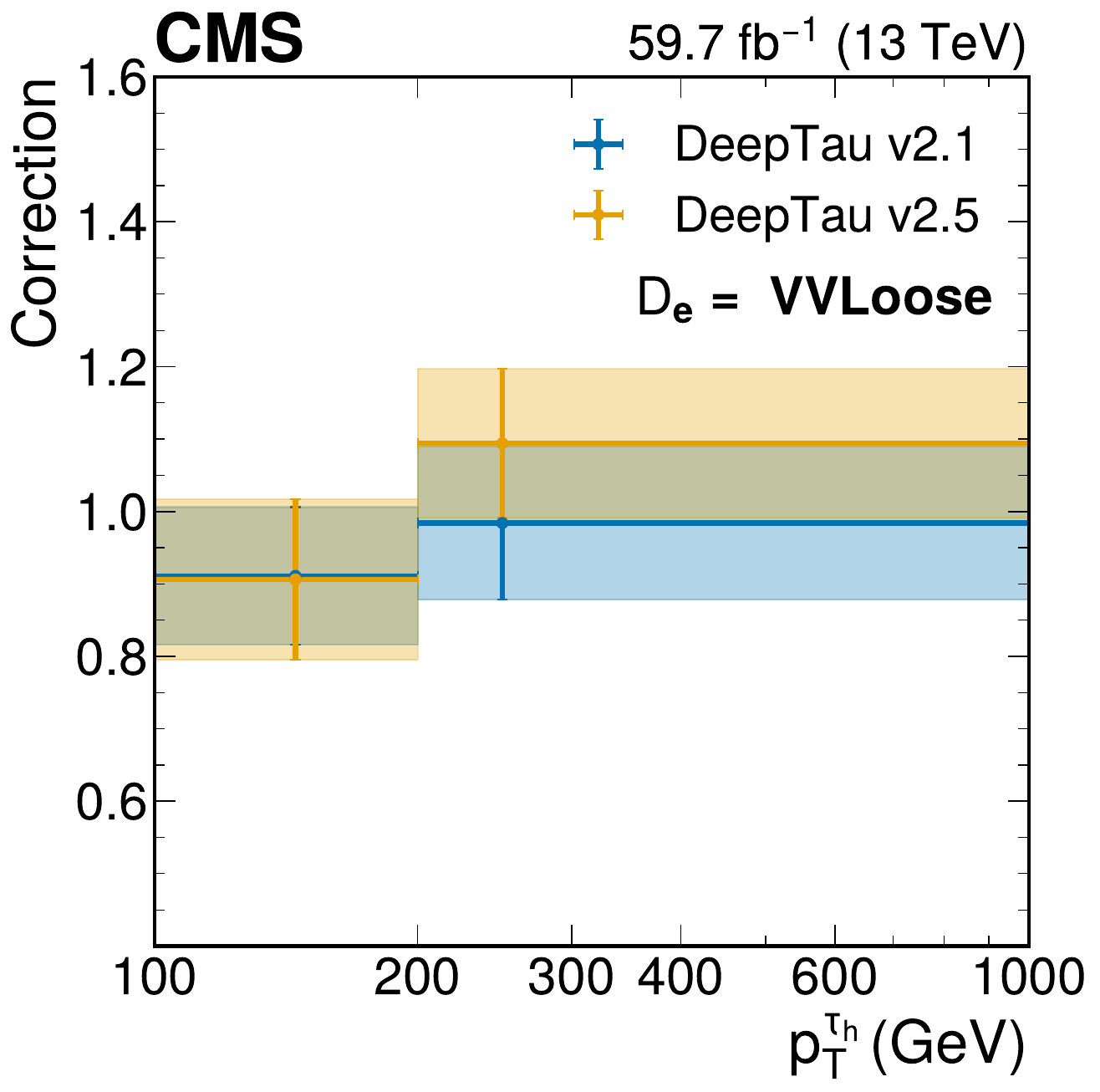}
    \includegraphics[width=\cmsFigWidth]{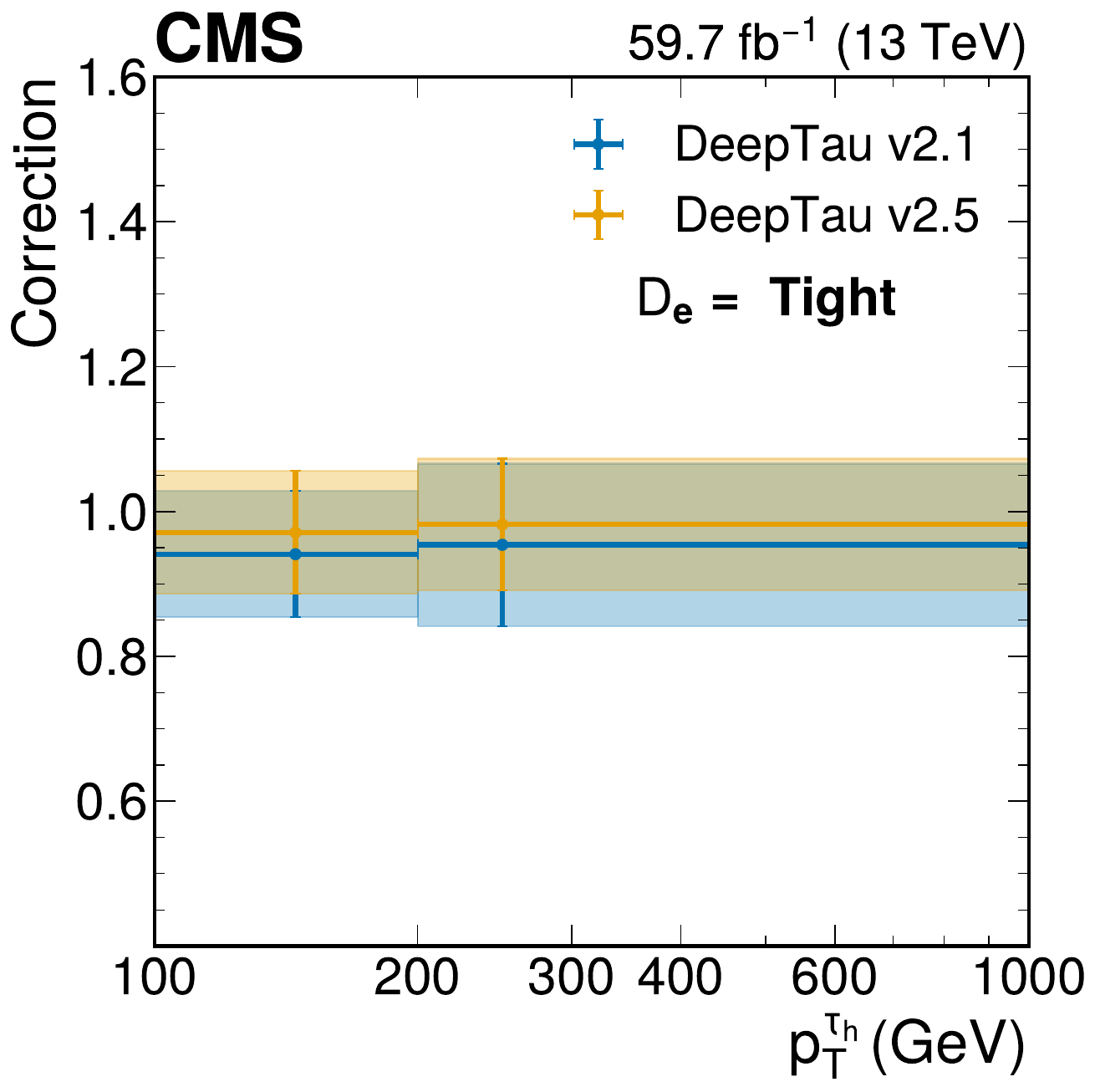}
    \includegraphics[width=\cmsFigWidth]{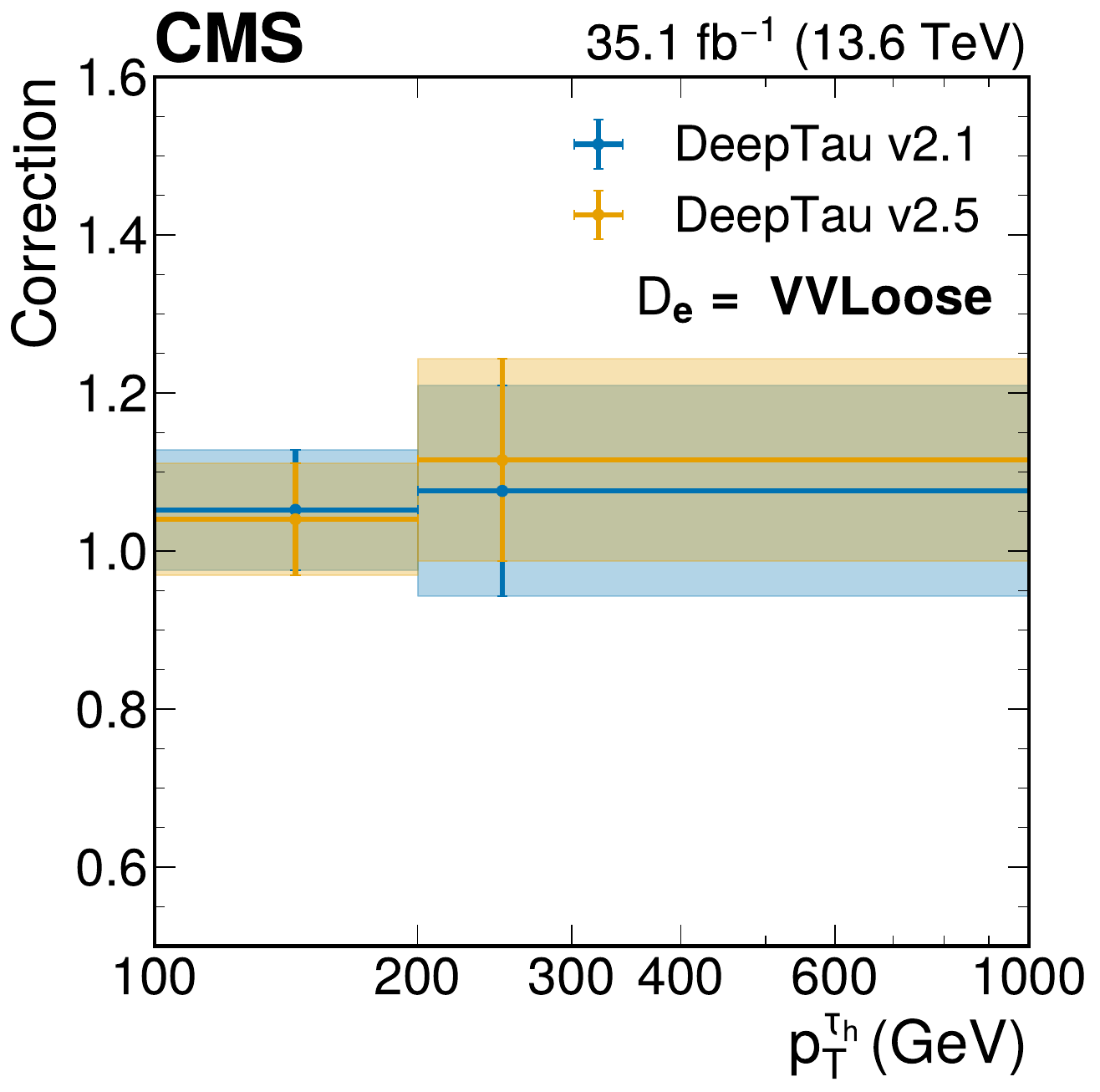}
    \includegraphics[width=\cmsFigWidth]{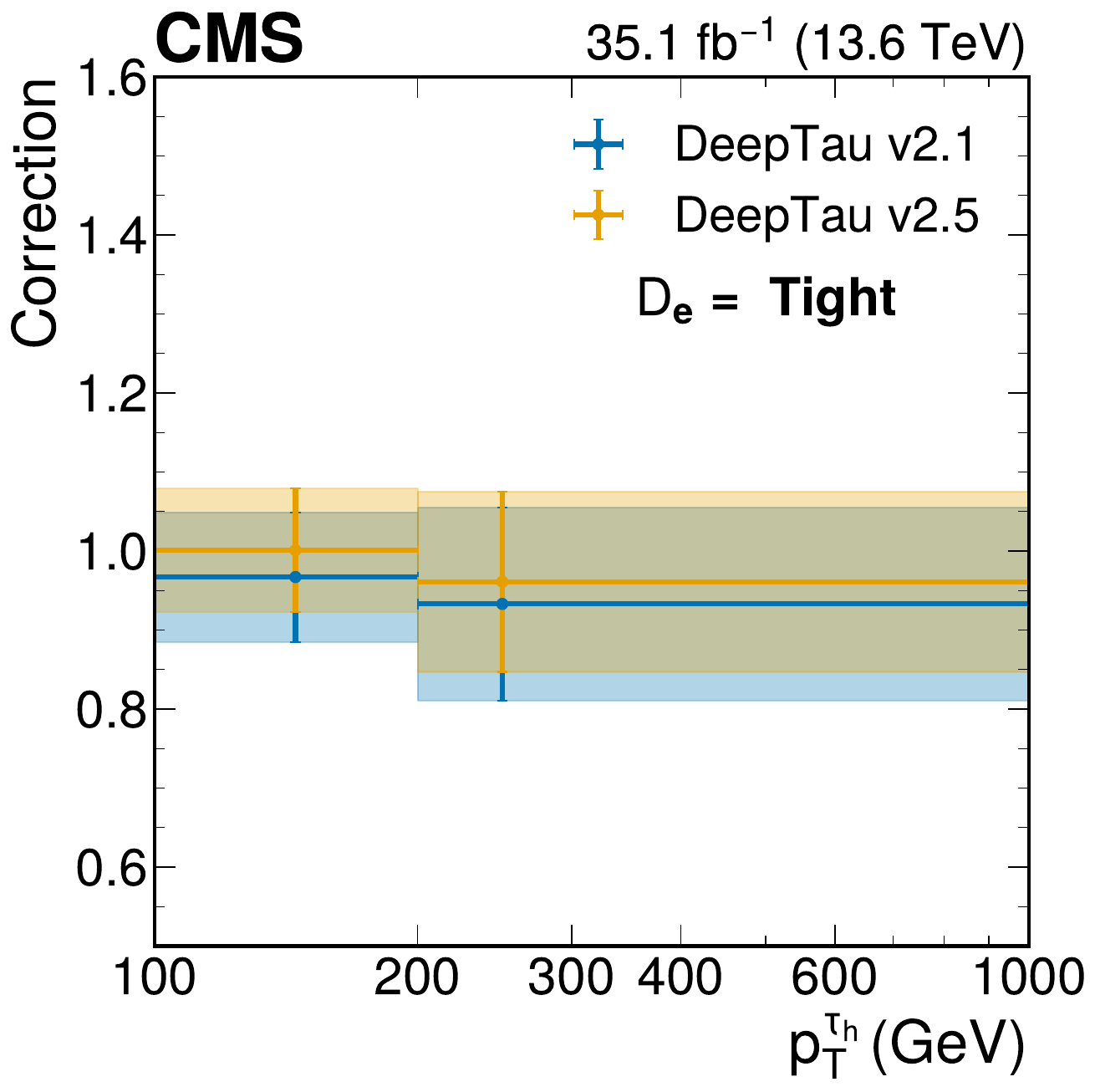}
    \caption{The high-\pt \tauh identification efficiency scale factors as a function of \tauh \pt for \Djet Medium, \Dm Tight and \De VVLoose (left) and Tight (right) discriminators (see Table~\ref{tab:wp-definition}). The scale factors are measured for the 2018 (top) and 2022 (bottom) datasets.}
    \label{fig:wtaunu_2022_sf}
\end{figure}

Measured scale factors are found to be consistent with unity within the measurement uncertainties, as shown in Fig.~\ref{fig:wtaunu_2022_sf}. For the \pt bin of $100<\pt<200\GeV$ ($\pt>200\GeV$) the uncertainty is 8--12\% (10--16\%) depending on the chosen working point of \DeepTau discriminator and data-taking period. It should be noted that the domain adaptation of the updated version of the \DeepTau tagger does not specifically target high-\pt tau leptons. As a consequence, equally good consistency of scale factors with unity is observed for both \DeepTau~v2.1 and \DeepTau~v2.5.

\begin{figure}
    \centering
    \includegraphics[width=\cmsFigWidth]{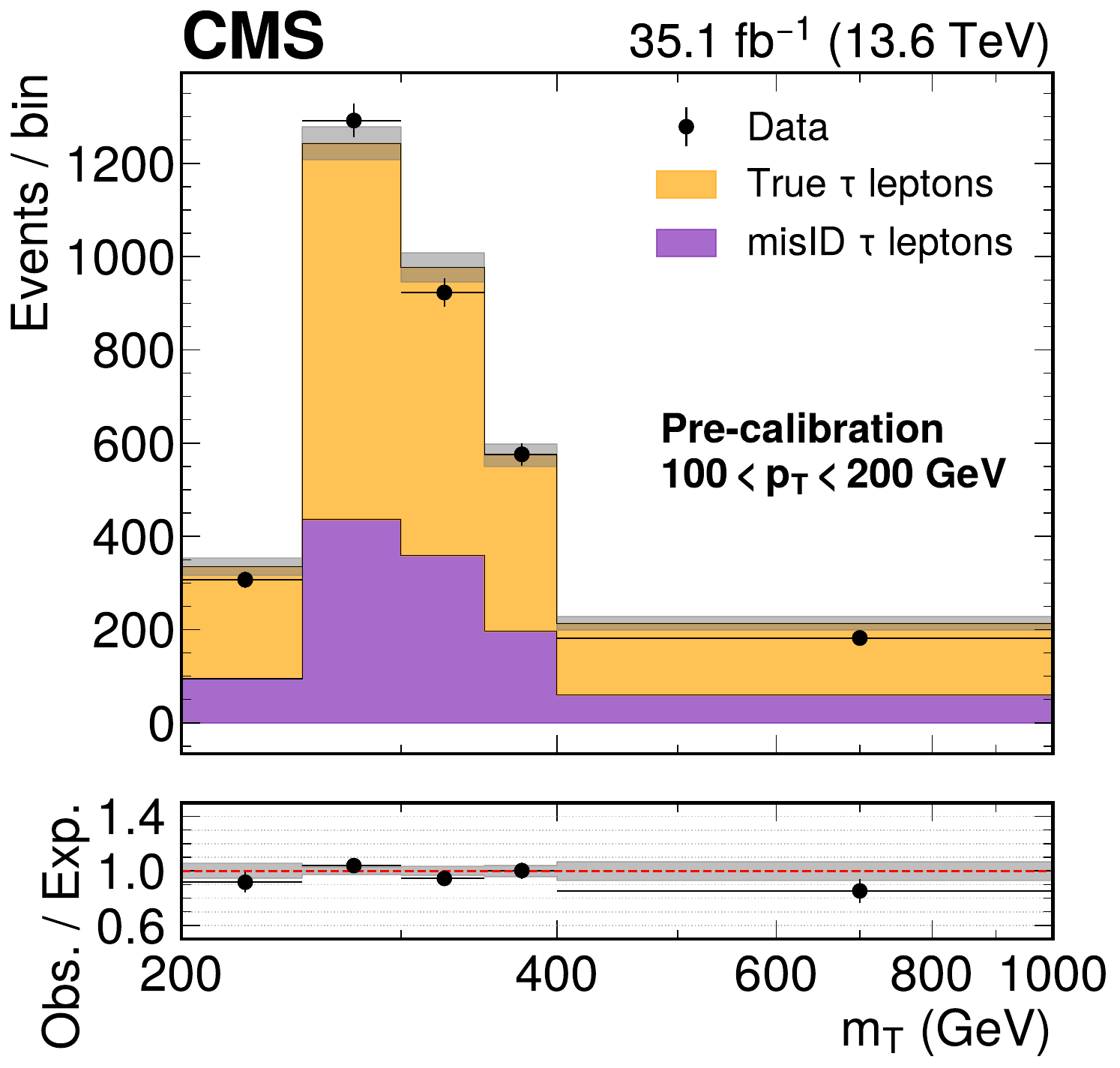}
    \includegraphics[width=\cmsFigWidth]{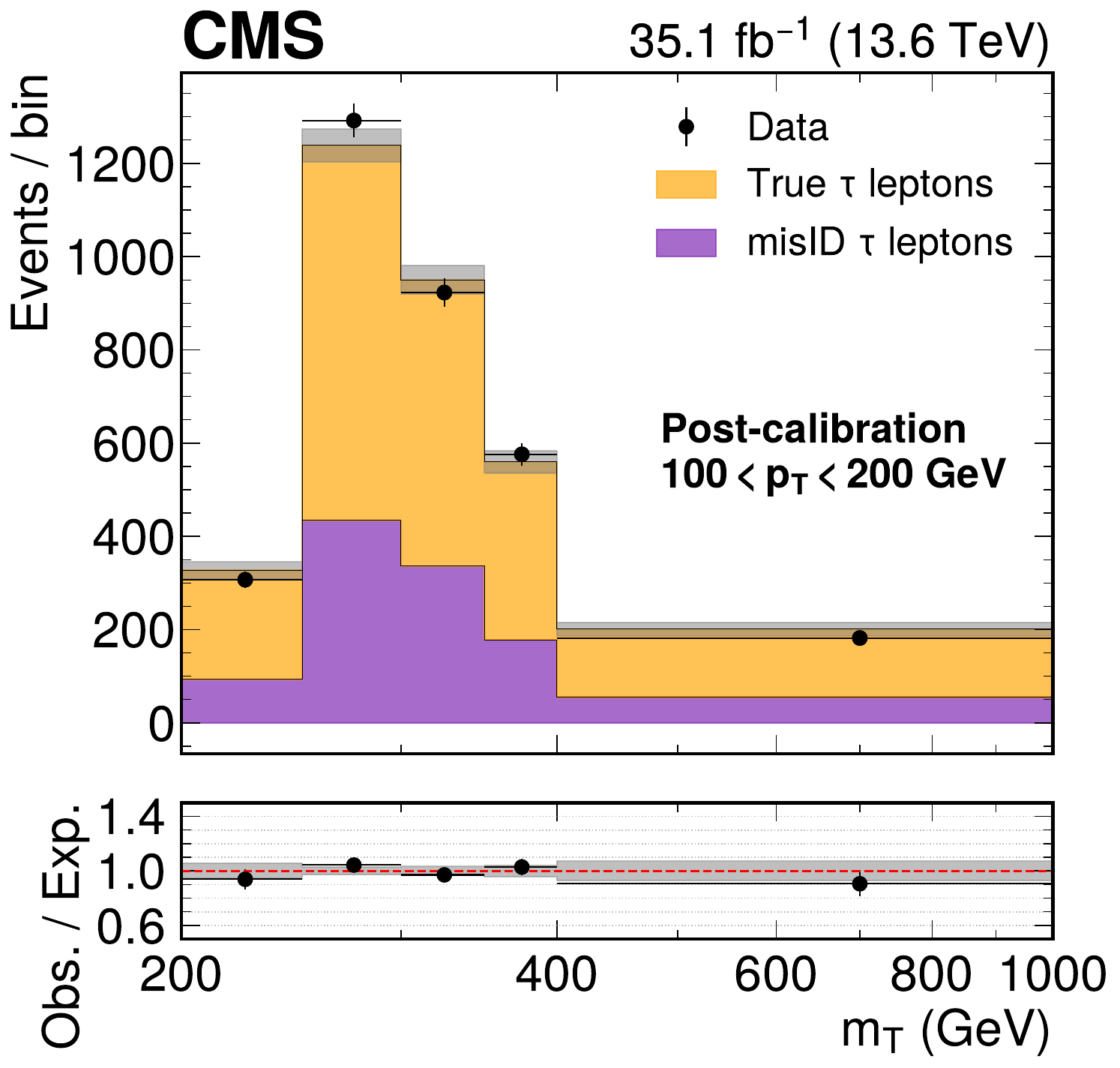}
    \includegraphics[width=\cmsFigWidth]{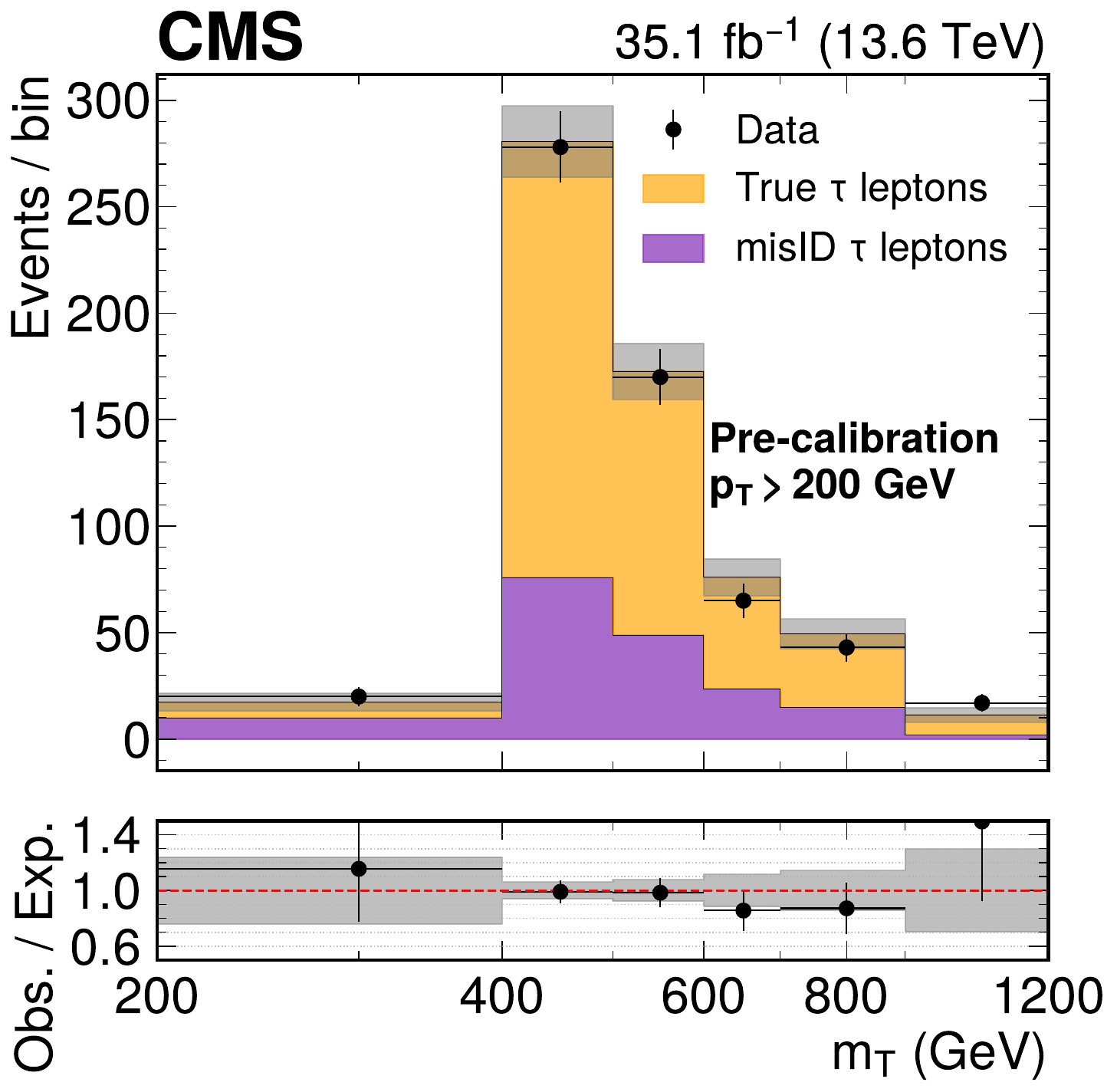}
    \includegraphics[width=\cmsFigWidth]{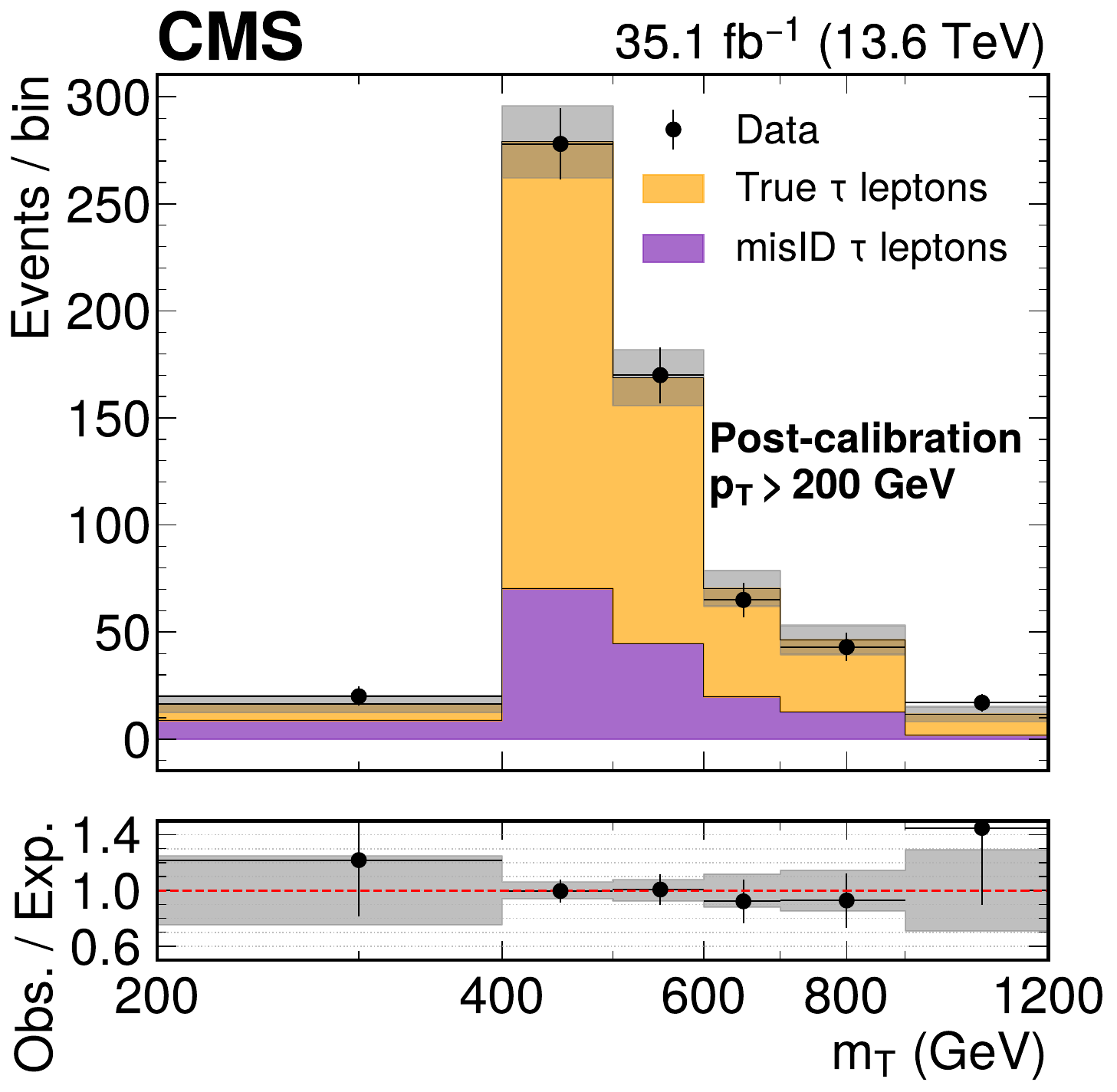}
    \caption{Prefit (left plots) and postfit (right plots) distribution of $\mT(\pt^{\tauh}, \ptmiss)$ for \pt bins of $100<\pt<200\GeV$ (upper plots) and $\pt>200\GeV$ (lower plots) in the 2022 dataset. Distributions are obtained for a combination of \Djet Medium, \Dm Tight and \De Tight discriminators (see Table~\ref{tab:wp-definition}).}
    \label{fig:wtaunu_2022}
\end{figure}

Figure~\ref{fig:wtaunu_2022} illustrates the measurement performed in the 2022 dataset. It presents distributions of $\mT(\pt^{\tauh}, \ptmiss)$ obtained before and after applying the maximum likelihood fit. Results are shown for a representative choice of the \DeepTau working points: Medium \Djet, Tight \Dm, and Tight \De. Good agreement between data and simulation is observed for high-\pt tau leptons before corrections, with dedicated scale factors providing further refinement.

\section{Summary}\label{sec:summary}

In this paper, the newly deployed version of the \DeepTau algorithm, v2.5, used to discriminate \tauh candidates from quark or gluon jets and electrons and muons, has been introduced.
This deep convolutional neural network exhibits improved performance with respect to its predecessor, reducing the jet misidentification rate by 30--50\% for a given \tauh reconstruction and identification efficiency.
The implementation of domain adaptation by backpropagation has reduced performance discrepancies between collision data and simulation, decreasing the necessary residual corrections by 5\%.
The domain adaptation was introduced by including an adversarial subnetwork in the gradient calculation of the neural network. This adversarial subnetwork was designed to discriminate between collision data and simulations, running in parallel with the \tauh classification task.
The \DeepTau algorithm, trained using both collision data and simulated samples, is able to maximize the \tauh classification performance, while minimizing the data-simulation discrepancies.

The \DeepTau~v2.5 algorithm was trained on simulated proton-proton collision data corresponding to the 2018 data-taking conditions, as well as on real collision data collected during the same year containing $\PZ \to \PGt\PGt$ decays, which was used for domain adaptation. The algorithm has been validated using 2018 and 2022 collision data. The observed \tauh efficiencies were found to agree with the expected efficiencies from simulated events within 10\% for 2018 and 15\% for 2022. This agreement is improved with respect to the previous iteration of the algorithm and confirms the effectiveness of domain adaptation. The algorithm has been introduced to be used in CMS physics analyses using data recorded from 2022 onwards.

\begin{acknowledgments}
We congratulate our colleagues in the CERN accelerator departments for the excellent performance of the LHC and thank the technical and administrative staffs at CERN and at other CMS institutes for their contributions to the success of the CMS effort. In addition, we gratefully acknowledge the computing centres and personnel of the Worldwide LHC Computing Grid and other centres for delivering so effectively the computing infrastructure essential to our analyses. Finally, we acknowledge the enduring support for the construction and operation of the LHC, the CMS detector, and the supporting computing infrastructure provided by the following funding agencies: SC (Armenia), BMBWF and FWF (Austria); FNRS and FWO (Belgium); CNPq, CAPES, FAPERJ, FAPERGS, and FAPESP (Brazil); MES and BNSF (Bulgaria); CERN; CAS, MoST, and NSFC (China); MINCIENCIAS (Colombia); MSES and CSF (Croatia); RIF (Cyprus); SENESCYT (Ecuador); ERC PRG, TARISTU24-TK10 and MoER TK202 (Estonia); Academy of Finland, MEC, and HIP (Finland); CEA and CNRS/IN2P3 (France); SRNSF (Georgia); BMFTR, DFG, and HGF (Germany); GSRI (Greece); NKFIH (Hungary); DAE and DST (India); IPM (Iran); SFI (Ireland); INFN (Italy); MSIT and NRF (Republic of Korea); MES (Latvia); LMTLT (Lithuania); MOE and UM (Malaysia); BUAP, CINVESTAV, CONACYT, LNS, SEP, and UASLP-FAI (Mexico); MOS (Montenegro); MBIE (New Zealand); PAEC (Pakistan); MES, NSC, and NAWA (Poland); FCT (Portugal); MESTD (Serbia); MICIU/AEI and PCTI (Spain); MOSTR (Sri Lanka); Swiss Funding Agencies (Switzerland); MST (Taipei); MHESI (Thailand); TUBITAK and TENMAK (T\"{u}rkiye); NASU (Ukraine); STFC (United Kingdom); DOE and NSF (USA).

\hyphenation{Rachada-pisek} Individuals have received support from the Marie-Curie programme and the European Research Council and Horizon 2020 Grant, contract Nos.\ 675440, 724704, 752730, 758316, 765710, 824093, 101115353, 101002207, 101001205, and COST Action CA16108 (European Union); the Leventis Foundation; the Alfred P.\ Sloan Foundation; the Alexander von Humboldt Foundation; the Science Committee, project no. 22rl-037 (Armenia); the Fonds pour la Formation \`a la Recherche dans l'Industrie et dans l'Agriculture (FRIA) and Fonds voor Wetenschappelijk Onderzoek contract No. 1228724N (Belgium); the Beijing Municipal Science \& Technology Commission, No. Z191100007219010, the Fundamental Research Funds for the Central Universities, the Ministry of Science and Technology of China under Grant No. 2023YFA1605804, the Natural Science Foundation of China under Grant No. 12061141002, 12535004, and USTC Research Funds of the Double First-Class Initiative No.\ YD2030002017 (China); the Ministry of Education, Youth and Sports (MEYS) of the Czech Republic; the Agence Nationale de la Recherche grant ANR-22-CE31-0002 and the Interdisciplinary Thematic Institute in Quantum Science and Nanomaterials (QMat) (France); the Shota Rustaveli National Science Foundation, grant FR-22-985 (Georgia); the Deutsche Forschungsgemeinschaft (DFG), among others, under Germany's Excellence Strategy -- EXC 2121 ``Quantum Universe" -- 390833306, and under project number 400140256 - GRK2497; the Hellenic Foundation for Research and Innovation (HFRI), Project Number 2288 (Greece); the Hungarian Academy of Sciences, the New National Excellence Program - \'UNKP, the NKFIH research grants K 131991, K 133046, K 138136, K 143460, K 143477, K 146913, K 146914, K 147048, 2020-2.2.1-ED-2021-00181, TKP2021-NKTA-64, and 2021-4.1.2-NEMZ\_KI-2024-00036 (Hungary); the Council of Science and Industrial Research, India; ICSC -- National Research Centre for High Performance Computing, Big Data and Quantum Computing, FAIR -- Future Artificial Intelligence Research, and CUP I53D23001070006 (Mission 4 Component 1), funded by the NextGenerationEU program (Italy); the Latvian Council of Science; the Ministry of Education and Science, project no. 2022/WK/14, and the National Science Center, contracts Opus 2021/41/B/ST2/01369, 2021/43/B/ST2/01552, 2023/49/B/ST2/03273, and the NAWA contract BPN/PPO/2021/1/00011 (Poland); the Funda\c{c}\~ao para a Ci\^encia e a Tecnologia, grant CEECIND/01334/2018 (Portugal); the National Priorities Research Program by Qatar National Research Fund;  MICIU/AEI/10.13039/501100011033, ERDF/EU, "European Union NextGenerationEU/PRTR", and Programa Severo Ochoa del Principado de Asturias (Spain); the Chulalongkorn Academic into Its 2nd Century Project Advancement Project, the National Science, Research and Innovation Fund program IND\_FF\_68\_369\_2300\_097, and the Program Management Unit for Human Resources \& Institutional Development, Research and Innovation, grant B39G680009 (Thailand); the Kavli Foundation; the Nvidia Corporation; the SuperMicro Corporation; the Welch Foundation, contract C-1845; and the Weston Havens Foundation (USA).    
\end{acknowledgments}

\bibliography{auto_generated}

\appendix
\numberwithin{figure}{section}
\numberwithin{table}{section}

\section{Loss Function}
\label{loss_appendix}

The loss function is defined as: 
\begin{equation}\begin{aligned}\label{class_loss}
   L(\textbf{y}^\text{true}, \textbf{y}^\text{pred})
   =& \underbrace{\kappa_\PGt H_\PGt(\textbf{y}^\text{true}, \textbf{y}^\text{pred}; {\omega})}
       _{\text{Separation for all }\alpha} \\
   &+ \underbrace{(\kappa_\Pe + \kappa_\PGm + \kappa_{j}) \overline{F}_\text{cmb}(1-y_\PGt^\text{true}, 1-y_\PGt^\text{pred}; \gamma_\text{cmb})}
       _{\text{Focused separation of }\Pe, \PGm,\text{ jet from \tauh}} \\
   &+ \underbrace{\kappa_F \sum_{i\in \{\Pe, \PGm, j\}}\kappa_i \hat{\theta}(y_\PGt - 0.1) \overline{F}_i(y_i^\text{true}, y_i^\text{pred}; \gamma_i)}
       _{\text{Focused separation of \tauh from }\Pe, \PGm,\text{ jet for }y_\PGt>0.1}, 
\end{aligned}\end{equation}
where $\textbf{y}^\text{pred}$ and $\textbf{y}^\text{true}$ are the predictions and generator-level truth, respectively; $H_\PGt$ is the categorical cross-entropy loss with ${\omega}$ a varying parameter for sample normalization; $\overline{F}_x$ is the normalized focal loss; $\hat{\theta}$ is a smoothened step function that approaches 1 for $y_\PGt>0.1$ and 0 for $y_\PGt<0.1$. This step function disregards the discrimination between \Pe, \PGm, and jets when the probability of a genuine \tauh is low. The default values of the $\kappa$ and $\gamma$ terms are given in \autoref{tab:loss_cst}.

The $\kappa$ values affect the relative importance of predicting each class correctly. During the domain adaptation training, these constants were set to $\kappa_\Pe=2$, $\kappa_\PGm=5$, $\kappa_\PGt=6$, and $\kappa_{j}=1$ in order to reduce the degraded performance of \tauh classification.

\begin{table}[htb]
    \centering
	\topcaption{Default values of the parameters used in the classification loss function for \DeepTau training.}
	\begin{tabular}{lcl}
		Parameter & Value & Emphasis on \\
		\hline
		$\kappa_\Pe$   & 0.4 & \Pe separation\\
		$\kappa_\PGm$  & 1.0 & \PGm separation\\
		$\kappa_\PGt$  & 2.0 & \tauh separation\\
		$\kappa_{j}$   & 0.6 & Jet separation\\
		[\cmsTabSkip]
		$\kappa_F$     & 5.0 & High \tauh identification efficiency\\
		[\cmsTabSkip]
		$\gamma_\Pe$   & 2.0 &  \Pe separation\\
		$\gamma_\PGm$  & 2.0 &  \PGm separation\\
		$\gamma_{j}$   & 2.0 &  Jet separation\\
		$\gamma_\text{cmb}$ & 0.5 &  \Pe, \PGm, jet separation combined\\
	\end{tabular}
    \label{tab:loss_cst}
\end{table}

\cleardoublepage \section{The CMS Collaboration \label{app:collab}}\begin{sloppypar}\hyphenpenalty=5000\widowpenalty=500\clubpenalty=5000
\cmsinstitute{Yerevan Physics Institute, Yerevan, Armenia}
{\tolerance=6000
A.~Hayrapetyan, V.~Makarenko\cmsorcid{0000-0002-8406-8605}, A.~Tumasyan\cmsAuthorMark{1}\cmsorcid{0009-0000-0684-6742}
\par}
\cmsinstitute{Institut f\"{u}r Hochenergiephysik, Vienna, Austria}
{\tolerance=6000
W.~Adam\cmsorcid{0000-0001-9099-4341}, J.W.~Andrejkovic, L.~Benato\cmsorcid{0000-0001-5135-7489}, T.~Bergauer\cmsorcid{0000-0002-5786-0293}, M.~Dragicevic\cmsorcid{0000-0003-1967-6783}, C.~Giordano, P.S.~Hussain\cmsorcid{0000-0002-4825-5278}, M.~Jeitler\cmsAuthorMark{2}\cmsorcid{0000-0002-5141-9560}, N.~Krammer\cmsorcid{0000-0002-0548-0985}, A.~Li\cmsorcid{0000-0002-4547-116X}, D.~Liko\cmsorcid{0000-0002-3380-473X}, M.~Matthewman, I.~Mikulec\cmsorcid{0000-0003-0385-2746}, J.~Schieck\cmsAuthorMark{2}\cmsorcid{0000-0002-1058-8093}, R.~Sch\"{o}fbeck\cmsAuthorMark{2}\cmsorcid{0000-0002-2332-8784}, D.~Schwarz\cmsorcid{0000-0002-3821-7331}, M.~Shooshtari\cmsorcid{0009-0004-8882-4887}, M.~Sonawane\cmsorcid{0000-0003-0510-7010}, W.~Waltenberger\cmsorcid{0000-0002-6215-7228}, C.-E.~Wulz\cmsAuthorMark{2}\cmsorcid{0000-0001-9226-5812}
% ...
\par}
\vskip\cmsinstskip
\dag:~Deceased\\
$^{1}$Also at Yerevan State University, Yerevan, Armenia\\
$^{2}$Also at TU Wien, Vienna, Austria\\
% ...
\end{sloppypar}
\end{document}